\documentclass[reqno]{amsart}
\usepackage{algorithm}
\usepackage{booktabs}
\usepackage[subrefformat=parens,labelformat=parens]{subcaption}

\usepackage[english]{babel}
\usepackage{graphicx}
\usepackage[breaklinks=true,colorlinks,linkcolor={blue},citecolor={red},urlcolor={purple},]{hyperref}
\usepackage{amssymb}
\usepackage{amsmath}
\usepackage[foot]{amsaddr}
\usepackage{amsfonts}
\usepackage{mathtools}
\usepackage{algpseudocode}
\usepackage{algorithm}
\usepackage[top=1in, bottom=1.25in, left=1.25in, right=1.25in]{geometry}
\usepackage [autostyle, english = american]{csquotes}
\MakeOuterQuote{"}
\usepackage{overpic}

\usepackage{verbatim}
\usepackage{mathrsfs}
\usepackage[svgnames]{xcolor}
\usepackage{dsfont}

\usepackage{esvect}
\usepackage{enumitem}

\usepackage{gensymb} % Pro­vides generic com­mands \de­gree, \cel­sius, \pert­hou­sand, \mi­cro and \
\usepackage{physics}

\newtheoremstyle{remarkStyle}{\parskip}{}{}{}{\bfseries}{.}{.5em}{}
\theoremstyle{plain} 
\theoremstyle{plain} 
\theoremstyle{remarkStyle} 
\theoremstyle{remarkStyle} 
\theoremstyle{plain} 
\theoremstyle{plain} 
\theoremstyle{plain} 
\theoremstyle{remarkStyle}

\graphicspath{{images/}}
\usepackage{float}

\usepackage[sort,nocompress]{cite}
\usepackage{todonotes}
\usepackage{multirow}
\usepackage{booktabs}
\def\rb{\mathrm{rb}}

\def\train{\mathrm{train}}
\def\test{\mathrm{test}}
\def\mubold{\boldsymbol{\mu}}

\def\tawss{\mathrm{TAWSS}}
\def\osi{\mathrm{OSI}}
\def\R{\mathbb{R}}

\usepackage[foot]{amsaddr}
\makeatletter
\renewcommand{\email}[2][]{%
  \ifx\emails\@empty\relax\else{\g@addto@macro\emails{,\space}}\fi%
  \@ifnotempty{#1}{\g@addto@macro\emails{\textrm{(#1)}\space}}%
  \g@addto@macro\emails{#2}%
}
\makeatother

\begin{document}
\title[A Reduced Order Model formulation for left atrium flow]{A Reduced Order Model formulation for left atrium flow: an Atrial Fibrillation case}
% \author{Author 1}
% \author{Author 2}
% \address[A1,A2]{Addresses of Author 1 and 2}
% \email[A1,A2]{emails of Authors 1 and 2}

% \author{Author 3}
% \author{Author 4}
% \address[A3,]{Addresses of Author 3 and 4}
% \email[A3,A4]{emails of Authors 3 and 4}
% \email[A5,A6]{emails of Authors 3 and 4}
\author{Caterina Balzotti\textsuperscript{*}}
\author{Pierfrancesco Siena\textsuperscript{*}}
\author{Michele Girfoglio\textsuperscript{*}}
\author{Giovanni Stabile\textsuperscript{*}}
\author{Jorge Due\~{n}as-Pamplona\textsuperscript{$\dagger$}}
\author{Jos\'e Sierra-Pallares\textsuperscript{$\ddagger$}}
\author{Ignacio Amat-Santos\textsuperscript{$\ddagger$}\textsuperscript{$\S$}}
\author{Gianluigi Rozza\textsuperscript{*}}
\address[*]{Scuola Internazionale Superiore di Studi Avanzati (SISSA), Mathlab, Italy}
\address[$\dagger$]{Departamento de Ingeniería Energética, Universidad Politécnica de Madrid, Spain}
\address[$\ddagger$]{Departamento de Ingeniería Energética y Fluidomecánica, Universidad de Valladolid, Spain}
\address[$\S$]{Clinical University Hospital of Valladolid, Spain}
\email[*]{cbalzott@sissa.it, psiena@sissa.it, mgirfogl@sissa.it, gstabile@sissa.it, grozza@sissa.it}
\email[$\dagger$]{jorge.duenas.pamplona@upm.es}
\email[$\ddagger$]{jsierra@uva.es}
\email[$\S$]{ijamat@gmail.com}

\begin{abstract}
A data-driven Reduced Order Model (ROM) based on a Proper Orthogonal Decomposition - Radial Basis Function (POD-RBF) approach is adopted in this paper for the analysis of blood flow dynamics in a patient-specific case of Atrial Fibrillation (AF). The Full Order Model (FOM) is represented by incompressible Navier-Stokes equations, discretized with a Finite Volume (FV) approach. Both the Newtonian and the Casson's constitutive laws are employed. The aim is to build a computational tool able to efficiently and accurately reconstruct the patterns of %primary variables \textcolor{red}{Michele: which? pressure and velocity? To be specified} and 
relevant hemodynamics indices related to the stasis of the blood in  % including the moments of the age, the washout, the wall shear stress and the oscillating shear index. 
a physical parametrization framework including the cardiac output in the Newtonian case and also %inlet velocity, 
the plasma viscosity and the hematocrit in the non-Newtonian one. %The main interest of this work related to the use of the FV method in combination with a data-driven ROM for a complex biomedical engineering involving blood flow in a chardiac chamber, which exhibits more difficulties with respect to vessels-like domains such coronary or aorta arteries. 
Many FOM-ROM comparisons are shown to analyze the performance of our approach as regards errors and computational speed-up. \\ \\%\vspace{0.1cm}
\textbf{Keywords}: Reduced order model, haemodynamics, cardiovascular flows, left atrium, data-driven techniques, patient-specific configurations.
\end{abstract}

\maketitle

\section{Introduction and motivation}
\label{sec:intro}
Atrial Fibrillation (AF) is the most common type of cardiac arrhythmia, affecting around 1.5\% of the population (more than 35 million people worldwide \cite{Benjamin2019}). Its incidence seems to be correlated with age, since 8\% of individuals above 80 years old are affected \cite{Go2001}. An AF episode causes the Left Atrium (LA) to contract in an irregular and ineffective way, being typically triggered by irregular electrical impulses coming from the Pulmonary Vein (PV) roots. This abnormal contraction pattern seems to be related with stroke incidence due to thrombus  formation within the Left Atrial Appendage (LAA) \cite{Wolf1991}. 

The LAA is a cavity which results from LA embryonic development, having a protruding and trabeculated morphology \cite{AlSaady1999}. When a patient develops AF, the LAA natural contractility reduces dramatically, making it prone to thrombi formation \cite{Seo2016, Goette2016}. Due to this, the LAA has received a lot of attention from both clinical and biomedical engineering fields. Although a large number of studies have attempted to delve deeper into the relationship between stroke risk and LAA morphology in AF patients, the underlying mechanisms are still not well understood. Some of these previous studies have suggested certain morphologies to be associated with a lower risk of stroke \cite{Yaghi2020, DiBiase2012, Lee2015}, while others could not find a correlation between the two phenomena \cite{Khurram2013, Nedios2014}. Other clinical studies associated stroke risk with certain atrial geometric parameters such as LAA volume \cite{Korhonen2015, Beinart2011}, LAA ostium area \cite{Lee2015, Lee2017, Beinart2011}, number of lobes \cite{Yamamoto2014}, and LAA depth \cite{Beinart2011}. Recent studies also drew attention to the importance of the atrial flow \cite{Lee2017}, and the PV morphology \cite{Polaczek2019}. Although everything seems to point to the existence of a mechanistic relationship between the thrombosis risk and the LA anatomy and flow, this one remains unknown at this time.

Recent advances in medical imaging have made it possible to apply Computational Fluid Dynamics (CFD) techniques to the study of LA flows. Early research on LA flow dynamics were part of whole left-heart simulations \cite{Chnafa2014, Vedula2015}. More recent CFD studies have presented numerical analyses of the LA flow patterns \cite{Chnafa2014, Otani2016,Lantz2019}, focused on investigating blood stasis, as it is considered a necessary thrombogenic factor. Some of them also studied specifically the LAA stasis in AF conditions \cite{Masci2019, GarciaIsla2018, Bosi2018, GarciaVillalba2021, DuenasPamplona2021a, DuenasPamplona2021}. All of them have helped to provide insights into the AF phenomenon and calculate otherwise inaccessible parameters such as residence times and shear stress that are related to blood stasis. Other studies have showed that the residence time and flow patterns for flexible-wall and rigid-wall simulations are very similar in case of impaired atrial function, especially when both the reservoir and booster functions are decreased \cite{GarciaVillalba2021,DuenasPamplona2021a}. An interesting approach is used by Dueñas-Pamplona et al. \cite{DuenasPamplona2022}, who developed a morphing technique to study the risk of long-term stasis due to geometrical parameters. To reproduce the most critical AF case (in the absence of atrial contraction) the atrium was kept rigid, regardless of the atrial function at the time of medical imaging. Results show the enormous influence of cardiac output in the blood age indices, and the relatively minor role played by the PV orientation. To our knowledge, this is the only study that attempts to parametrize the LA flow problem. %, including geometrical parameters employing a rigging and morphing technique. 
On the other hand, Gonzalo et al. \cite{Gonzalo2022} have drawn attention to the fact that the non-Newtonian blood rheology can impact the left atrial stasis in patient-specific simulations, submitting that hematocrit-dependent non-Newtonian blood rheology should be considered when calculating patient-specific blood stasis indices by CFD \cite{Gonzalo2022}. 

Reduced Order Models (ROMs) \cite{hesthaven2016book,benner2020book,degruyter2,degruyter1,aromabook} have become increasingly important in hemodynamics applications due to the complex and multiscale nature of the cardiovascular system. ROM techniques allow for the creation of simplified models able to capture the essential features of the system and to significantly reduce the computational cost with respect to the standard CFD models based on classic discretization techniques, such Finite Volume (FV) or Finite Elements (FE) (hereinafter referred to as Full Order Model (FOM)). %These reduced models can be used to simulate and analyze various cardiovascular phenomena such as blood flow, arterial stiffness, and cardiac function. 
ROMs can also aid in the development of personalized medicine, as they enable the creation of patient-specific models that can help clinicians to better understand and treat cardiovascular diseases. 

Classic projection-based ROMs were already %Both intrusive and non-intrusive ROMs are already 
adopted to speed up patient-specific cases or idealized cardiovascular benchmarks. A Proper Orthogonal Decomposition (POD)-Galerkin strategy is employed in combination with a FV full order solver in \cite{buoso2019reduced}. The aim is to predict the pressure drop along an idealized vessel in a geometrical parameter setting. The same ROM approach but within a FE environment is adopted in \cite{zainib2021reduced,ballarin2016fast,ballarin2017numerical} for the study of the blood flow patterns in patient-specific configurations of % a patient-specific configuration of 
coronary artery bypass grafts where a physical parametrization involving the Reynolds number as well as an efficient centerlines-based geometrical parametrization are employed. %Within the same context, in \cite{zainib2021reduced%Otherwise, the ROM is combined with the finite element method in an optimal control framework, and the physical parametrization involves the Reynolds number. Many geometries of coronary arteries are analyzed also in \cite{ballarin2016fast,ballarin2017numerical}, where an efficient centerlines-based
%parametrization is added in a POD-Galerkin and finite element environment. 

More recently, the combination of data-driven ROMs and FV method is becoming particularly appealing, due to the diffusion in the biomedical engineering community of commercial codes relying on FV schemes: see, e.g., \cite{caruso2015computational,vignali2021fully,benim2011simulation}. In \cite{girfoglio2022non} a POD with interpolation by Radial Basis Function (RBF) approach is adopted for the investigation of the hemodynamics in the aortic arch in presence of a left ventricular assist device. The authors report a speed up of $\mathscr{O}(10^{6})$ associated with an error less than $15\%$ which represents a promising result in such a direction. For this reason, in \cite{balzottidata2022, siena2023fast,siena2023data} a similar approach where the interpolation is carried out by feed-forward Neural Network instead of RBF is proposed for the analysis of the blood flow patterns in coronary artery bypass grafts. While in \cite{balzottidata2022} only one physical parameter is introduced (the Reynolds number), in \cite{siena2023data} a geometrical parametrization setting (with respect to the diameter of an isolated stenosis) is also considered. In such works, a speed up of $\mathscr{O}(10^{5})$ and an average error below $5\%$ are obtained. % even if it requires deep NNs to obtain satisfactory results.%the authors propose a similar approach, the POD-Neural Network (NN) one, based on a combination of machine learning techniques. 

Concerning the problem addressed in this paper, some recent works \cite{saiz2022unsupervised,pons2022joint} use machine learning-based models to infer LAA blood stasis from LAA geometry stasis. Their goal is to elaborate the huge amount of information coming from the data in order to identify patients with AF at the highest risk of thrombus formation. Unlike these works, mainly based on data processing, our aim is to retain the physical problem and build a cooperation between ROM, CFD and data-driven techniques in order to build an efficient computational tool able to achieve faithful solutions.
Following this research line and its encouraging results, the present work tries to extend the use of data-driven ROM approaches to the study of the blood flow in the LAA portion in a physical parametrization setting. This is of course a more complex case and a step forward with respect to vessel-like structures addressed in the previous works \cite{balzottidata2022,siena2023data,siena2023fast,girfoglio2021non,girfoglio2022non}. %The strong nonlinear behavior of the solutions can be problematic for classic ROMs, however, in our opinion, a data-driven ROM can lead to good results. 
Indeed, to the best of our knowledge, this work represents the first study about the application of ROM to the haemodynamics in the LA. For sake of completeness, we mention \cite{fresca2021pod} where a ROM for the propagation of the electrical signal in the heart is analyzed. However, % The authors overcome the issue related to nonlinear behavior of the solutions with a deep learning ROM. Furthermore, 
the authors do not address the fluid dynamics of the problem and furthermore use an idealized LA domain while a patient-specific one is used in this work, which represents an additional difficulty introduced in our research.
%Another fundamental difference with \cite{fresca2021pod} is their use of an idealized LA domain, while a patient-specific one represents an additional difficulties introduced in our work.

%\cate{Finally, in the recent works \cite{saiz2022unsupervised,pons2022joint} the authors propose machine learning-based models to infer LAA blood stasis from LAA geometry stasis. The goal of these studies is to elaborate the huge amount of information coming from the data in order to identify patients with AF at the highest risk of thrombus formation.}

%\gs{I believe in the introduction is missing something ROM for cardiovascular applications. The first part I think it is perfect, the problem is summarized from a medical point of view, then CFD is works are recalled. In the last part we should go from CFD to ROMs, report similar efforts and summarize what we do in this journal.}
% ROM stuff

The paper is organized as follows. In Section~\ref{sec:problem_formulation} and \ref{sec:rom}, the FOM and the ROM adopted are explained in detail, along with the hemodynamics indices of interest and the techniques chosen for each algorithm. 
Then, Section \ref{sec:numerical} is reserved for error and efficiency analysis of our ROM approach. It also includes some FOM-ROM qualitative comparisons and clinical considerations on the patterns obtained. Finally, Section \ref{sec:conclusions} is dedicated to draw conclusions and some possible extensions of this work. % to improve furthermore the results.
\section{The Full Order Model}%
\label{sec:problem_formulation}

%\subsection{The full order model}\label{sec:fom}
The FOM employed in this paper is similar to the one proposed by Dueñas-Pamplona et al.\ in \cite{DuenasPamplona2021}. It consists of parametrized Navier-Stokes equations for the blood flow in a patient specific domain $\Omega$ %=\Omega(\boldsymbol{\mu})$ 
over a time interval of interest $(t_0,T]$:%, considering the blood as a Newtonian fluid:
\begin{equation}
	\begin{cases} 
	\rho \partial_t \mathbf v (\mathbf{x}, t; \boldsymbol{\mu})+ \rho \nabla \cdot (\mathbf v  (\mathbf{x}, t; \boldsymbol{\mu}) \otimes \mathbf v  (\mathbf{x}, t; \boldsymbol{\mu})) - \nabla \cdot \mathbb T  (\mathbf{x}, t; \boldsymbol{\mu}) = 0   & \quad \text{in} \quad \Omega \times (t_0,T],\\ %  & \quad \text{in} \quad \Omega\times (0,T], \\
	\nabla \cdot \mathbf v  (\mathbf{x}, t; \boldsymbol{\mu}) = 0 & \quad \text{in} \quad \Omega \times (t_0, T], %, \\
    %\mathbf v = \mathbf{v}_{in}, &\quad \text{on} \quad \Gamma_i, \\
    %\nabla \mathbf{v} \cdot \mathbf{n} = 0 \quad \text{on} \quad \partial  \Omega \smallsetminus \Gamma_i%& \quad \text{in} \quad \Omega\times (0,T],	
	\end{cases}  \label{N-S}
\end{equation}	
%\begin{eqnarray}
%   \label{eq:NS1}\nabla \cdot \mathbf{v}  &=& 0, \quad \mbox{on} \quad \Omega, \\
%    \label{eq:NS2}\frac{\partial \mathbf{v}}{\partial t} + \mathbf{v} \cdot \nabla \mathbf{v} &=& - \frac{\nabla p}{\rho } + \frac{\mu}{\rho}\nabla^2 \mathbf{v} \quad \mbox{on} \quad \Omega.
%\end{eqnarray}
where $\mathbf{v} = \mathbf{v}(\mathbf{x}, t ;\boldsymbol{\mu})$ and $p = p(\mathbf{x}, t;\boldsymbol{\mu})$ are the velocity and the pressure depending on the physical parameter vector $\boldsymbol{\mu}$, respectively. In addition, $\rho = 1050$ kg/m$^3$ is the blood density. Problem \eqref{N-S} is endowed with initial data $\mathbf{v}(\mathbf{x}, t_0 ;\boldsymbol{\mu}) = \mathbf{0}$ %_0(\mathbf{x} ;\boldsymbol{\mu})$ in $\Omega \times \{t_0\}$ \textcolor{red}{non è semplicemente at rest?} 
and suitable boundary conditions reported in Section \ref{sec:geo}. 
$\mathbb T$ is the Cauchy stress tensor whose constitutive relation has the form:
\begin{equation}
		\mathbb T =-p\mathbb I + \mathbb{T}_d,
		\label{constitutive}
\end{equation} %$\mu = 0.0035 $ Pa $\cdot$ s is the blood viscosity and $\Omega$ is the patient-specific domain of interest
where the deviatoric component, $\mathbb{T}_d$, depends on the fluid model.
It is known that, even if plasma is approximately Newtonian, whole blood could exhibit significant non-Newtonian features \cite{chien1967blood}. Many works compare Newtonian and non-Newtonian models, showing that the Newtonian one is in general admissible. However, some differences can be found locally for some portions of the domain and/or during specific time instances of the cardiac cycle \cite{johnston2004non}. For such a reason, in this work both models, Newtonian and non-Newtonian, are considered.

For a Newtonian fluid, the tensor $\mathbb{T}_d$ is:
\begin{equation}
		\mathbb{T}_d = 2\mu\mathbb{D(\mathbf{v})},
		\label{Newtonian}
\end{equation}
with $\mu = 0.0035$ Pa$\cdot$s the constant dynamic viscosity and $\mathbb{D(\mathbf u)}=\frac{\nabla \mathbf u + \nabla \mathbf u ^T}{2}$ the strain rate tensor. 

On the other hand, the non-Newtonian behavior of the blood is handled with the Casson's model \cite{drapaca2018comparison}:
\begin{equation}
		\mathbb{T}_d =2\mu(J_2)\mathbb{D(\mathbf{v})},
		\label{Casson}
\end{equation}
and
\begin{equation}
    \mu (J_2) = \left[ (\bar\eta^2 J_2)^{1/4} + \Big(\frac{\tau_y}{2}\Big)^{1/2} \right]^{1/2} J_2^{-1/2},
\end{equation}
where $J_2$ is the second invariant of $\mathbb{D(\mathbf{v})}$, $\bar\eta = \eta / (1- H )^{2.5}$ with $\eta$ the plasma viscosity and $H$ the hematocrit, and $\tau_y = (a_1 +a_2 H )^3$ where for blood $a_1=0$ and $a_2=0.625$ \cite{fung2013biomechanics}. For more details % a deeper study about the relation with the hematocrit and the plasma viscosity 
we refer to \cite{errill1969rheology}. We note that the plasma viscosity $\eta$ and the hematocrit $H$ will be treated as parameters in our ROM framework. % underline the presence of two additional parameters in the Casson's model involved in the reduction step, respectively the plasma viscosity $\eta$ and the hematocrit $H$ \textcolor{red}{Punto 2}.  

Equations above are solved using the FV method implemented in OpenFOAM(R) 2204 \cite{openFoam}. Direct numerical simulation of the blood flow is used here assuming that the flow is not turbulent. %\gs{add citation to the OF library} 
Second order schemes are adopted for space discretization and Euler implicit time scheme is used for time discretization. More details about the FV discretization of Navier-Stokes equations can be found in \cite{girfoglio2022non}. 

\subsection{Patient-specific geometry and boundary conditions} \label{sec:geo}
\begin{figure}[h]
    \centering
    %\subfloat[LA patient-specific geometry]{\captionsetup{width=.8\linewidth}
    \includegraphics[height=4.5cm]{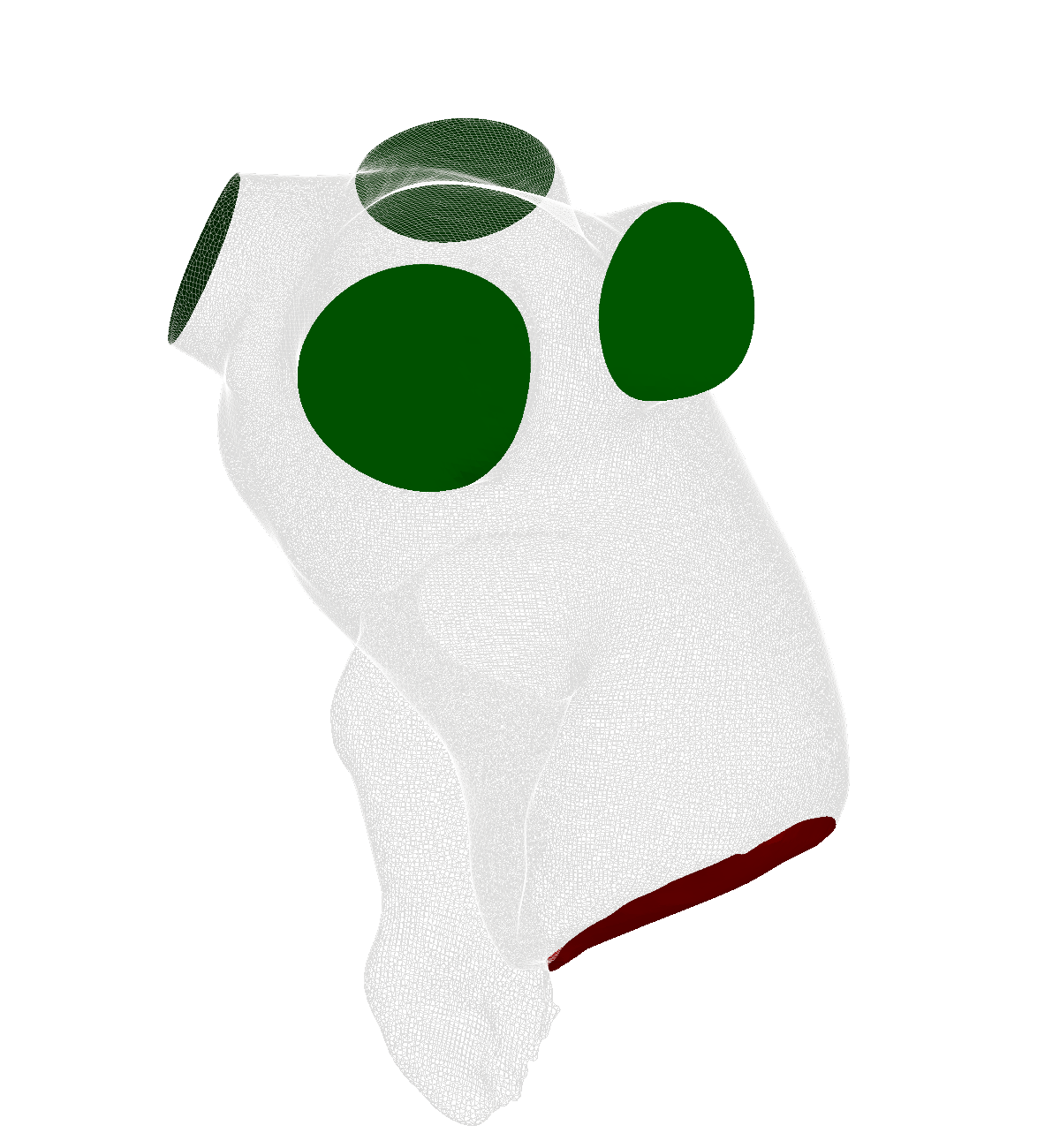}
    %}
    \qquad
    %\subfloat[Flow boundary conditions]{\captionsetup{width=.8\linewidth}
    \includegraphics[width=0.35\columnwidth]{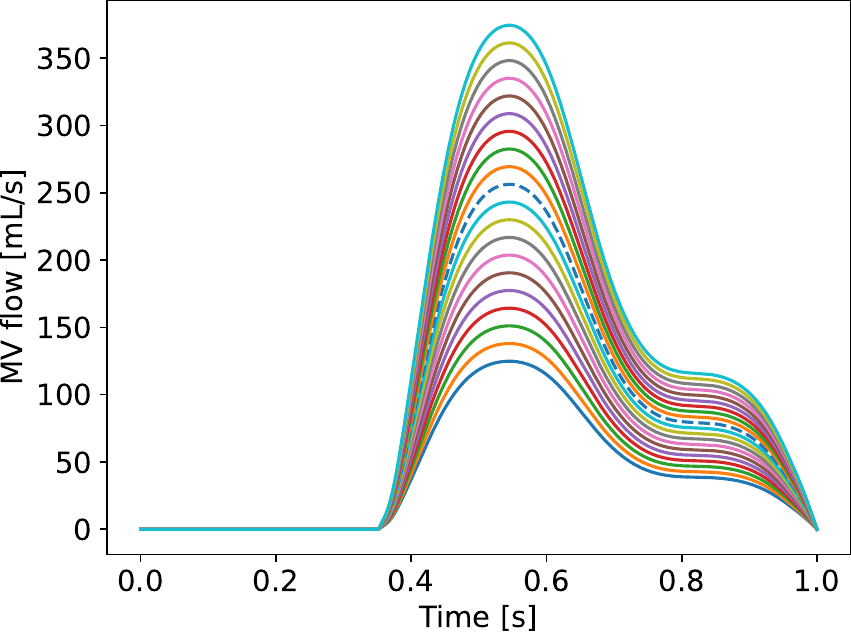}
    %}
    \caption{Left: LA patient-specific geometry; the green surfaces are the PV inlets while the red region is the MV outlet. Right: boundary conditions for MV flow for different scaling factors ranging in [0.5, 1.5] (the dashed line corresponds to 1).}
    \label{fig:geometry}
\end{figure}

%\textcolor{red}{Michele: I seem that some abbreviations, already defined, are defined again, such as LA, LAA and PV} \pier{yes I check everything.} 
In this research, the same patient geometry reported by Due\~{n}as-Pamplona et al.\ in \cite{DuenasPamplona2022} is used as test configuration. The patient had a history of paroxysmal AF, but no previous stroke or transient ischemic attack, and underwent CT-imaging and Doppler transesophageal echocardiography at the Puerta de Hierro Hospital in Madrid. The images were obtained from the left atrial endocardial surface, including the LAA and the PVs, during normal sinus rhythm, using a SIEMENS Sensation 64 scanner with specific scanning parameters. The resulting DICOM images were manually segmented with the aid of the non-commercial code 3D-Slicer to extract the three-dimensional surface of the endocardium. The PVs were removed beyond the first bifurcation using Autodesk MeshMixer, and the LA endocardial surface geometry, including the LAA, was provided. In Figure \ref{fig:geometry} (left) the patient-specific geometry is reported.

Walls are assumed to be completely rigid due to the patient's condition.  %\gs{Maybe it is interesting to put a figure with the test geometry before is cited in the next section, or reuse the same figure}
Regarding inflow and outflow boundary conditions, those reported in\cite{DuenasPamplona2022} are also used. They consist of Mitral Valve (MV) velocities during sinus rhythm after CT imaging. These blood velocities were used to establish the simulation boundary conditions, assuming each PV carries out the same flow-rate. In order to clarify the inflow and outflow boundaries, in Figure \ref{fig:geometry} (left) are shown in red and green respectively the MV and the PVs. In addition, in the right panel we report the time evolution of the MV flow-rate for different values of the scaling factor $f$ in the interval $[0.5, 1.5]$. Note that the scaling factor $f$ belongs to the parameter space of our problem. %is one of the parameters of our problem. % As we will see after the scaling factor belongs %(\textcolor{red}{ Metterei una figura della geometria, magari con le freccie per indicare inlet e outlet Figure ... e anche la figura delle BC} )

\subsection{Hemodynamics indices}\label{sec:indices}
Now we are going to introduce some relevant indices associated with the stasis of the blood flow \cite{DuenasPamplona2021a}. They represent fundamental quantities for the medical community because high residence times of the blood flow are related to thrombus formation.

\subsubsection{Age distribution}
Following \cite{Sierra2017}, the first and second moments of age, $m_1$ and $m_2$, 
%the mean age of the blood $m_1$ and the first moment of age $m_2$ 
are computed as follows: 
\begin{equation}
    m_k(\mathbf{x}, t; \boldsymbol{\mu})= \int_{-\infty}^{t} t^k\phi(\mathbf{x}, t;\boldsymbol{\mu}) dt, \quad k=1,2, \label{eq:age_dist}
\end{equation}
where $\phi(\mathbf{x}, t;\boldsymbol{\mu})$ is the blood age distribution.  In particular, $m_1$ represents the mean age of blood and describes the time needed for a given blood particle to reach another position in the computational domain, so large values of $m_1$ denote stasis of the blood. The second moment $m_2$ does not have any physical meaning. However we have decided to consider it as a further variable to be taken into account to show the versatility of our ROM approach. %which i able to reconstruct also the following higher-order moments accurately \textcolor{red}{Punto 7}. 
%In general such indices can represent fundamental quantities for the generation of variables related to the blood stasis.
It should be noted that once $m_k$ is known, it would be  possible to compute  also $\phi$  \eqref{eq:age_dist}: see, e.g., \cite{john2007techniques, Sierra2017}. %\textcolor{red}{Punto 8}. 
For a laminar flow, we have for $k = 1, 2$: % in the domain of these indices is described by:
\begin{equation}
\begin{cases}
    \rho\frac{\partial m_k}{\partial t}  + \rho\nabla \cdot (\mathbf{v} m_k) = \nabla \cdot \big[ \mu_{m_k} \nabla m_k\big] + \rho k m_{k-1}   & \quad \text{in} \quad \Omega \times (t_0, T], \\
    m_k = 0 & \quad \text{on} \quad \Gamma_i \times (t_0, T], \\
    \nabla m_k \cdot \mathbf{n} = 0, &\quad \text{on} \quad \partial  \Omega \smallsetminus \Gamma_i \times (t_0, T], %\\
    %m_k = 0 & \quad t=0,
    
    \label{moment}
\end{cases} %\quad k = 1,2,
\end{equation}
and initial data $m_k = 0$, where $\mathbf{n}$ is the unit normal outward vector to the boundary, $\Gamma_i$ denotes the inflow boundaries (i.e.\ the PV inlets, see Figure \ref{fig:geometry}), $\mu_{m_k} = 10^{-10}\, \text{kg}/(\text{m} \cdot \text{s})$ is the mass diffusivity of the moment $m_k$ \cite{DuenasPamplona2021a} and $m_0=1$ \cite{Sierra2017}. 

Another variable related to the blood age is the washout.  It is computed by solving a scalar transport equation similar to \eqref{moment} but without the source term 
\begin{equation}
\begin{cases}
    \rho \frac{\partial \varphi}{\partial t}  + \rho \nabla \cdot (\mathbf{v} \varphi) = \nabla \cdot \big[ \mu_{\varphi} \nabla \varphi\big] & \quad \text{in} \quad \Omega \times (t_0, T], \\
    \varphi = 0 & \quad \text{on} \quad \Gamma_i \times (t_0, T], \\
    \nabla \varphi \cdot \mathbf{n} = 0 &\quad \text{on} \quad \partial  \Omega \smallsetminus \Gamma_i \times (t_0, T], %\\
    %\varphi = 1 & \quad %t=0,
    \label{washout}
\end{cases}
\end{equation}
and initial data $\varphi = 1$, where $\mu_{\varphi} = 10^{-10} \text{kg}/(\text{m} \cdot \text{s})$. %$\varphi$ is set to 1 for the entire computational domain and the incoming flow is $\varphi = 0$ \textcolor{red}{Punto 9 mettiamo valore $\Gamma$}. 
The washout is the residual value of $\varphi$ as the dynamics evolves over the cardiac cycle. It implies zones with high stasis and therefore high age. More details could be found in % about these indices and several methods to reconstruct the blood age distribution can be found in 
\cite{john2007techniques, Sierra2017}. 

Equations \eqref{moment} and \eqref{washout} are coupled with system \eqref{N-S} but, in order to reduce the computational cost, we adopt a segregated algorithm, so they are solved \emph{after} the system \eqref{N-S}. In other words, $m_k$ and $\varphi$ are treated as passive scalars. Equations \eqref{moment} and \eqref{washout} are employed by using second order schemes for space discretization and Euler scheme for time discretization.

\subsubsection{Wall Shear Stress and Oscillating Shear Index}

The Wall Shear Stress (WSS) can be defined as follows: %is related to the blood velocity, and therefore to its stasis as follows:
\begin{equation}
    \text{WSS} = \mathbb{T}_d \cdot \mathbf{n},  \quad \mbox{on} \quad \partial\Omega.%\tau(\mathbf{u}) \cdot \mathbf{n}, \quad \mbox{on} \quad \partial\Omega,
\end{equation}
%where %$\tau (\mathbf{u}) = \frac{\mu}{\rho}(\nabla \mathbf{u} + \nabla \mathbf{u}^T)$ is the stress tensor \textcolor{red}{Michele: maybe it would be better to link this definition with the Cauchy stress tensor one} and 
%$\mathbf{n}$ is the unit normal outward vector to the boundary.
%\textcolor{red}{è su tutto il boundary? o solo sul wall?} 
The interest is on the Time Averaged Wall Shear Stress (TAWSS)  representing the mean effect of the WSS on the entire cardiac cycle:
\begin{equation}
    \text{TAWSS} = \frac{1}{T}\int_0^T \|\text{WSS}\| dt.
\end{equation}
Low values of TAWSS correspond to stasis regions.

Another important quantity is the Oscillating Shear Index (OSI):
\begin{equation}
    \text{OSI} = \frac{1}{2} \left[ 1 -  \frac{\|\int_0^T \text{WSS}dt\|}{\int_0^T \|\text{WSS}\| dt} \right].
\end{equation}
It is related to the oscillations of the flow and ranges from 0, when the flow is unidirectional, to $0.5$, when the direction of the flow is totally reversed. The OSI is a useful indicator in cardiovascular problems because it is correlated with the intimal thickness of the wall and the restenosis process \cite{ku1985pulsatile}. 

%\textcolor{red}{Michele: Here I would refer the reader to LVAD paper if he/she is interested in having more details about the FV discretization of NSE system} %\pier{The boundary conditions adopted for the velocity derive from a realistic and time dependent microvascular (MV) flow \textcolor{red}{metterei una citazione per MV flow e la figura delle BC}. A scaling factor is adopted to obtain the curves in Figure ... and to consider several flow cases. This constant is the physical parameter involved in the reduction problem.}
%\textcolor{red}{dopo metto meglio che BC sono usate e metti meglio $\Omega in NS$}
%\gs{I think it won't be really helpful for the community of doctors but I think it would make sense to add more details and to summarize the numerical setting that is used (i.e. space discretization, time discretization, laminar vs turbulent etc)}

\section{The reduced order model}\label{sec:rom}
The basic assumption of ROM for a partial differential equations problem depending on time  $t$ and parameter vector $\mubold$ is that any solution can be represented as a linear combination of a reduced number of
global basis functions, that depend exclusively on space $\mathbf{x}$, with the weights of the linear combination
depending only on $t$ and $\mubold$. For a generic variable $\Phi$ this is written as:
\begin{align}
\Phi(\mathbf{x}, t; \mubold) \approx 
\Phi^{\rb}(\mathbf{x}, t; \mubold) = \sum_{l=1}^{L}\alpha_{l}(t, \mubold)\boldsymbol{\ell}_{l} (\mathbf{x}), \label{eq:generic_var}
\end{align}
where $\Phi^{\text{rb}}$ is the reduced order approximation of $\Phi$, 
$L$ is the number of basis functions, the $\boldsymbol{\ell}_l$ are
the basis functions and the $\alpha_l$ are the weights of the linear combination (the so-called modal coefficients). In this work, we are interested in the reconstruction of all the hemodynamics indices introduced in Section~\ref{sec:indices}, so we have $\Phi = \{m_1, m_2, \varphi, \text{TAWSS}, \text{OSI}\}$. It should be noted that TAWSS and OSI are steady-state variables. In this case, equation \eqref{eq:generic_var} becomes:
\begin{align}
\Phi(\mathbf{x}; \mubold) \approx 
\Phi^{\rb}(\mathbf{x}; \mubold) = \sum_{l=1}^{L}\alpha_{l}(\mubold)\boldsymbol{\ell}_{l} (\mathbf{x}). \label{eq:generic_var2}
\end{align}
However, for the sake of good order, in the following explanation we refer to a time and parameter dependent variable.
%In this section, we briefly recall the main features of our ROM approach. %of ROMs and we refer to \cite{benner2020book, hesthaven2016book, manzoni2012MJM, rozza2008ACME} for a comprehensive description. 

We use the POD-RBF technique, which is divided into the following two phases:
\begin{description}
\item[Offline] given a set of physical parameters values and time instances, the corresponding high fidelity solutions (the so-called snapshots) %\gs{FOM is not introduced before} 
are computed and collected into a matrix. The POD algorithm is used to extract the reduced basis space from the snapshots matrix. Then, the snapshots are projected onto the POD space by obtaining the corresponding modal coefficients. Finally, a RBF interpolation algorithm is used to compute a map between the parameters and the modal coefficients. 
\item[Online] given a set of new physical parameters values, the corresponding modal coefficients are computed via the RBF function and the approximated solution is recovered as a linear combination between these coefficients and the POD reduced basis (see equation \eqref{eq:generic_var}). 
\end{description}
The offline-online procedure is summarized in Algorithm \ref{alg:ROM}. % for a generic parameter and time dependent variable $\Phi$. 
The implementation of the ROM relies on the Python library EZyRB \cite{demo2018JOSS}.
\begin{algorithm}
\caption{Description of the main steps of the Offline/Online framework}
\label{pod-alg}
\raggedright
%\text{\textbf{Task:} Given a new time istance the vector $(t, \mubold)$ find the reduced solution $\Phi^{\rb}(t, \mubold)$}.\\
\text{\textbf{Offline stage} }
\begin{algorithmic}[1]
\State Compute the finite volume solutions $\Phi(\mathbf{x}, t_i; \mubold_j)$ for $i = 1,\dots, N_t$ and $j = 1,\dots, N_p$.
\State Generate the reduced basis
$\mathcal{B} = [\boldsymbol{\ell}_{1}|\dots|\boldsymbol{\ell}_{L}]$ with the POD algorithm.
\State Interpolate with the RBF technique the coefficients $\alpha_{l}(t_i, \mubold_j) = \left(\mathcal{B}^{T}\Phi(\mathbf{x}, t_i; \mubold_j)\right)_{l}$ for $l = 1, \dots, L$.
\end{algorithmic}
\text{\textbf{Online stage} }
\begin{algorithmic}[1]
\State Estimate the coefficients for new parameters values $\boldsymbol{\alpha}(t, \mubold_{\text{new}})$ by RBF.
\State Compute the reduced solution $\Phi^{\rb}(\mathbf{x}, t; \mubold_{\text{new}}) = \sum_{l=1}^{L}\alpha_{l}(t, \mubold_{\text{new}})\boldsymbol{\ell}_{l}(\mathbf{x})$.
\end{algorithmic}
\label{alg:ROM}
\end{algorithm}

%\gs{it could help to have both the offline and the online stage in an algorithm environment}

\subsection{The proper orthogonal decomposition}\label{sec:POD}
Let $\mathcal{K} = \{\boldsymbol{\mu}_1, \dots, \boldsymbol{\mu}_{N_p}\}$ be a finite dimensional training set of samples chosen inside the
parameter space $\mathcal P$ and for each time $t_k \in \mathcal{T} = \{t_1, \dots, t_{N_t}\} \subseteq (t_0, T]$.
%Let us consider a set  $\mathcal{\widehat P} = (t_0,T] \times \mathcal{P}$ of physical parameters. %$\xibold$. 
We refer to the FOM snapshots %\eqref{N-S}--\eqref{washout}, 
$\Phi(\mathbf{x}, t_i; \mubold_j)$ for $i = 1,\dots, N_t$ and $j = 1,\dots, N_p$. %with $\Phi = \{m_1, m_2, \varphi\}$. %We also are interested in the reconstruction of steady state variables, $\Phi = \{\text{TAWSS}, \text{OSI}\}$ with $\Phi(\mubold_j)$ for $j = 1,\dots, N_p$. However, for the sake of good order, in the following explanation we refer to a time and parameter dependent variable. %\{\Phi(t, \mubold)\}_{(t, \mubold)\in\mathcal{\widehat P}}$.

The POD \cite{atwell2001MCM, gunzburger2002SIAM, kunisch2002JNA, volkein2011notes, willcox2002AIAA} is one of the most common techniques used to extract the essential information from the space generated by the solution manifold. Let $N_{s} = N_t \cdot N_p$ be the dimension of this space: the goal of the POD is to construct a reduced basis space of dimension $L\ll N_{s}$ which is optimal in the least-square sense. More precisely, the POD algorithm builds the reduced basis space which minimizes the quantity
\begin{equation*}
\sqrt{\sum_{(t, \mubold)\in\mathcal{T} \times \mathcal{K}}\inf_{\Phi^{\rb}\in V^{\rb}}\|\Phi(\mathbf{x}, t; \mubold)-\Phi^{\rb}(\mathbf{x}, t; \mubold)\|^{2}},
\end{equation*}
among all $L$-dimensional subspaces $V^{\rb}$ spanned by the FOM solutions \cite{hesthaven2016book}.  % $\mathrm{span}\{\Phi(t, \mubold)\}_{(t, \mubold)\in\mathcal{\widehat P}}$, 
%where $\|\cdot\|$ is the Frobenius norm and $\Phi^{\rb}$ is the reduced solution \cite{hesthaven2016book}.

Let %$\mathcal{\widehat M}=\{(t_i, \mubold_j) \mid i = 1, \dots, N_t$  \mbox{and} $j = 1, \dots N_p\}$ and 
$N_{h}$ %be a set of physical parameters chosen into the space $\mathcal{\widehat P}$ and 
be the number of the mesh cells. 
%For each $(t_i, \mubold_j)\in\mathcal{\widehat M}$ we solve the FOM and we collect these 
We collect the FOM solutions %\emph{high fidelity solutions} 
into the snapshot matrix $S^{\Phi}$ given by
\begin{equation}\label{eq:snapMatrix}
	% S^{\Phi} = \begin{pmatrix}
	% \Phi_{1}(\xibold_{1}) & \Phi_{1}(\xibold_{2}) & \dots & \Phi_{1}(\xibold_{N_{s}})\\
	% \vdots & \vdots & \ddots &\vdots \\
	% \Phi_{N_{h}}(\xibold_{1}) & \Phi_{N_h}(\xibold_{2}) & \dots & \Phi_{N_{h}}(\xibold_{N_{s}})\\
	% \end{pmatrix},
    S^{\Phi} = \begin{pmatrix}
	\Phi_{1}(\mathbf{x}, t_{1};\mubold_{1}) & \dots & \Phi_{1}(\mathbf{x},t_{N_t};\mubold_{1}) & \Phi_{1}(\mathbf{x}, t_{1};\mubold_{2}) & \dots & \Phi_{1}(\mathbf{x},t_{N_t};\mubold_{N_p})\\ 
    \Phi_{2}(\mathbf{x},t_{1};\mubold_{1}) & \dots & \Phi_{2}(\mathbf{x},t_{N_t};\mubold_{1}) & \Phi_{2}(\mathbf{x},t_{1};\mubold_{2}) & \dots & \Phi_{2}(\mathbf{x}, t_{N_t}; \mubold_{N_p})\\ 
    \vdots & \ddots & \vdots & \vdots & \ddots & \vdots\\
    \Phi_{N_h}(\mathbf{x}, t_{1};\mubold_{1}) & \dots & \Phi_{N_h}(\mathbf{x}, t_{N_t};\mubold_{1}) & \Phi_{N_h}(\mathbf{x}, t_{1};\mubold_{2}) & \dots & \Phi_{N_h}(\mathbf{x}, t_{N_t};\mubold_{N_p})\\
	\end{pmatrix},
\end{equation}
whose dimension is $N_{h}\times N_{s}$. Since $S^{\Phi}$ is usually not squared, we introduce its rank $R\leq\min\{ N_{h}, N_{s}\}$. By applying the Singular Value Decomposition (SVD) to $S^{\Phi}$, we can rewrite it as
\begin{equation*}
	S^{\Phi} = \mathcal{L} \Sigma \mathcal{R}^{T},
\end{equation*} 
where $\mathcal{L} = [\boldsymbol{\ell}_{1}| \dots |\boldsymbol{\ell}_{N_{h}}] \in\R^{N_{h}\times N_{h}}$ and  $\mathcal{R}= [\boldsymbol{r}_{1}| \dots |\boldsymbol{r}_{N_{s}} ] \in\R^{N_{s}\times N_{s}}$ are orthogonal matrices whose columns are the left and right singular vectors, respectively, and $\Sigma\in\R^{N_{h}\times N_{s}}$ is a diagonal matrix with $R$ non-zero real singular values $\sigma_{1}\geq \sigma_{2} \geq\dots\geq \sigma_{R} > 0$.

As the rank $R$ is typically large, we are now concerned with reducing the size of the problem. We rely on the Schmidt-Eckart-Young theorem \cite{eckart1936P}, which states that the first $L$ left singular vectors of $S^{\Phi}$ are the POD bases of rank $L$, with $L<R$. Hence, the POD bases are the first $L$ columns of the matrix $\mathcal{L}$. In particular, %since $S^{\Phi}(S^{\Phi})^{T}\mathcal{L} = \mathcal{L}\Sigma\Sigma^{T}$, 
the $l$-th column of $\mathcal{L}$ is the eigenvector associated with $\mathcal{C}\boldsymbol{\ell}_{l}=\sigma_{l}^{2}\boldsymbol{\ell}_{l}$, where $\mathcal{C}=(S^{\Phi})^{T}S^{\Phi}$ is the snapshot correlation matrix. Therefore, $\boldsymbol{\ell}_{l}$ is given by
\begin{equation*}
	\boldsymbol{\ell}_{l} = \frac{1}{\sigma_{l}}S^{\Phi}\boldsymbol{c}_{l}, \quad \text{for} \quad l=1,\dots, L.
\end{equation*}
%for $l=1,\dots, L$. 
Then, the POD bases, also known as modes, are collected into the matrix
\begin{equation}\label{eq:podBasis}
	\mathcal{B} = [\boldsymbol{\ell}_{1}|\dots|\boldsymbol{\ell}_{L}].
\end{equation} 
It remains only to choose properly the value of $L$. A common choice is to define it as the smallest integer such that
\begin{equation}\label{eq:eigenEnergy}
	\frac{\sum_{l=1}^{L}\sigma_{l}^{2}}{\sum_{l=1}^{R}\sigma_{l}^{2}}\geq\varepsilon,
\end{equation}
for a given threshold $\varepsilon$ on the cumulative energy of the eigenvectors.

Once defined the POD basis matrix $\mathcal{B}$, the reduced solution $\Phi^{\rb}$ that approximates the truth one $\Phi$ is given by
\begin{equation*}
	\Phi^{\rb}(\mathbf{x}, t_i; \mubold_j) = \sum_{l=1}^{L}\alpha_{l}(t_i, \mubold_j)\boldsymbol{\ell}_{l}(\mathbf{x}), \quad \text{for} \quad i = 1,\dots, N_t \quad \text{and} \quad j = 1,\dots, N_p
\end{equation*}
and where $\alpha_{l}(t_i, \mubold_j) = \left(\mathcal{B}^{T}\Phi(t_i, \mubold_j)\right)_{l}$ is the $l$-th modal coefficient.

% \subsubsection{$k$-Neighbors Regressor}
% To conclude the offline phase we define a map $f:\mathcal{\widehat P}\mapsto\R^{L}$ from the parameters space to the modal coefficients one. To this end, we use the $k$-Neighbors Regressor ($\knr$) \cite{fix1989ISR, cover1967IEEE, scikit}, that is a supervised machine learning regression method.

% The $\knr$ algorithm requires a training set of couples $(\boldsymbol{x}_i, \boldsymbol{y}_i)$ and a metric to measure the distance between the points $\boldsymbol{x}_i$ and a query one $\boldsymbol{x}$. The predicted value $\boldsymbol{y}=f(\boldsymbol{x})$ associated with the query point $\boldsymbol{x}$ is given by the weighted average of the values of the $k$ nearest neighbors
% \begin{equation}\label{eq:knr}
%     f(\boldsymbol{x}) = \frac{1}{\sum_{j=1}^k 1/d(\boldsymbol{x}, \boldsymbol{x}_j)}\sum_{j=1}^k \frac{\boldsymbol{y}_j}{d(\boldsymbol{x}, \boldsymbol{x}_j)},
% \end{equation}
% where $d(\boldsymbol{x}, \boldsymbol{x}_j)$ is the distance between $\boldsymbol{x}$ and $\boldsymbol{x}_j$.

% As explained in Section \ref{sec:POD}, the POD algorithm gives us the modal coefficients $\boldsymbol{m}(\xibold_{i}) = \{m_{j}(\xibold_{i})\}_{j=1}^L$ corresponding to the sampled parameters $\xibold_{i}$, $i=1,\dots,N_{s}$. Thus, in our case, the training couples $(\boldsymbol{x}_i, \boldsymbol{y}_i)$ are $(\xibold_i, \boldsymbol{m}(\xibold_{i}))$ and $f(\xibold)=\boldsymbol{m}(\xibold)$ is the modal coefficient associated with a new parameter $\xibold$.

\subsection{Radial basis function interpolation}
The last step of the offline phase consists of the definition of a map $F:\mathcal{T} \times \mathcal{K} \mapsto\R^{L}$ from the parameters space to the modal coefficients one. To this end, we use the RBF interpolation \cite{buhmann2003book, forti2014IJCFD, skala2016IO}. %By RBF we refer to a smooth function $\psi$ which depends only on the Euclidean distance between its argument and a specific point $\boldsymbol c$, i.e. %\ there exists a function \textcolor{red}{Punto 12} \textcolor{blue}{un altro simbolo rispetto a x? Lascerei cosi pensandoci bene in questa piccola sezione.}$\widetilde\psi$ such that 
%$\psi(\boldsymbol{x})=\widetilde\psi(\|\boldsymbol{x}-\boldsymbol{c}\|)$.
%As explained in the previous section, the POD algorithm gives us the modal coefficients $\boldsymbol{m}(\xibold_{i}) = \{m_{j}(\xibold_{i})\}_{j=1,\dots,L}$ corresponding to the sampled parameters $\xibold_{i}$, $i=1,\dots,N_{s}$. 
In general given a set of $N_s$ couples $(\boldsymbol{x}_i, \boldsymbol{y}_i)$ and a query point $\boldsymbol{x}$, the RBF is defined as
\begin{align}
	%\label{eq:fRBF}&f(\mubold) = \sum_{j=1}^{L}w_{j}\tilde\varphi(\|\mubold-\mubold_{j}\|)+P(\mubold)\\
	%\nonumber&\text{subject to }f(\mubold_{i})=\boldsymbol{m}_{i}(\mubold_{i}) \text{ for } i=1,\dots,L,
	%\label{eq:fRBF}&f(\xibold) = \sum_{i=1}^{N_{s}}w_{i}\tilde\varphi(\|\xibold-\xibold_{i}\|)+P(\xibold)\\
	%\nonumber&\text{subject to }f(\xibold_{i})=\boldsymbol{m}(\xibold_{i}) \text{ for } i=1,\dots,N_{s},
    \label{eq:fRBF}%&
    F(\boldsymbol{x}) = \sum_{j=1}^{N_{s}}w_{j}\widetilde\psi(\|\boldsymbol{x}-\boldsymbol{x}_{j}\|)+P(\boldsymbol{x}),
	\quad \text{subject to }F(\boldsymbol{x}_{i})=\boldsymbol{y}_{i} \text{ for } i=1,\dots,N_{s},
\end{align}
where $w_{j}$ are weights to be determined and $P(\boldsymbol{x})$ is a polynomial required for stability reasons. For simplicity, we assume that $P$ is of degree 1. By adding the conditions
\begin{align*}
	\sum_{j=1}^{N_{s}}w_{j} = 0\qquad\text{and}\qquad \sum_{j=1}^{N_{s}}w_{j}\boldsymbol{x}_{j} = 0
\end{align*}
the weights $w_{j}$ and the polynomial $P$ are uniquely determined.

As explained in Section \ref{sec:POD}, the projection of the snapshots onto the POD space gives us the modal coefficients $\boldsymbol{\alpha}(t_i, \mubold_j) = \{\alpha_{l}(t_i, \mubold_j)\}_{l=1}^L$ corresponding to the samples $(t_i, \mubold_j)$, with $i=1,\dots,N_t$ and $j=1,\dots,N_p$. Thus, in our case, we have that  $(\boldsymbol{x}, \boldsymbol{y}) \equiv \left((t, \mubold), \boldsymbol{\alpha}(t, \mubold)\right)$. %and $f(t, \mubold)=\boldsymbol{\alpha}(t, \mubold)$ is the modal coefficient associated with a new instance $(t, \mubold_{new})$.

\section{Numerical results}
\label{sec:numerical}

%\subsection{Tests on the full order model}
In this section, we investigate the performance of our ROM model. %is devoted to the numerical results comparing the FOM and the ROM introduced in Sections \ref{sec:problem_formulation} and \ref{sec:rom} respectively. 
For what concerns the validation of the FOM model in an artificial setup and the independence of the numerical results from the mesh for the patient-specific case of interest, we refer to Appendix \ref{appendice}. 

\begin{figure}[ht!]
\includegraphics[trim={50 50 50 50}, width=0.25\columnwidth]{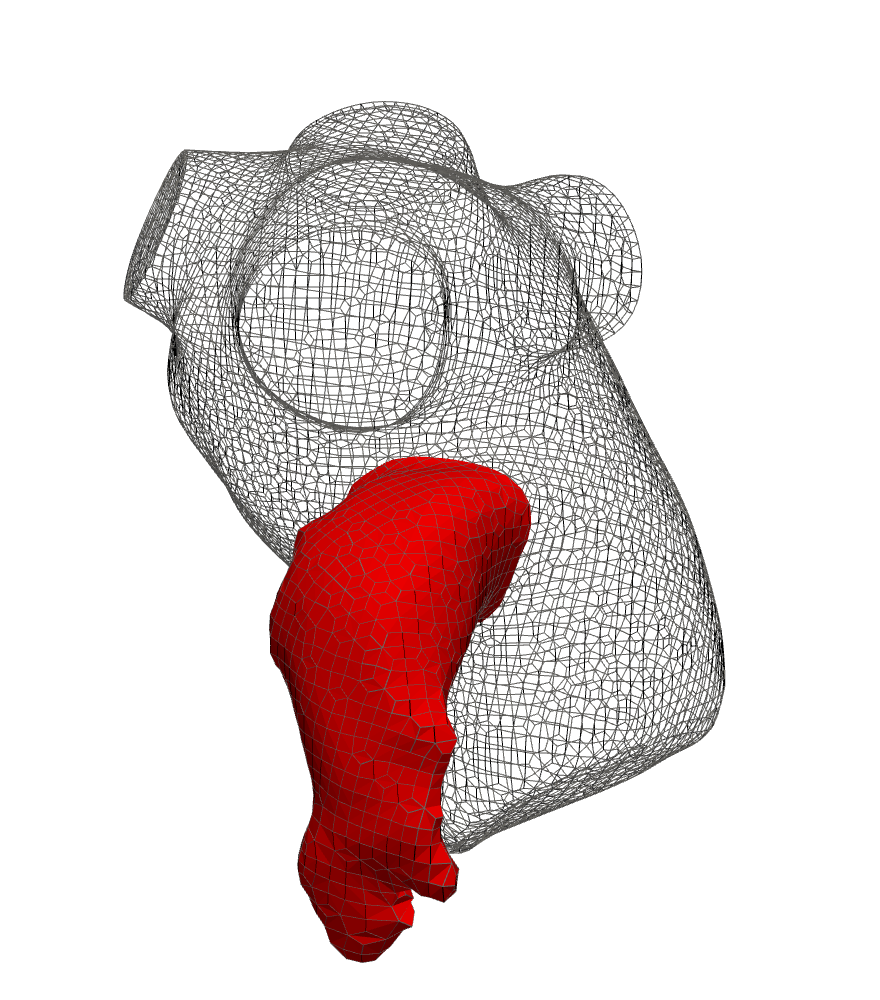}
\caption{Sketch of the mesh. The red portion corresponds to LAA.}
\label{fig:mesh}
\end{figure}

%Let's start with some details about the computational grid. % to build the high fidelity solutions that feed the ROM.
We consider a tetrahedral mesh inside the domain $\Omega$ and on its boundary $\partial\Omega$, consisting of 542.560 cells and 79.708 cells respectively, i.e. the Grid 2 reported in Table \ref{tab:gci_study}.
In Figure \ref{fig:mesh} we show a sketch of the mesh and we highlight the LAA portion on which we will focus our tests (the reason for this choice is discussed in Section~\ref{sec:intro}). %\textcolor{red}{verificare se è gia' stata menzionata prima sta cosa}. 
Since we are going to analyze the regime behaviour, i.e. downstream of the transient effects, we collect the FOM solutions corresponding to the fourth cardiac cycle whose period is 1.07 s. Therefore, the effective time interval is $(t_{0},T] = (4.33, 5.4]$ s which, for sake of convenience, we report to $(0, 1]$. We ran the FOM simulations with a time step $\Delta t=0.01\,$s collecting $N_t = 107$ time dependent snapshots. %We ran the FOM simulations with a time step of 1 millisecond and we collected, for each simulation, $N_t = 107$ time dependent snapshots, i.e.\ one every 10. %with $T_{0}=4.33\,$s and $T=5.4\,$s, and we normalize it to $(0,1]$. 
% \textcolor{red}{First of all we build the snapshot matrix. We have computed the high fidelity solutions $\Phi(t_{n},\mubold_{i})$ for a set $\mathcal{M}\subset\mathcal{P}$ of physical parameters $\mubold_{i}$, $i=1,\dots,N_{p}$, during the discrete times $t_{n}$, $n=1,\dots, N_{t}$. Since we are working with a dynamical system, we would like to predict not only the reduced solution for new parameters $\mubold\in\mathcal{P}$ but also at any time $t\in(0,T]$. 
% Therefore, for each $\mubold_{i}\in\mathcal{M}$ and time $t_{n}$, we collect the solutions into the snapshot matrix  \textcolor{red}{Punto 14}
% \begin{equation*}
% 	S^{\Phi} = \begin{pmatrix}
% 	\Phi_{1}(t_{1},\mubold_{1}) & \dots & \Phi_{1}(t_{N_t},\mubold_{1}) & \Phi_{1}(t_{1},\mubold_{2}) & \dots & \Phi_{1}(t_{N_t},\mubold_{N_p})\\ 
%     \Phi_{2}(t_{1},\mubold_{1}) & \dots & \Phi_{2}(t_{N_t},\mubold_{1}) & \Phi_{2}(t_{1},\mubold_{2}) & \dots & \Phi_{2}(t_{N_t},\mubold_{N_p})\\ 
%     \vdots & \ddots & \vdots & \vdots & \ddots & \vdots\\
%     \Phi_{N_h}(t_{1},\mubold_{1}) & \dots & \Phi_{N_h}(t_{N_t},\mubold_{1}) & \Phi_{N_h}(t_{1},\mubold_{2}) & \dots & \Phi_{N_h}(t_{N_t},\mubold_{N_p})\\
% 	\end{pmatrix}.
% \end{equation*}
% The latter has $N_{s}=N_{p}\cdot N_{t}$ columns, with $N_{t}=108$ and $N_{p}=20$ (resp.\ $N_{p}=30$) for Newtonian (resp.\ non-Newtonian) fluids. Note that the variables $\tawss$ and $\osi$ are time independent, thus their snapshot matrices have $N_p$ columns.}

%\textcolor{red}{We consider a time horizon $(0,T]$ divided into time steps $t_{n}=n\Delta t$.} %Hereafter we let $\mathcal{\widehat P}=(0,T]\times\mathcal{P}$ and therefore $\xibold=(t,\mubold)$. \textcolor{red}{Punto 12}
%As already discussed, we examine both Newtonian and non-Newtonian cases. 
To unify the notation, where necessary, hereinafter we will use the superscript $N$ to refer to the Newtonian model and the superscript $C$ to refer to the Casson's one. 
In the Newtonian case we have one only parameter represented by the scaling factor $f$ ranging in the interval [0.5, 1.5] (see Figure \ref{fig:geometry}), i.e.\ $\mubold^N = f$. On the other hand, in the non Newtonian case, we also consider the plasma viscosity $\eta \in [1.5\text{e}-05, 1.7\text{e}-05]$ and the hematocrit $H \in [35, 50]$, i.e. $\mubold^C =  \{f, \eta, H \}$. Specifically, the discrete set of $\mubold^N$ consists of 20 points obtained by a uniform sampling procedure. Instead, for the Casson's model, % for each of the three parameters involved, 
we get a distribution of 30 points for $\mubold^C$ using Latin hypercube sampling \cite{mckay2000comparison}. %The size of the snapshot matrix $S^\Phi$ in \eqref{eq:snapMatrix} is $N_{s}=N_{p}\cdot N_{t}$, where $N_{t}=108$ and $N_{p}=20$ (resp.\ $N_{p}=30$) for Newtonian (resp.\ non-Newtonian) case. 
%Note that the variables $\tawss$ and $\osi$ are time independent, thus their snapshot matrices have $N_p$ columns. 

%\subsection{Tests on the reduced order model}

%Denoting by $\mathcal{M}\subset\mathcal{P}$ a set of physical parameters $\mubold_{j}$, $j=1,\dots,N_{p}$,

For both the models, we split the initial database in a training set $\mathcal{K}_{\train}\subset\mathcal{K}$ and in a validation set $\mathcal{K}_{\test} = \mathcal{K}\setminus\mathcal{K}_\train$. All the snapshots belonging to the training set are stored in the matrix $S^\Phi$. The validation set is used to assess the accuracy of the ROM solution. In order to compare the results obtained with the two models, we choose similar test values for the scaling factor $f$, which is the only shared parameter. %belonging to both $\mathcal{P}^N$ and $\mathcal{P}^C$. 
%Indeed, we recall that the physical parameter for the Newtonian model is the cardiac output $f$, while for the Casson's model is a three-dimensional vector defined by cardiac output $f$, hematocrit $H$ and plasma viscosity $\eta$.
Therefore, we set $\mathcal{K}_\test^N=\{\mubold^N_3,\mubold^N_{7}\}$ and $\mathcal{K}_\test^C=\{\mubold^C_{1},\mubold^C_{14}\}$, where $\mubold^N_3=0.65$, $\mubold^M_7=0.87$, $\mubold^C_1=[0.65, 39.7, 1.62\mathrm{e}-03]$ and $\mubold^C_{14}=[0.88, 46.7, 1.66\mathrm{e}-03]$. %\textcolor{red}{Punto 16} \textcolor{red}{sequenza campioni ordinamento random? sentiamo Cate} 

To analyze the accuracy of the ROM we compute the relative error
\begin{equation}\label{eq:errors}
	\mathsf{e}(t, \mubold) = \frac{\|\Phi(t,\mubold)-\Phi^{\rb}(t,\mubold)\|}{\|\Phi(t,\mubold)\|},
\end{equation}
where $\mubold\in\mathcal{K}_{\test}$ and $\|\cdot\|$  is the Frobenius norm. %is the Frobenius norm. 
For the steady state variables, $\text{TAWSS}$ and $\text{OSI}$, the relative error \eqref{eq:errors} becomes:
\begin{equation}\label{eq:errors2}
	\mathsf{e}(\mubold) = \frac{\|\Phi(\mubold)-\Phi^{\rb}(\mubold)\|}{\|\Phi(\mubold)\|}.
\end{equation}

% therefore we have chosen the value closest to the Newtonian case for comparison.
%Furthermore, we stress that all the tests are performed on the appendage area.

\subsection{Choice of the number of modes}
\begin{figure}[h!]
\subfloat[Newtonian case: modes number]{\captionsetup{width=.8\linewidth}
\includegraphics[width=0.35\columnwidth]{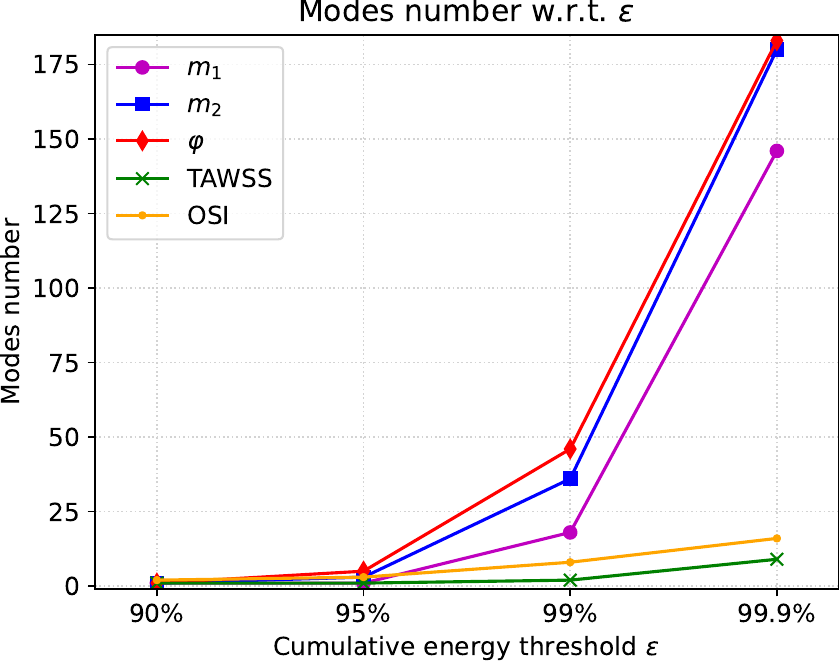}
}\quad
\subfloat[Casson's case: modes number]{
\includegraphics[width=0.35\columnwidth]{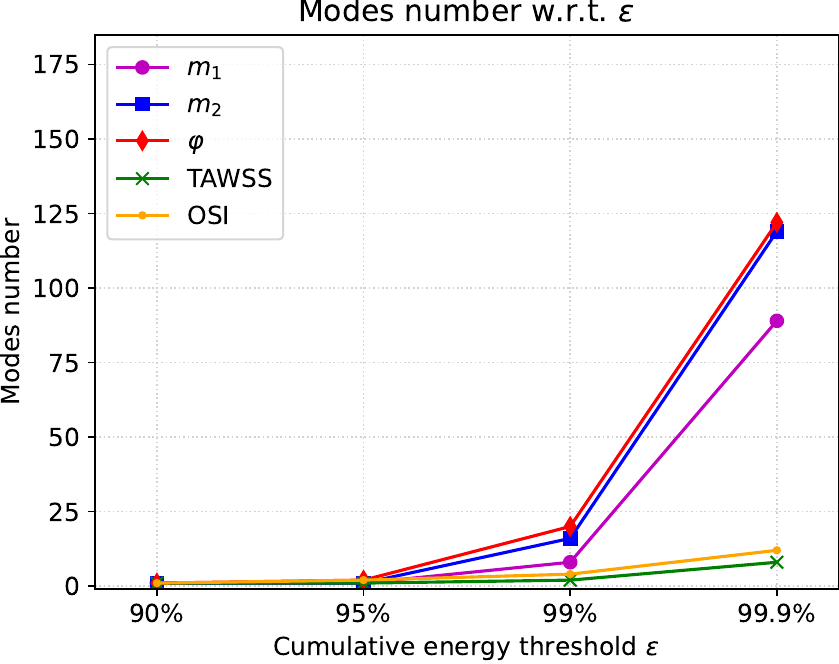}
}\\[1.1ex]
\subfloat[Newtonian case: relative error]{
\includegraphics[width=0.35\columnwidth]{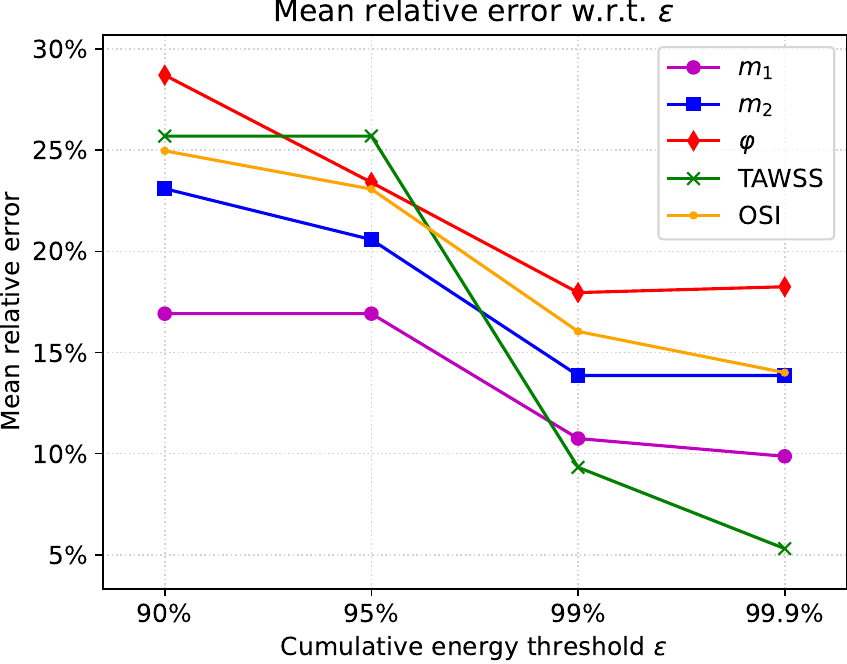}
}\quad
\subfloat[Casson's case: relative error]{
\includegraphics[width=0.35\columnwidth]{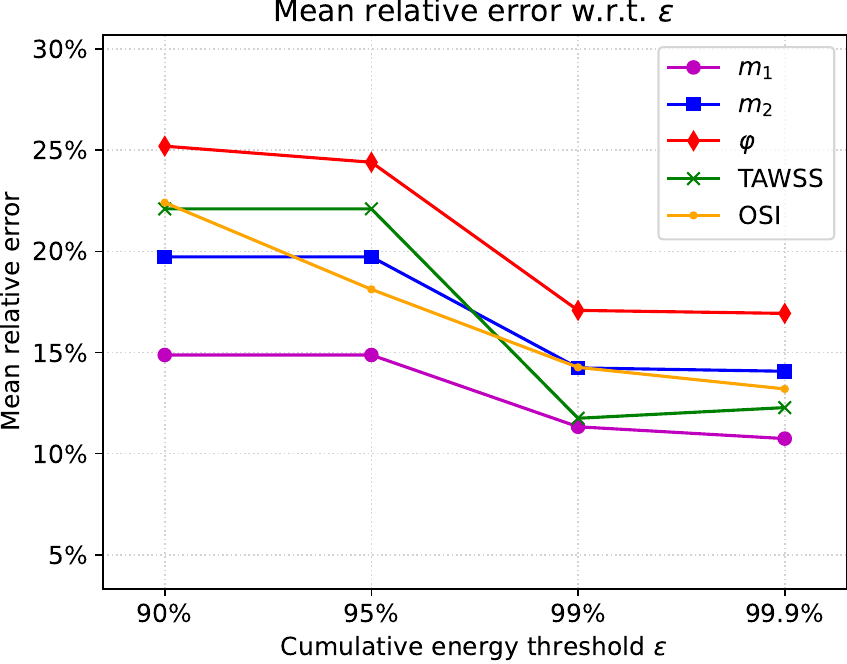}
}
\caption{Variation of the number of modes and of the relative error with respect to the cumulative energy threshold $\varepsilon$ for the Newtonian (first column) and non Newtonian (second column) case for all the variables involved. }% Domain: LAA \textcolor{red}{è solo su questo?}.}
\label{fig:threshold_choice}
\end{figure}

As explained in Section \ref{sec:POD}, the number of modes $L$ is generally selected through the cumulative energy threshold $\varepsilon$ in equation \eqref{eq:eigenEnergy} affecting the accuracy of the ROM approximation.
In Figure \ref{fig:threshold_choice} we plot the modes number as well as the relative error $\mathsf{e}$ (see equations \eqref{eq:errors} and \eqref{eq:errors2}), which is time average for $m_1$, $m_2$ and $\varphi$, against the value of $\varepsilon$ = $\{90\%, 95\%, 99\%,99.9\%\}$, for both the Newtonian and Casson's models. As expected, the relative error decreases at increasing of modes number, 
i.e.\ for higher values of $\varepsilon$. However for $\varepsilon\geq99\%$ the relative error reaches a plateau for most of the variables. On the contrary, the number of modes increases. Such increase is particularly pronounced for $m_1$, $m_2$ and $\varphi$ (and appears more evident in the Newtonian case, %In addition, we note that for $\varepsilon = 99\%$ and 99.9\% %\textcolor{red}{Almeno per 99.9 \% sarebbe utile scrivere nel testo esplicitamente quanti modi servono}, 
%the Casson's model requires a significantly lower number of modes with respect to the Newtonian one, 
although the error level exhibited by the two models is comparable). %Based on this results, in perspective it could be interesting to introduce, at least %At the aim to go more in depth about this, it could be interesting in a future work to move towards a nonlinear replace the POD with a non-line such as  as an encoder. \textcolor{red}{da qui}.
%Morever, both models require a 
 %much higher number of modes to reconstruct $m_1$, $m_2$ and $\varphi$ with respect to TAWSS and OSI. T
  This could be due to the fact that TAWSS and OSI derive from an average in time over the cardiac cycle whilst the other indices are time-dependent and therefore characterized by a richer modal content. %their greater complexity requires a larger number of modes. 
 We fix the value of $\varepsilon$ to $99\%$ for the following tests, as it guarantees a good accuracy and at the same time it allows to monitor the computational cost.

\subsection{ROM solutions}
\begin{figure}[h!]
\subfloat[Newtonian case: $m_{1}$]{\captionsetup{width=.8\linewidth}
\includegraphics[width=0.35\columnwidth]{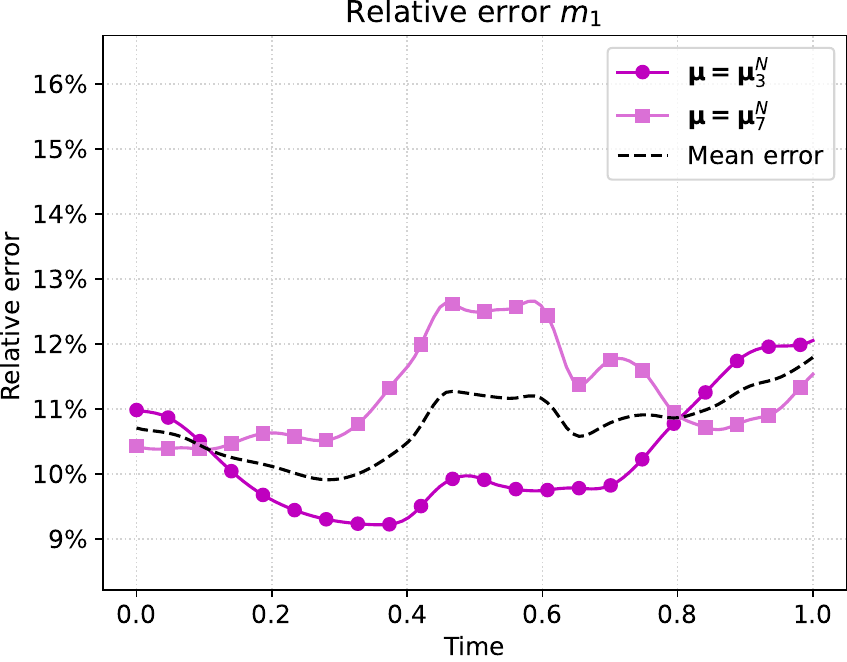}
}\quad
\subfloat[Casson's case: $m_{1}$]{\captionsetup{width=.8\linewidth}
\includegraphics[width=0.35\columnwidth]{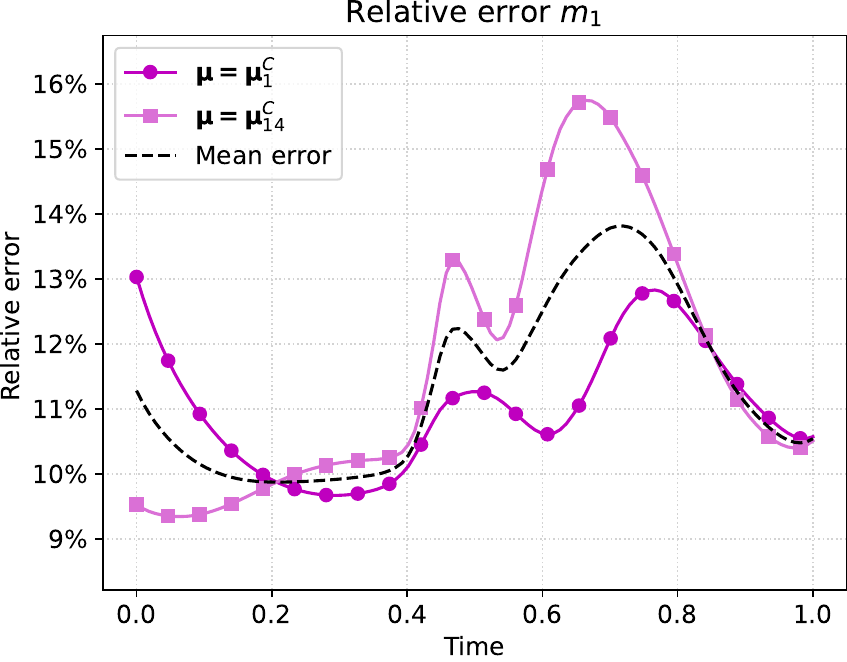}
}\\[1.1ex]
\subfloat[Newtonian case: $m_{2}$]{\captionsetup{width=.8\linewidth}
\includegraphics[width=0.35\columnwidth]{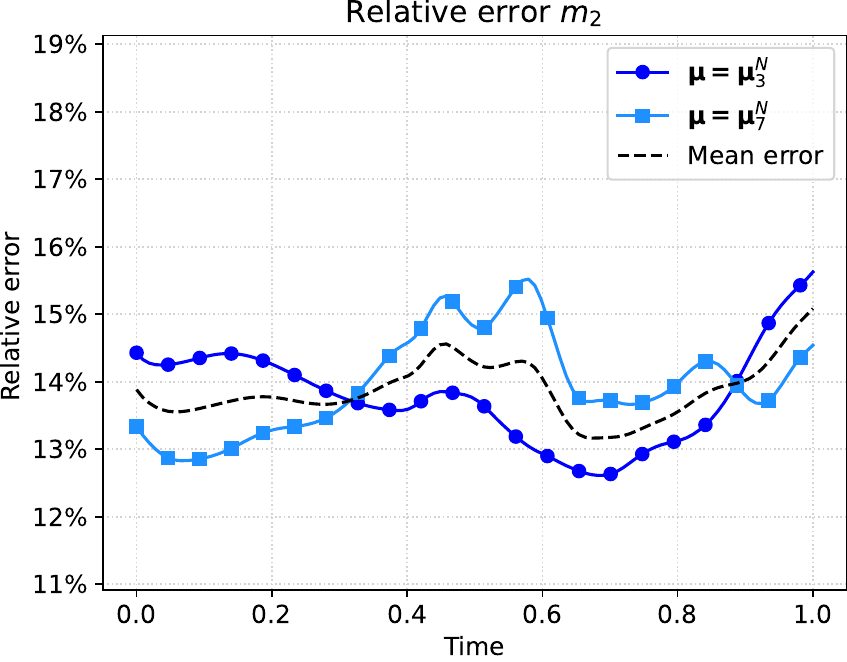}
}\quad
\subfloat[Casson's case: $m_{2}$]{\captionsetup{width=.8\linewidth}
\includegraphics[width=0.35\columnwidth]{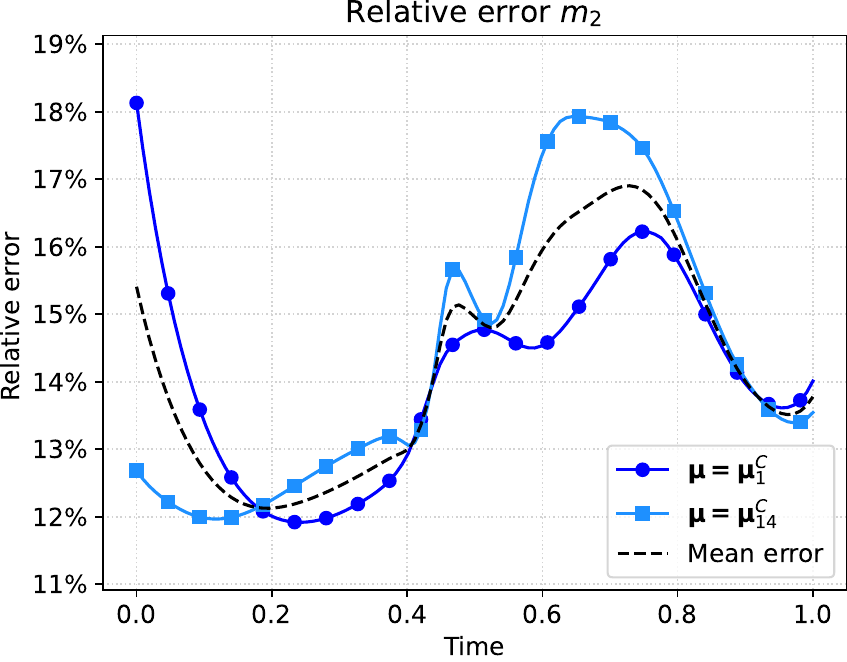}
}\\[1.1ex]
\subfloat[Newtonian case: $\varphi$]{\captionsetup{width=.8\linewidth}
\includegraphics[width=0.35\columnwidth]{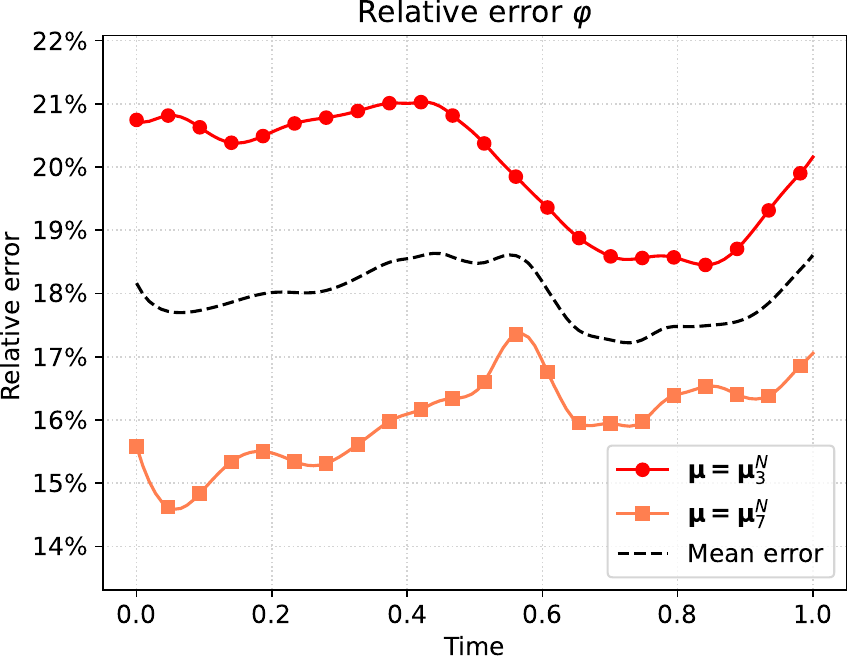}
}\quad
\subfloat[Casson's case: $\varphi$]{\captionsetup{width=.8\linewidth}
\includegraphics[width=0.35\columnwidth]{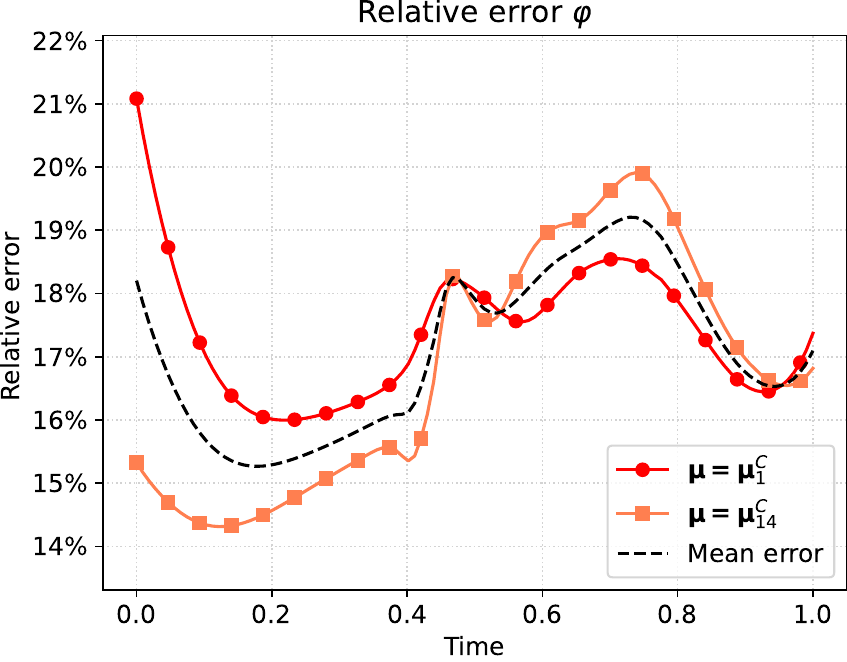}
}
\caption{Variation in time of the relative error corresponding to the test parameters values for the Newtonian (first column) and non Newtonian (second column) case for $m_1$, $m_2$ and $\varphi$.} %in $\mathcal{M}_\test$. Domain: LAA.}
\label{fig:rel_errors}
\end{figure}

Once set the cumulative energy threshold $\varepsilon=99\%$, we proceed with the ROM simulations. % for the variables $m_1$, $m_2$, $\varphi$, $\tawss$ and $\osi$ for both the Newtonian and Casson's models. %Let us briefly summarize the procedure. We perform the POD on the snapshot matrix $S^\Phi_\train$ and we apply the RBF interpolation algorithm to define a map between the parameter space $\mathcal{P}$ and the modal coefficients space. Then, we compute the approximate solutions \eqref{eq:ROMsolution} corresponding to the test parameters. 
In Figure \ref{fig:rel_errors} we show the variation in time of the relative error $\mathsf{e}(t,\boldsymbol{\mu})$ defined in equation \eqref{eq:errors} for $m_1$, $m_2$ and $\varphi$ associated with the parameters values in $\mathcal{K}_\test^N$ and $\mathcal{K}_\test^C$, together with their mean values. %The plots refer only to the time-dependent variables $m_1$, $m_2$ and $\varphi$.
Overall, we observe that the relative errors vary between 9\% and 21\% demonstrating a clinical relevance. %demonstrating a good accuracy from a clinical viewpoint. 
More specifically, 
in the Newtonian case (left plots), the relative errors %$\mathsf{e}(t,\mubold^N_{3})$ and $\mathsf{e}(t,\mubold^N_{7})$ 
are quite similar to each other. They, along with the associated mean value, do not exhibit large oscillations during the cardiac cycle.
 For the Casson's model (right plots) we observe that the error increases immediately after the opening of the MV (happening around $t=0.4$, see Figure \ref{fig:geometry}), so unlike the Newtonian case the error is affected by wide oscillations. %Furthermore the results obtained with the parameter $\mubold^C_1$ are more accurate than those associated with $\mubold^C_{14}$. 
 We also note  the increase of the error  during the first few time steps which might be due to the transient nature of the flow, as also
noted in other configurations including academic benchmarks \cite{girfoglio2019finite}. %It is more evident in the Casson's model. This behavior is related to the fact that we can obtain good results for the parameters values inside the set chosen to train the ROM, but usually data-driven approaches are not predictive, i.e. they are not able to generate a proper approximation outside the training set. %\textcolor{red}{Michele: maybe some further comments}

\begin{figure}[h!]
    \centering
    \subfloat[$m_1$, $t=0$]{\includegraphics[trim={50 10 50 10}, width=0.17\columnwidth]{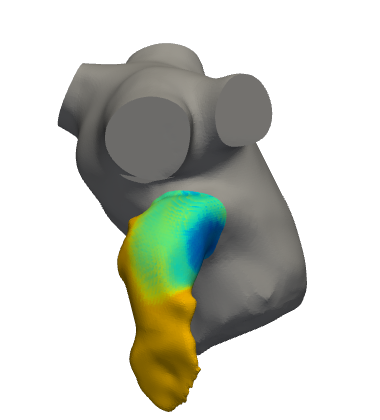}}\,
    \subfloat[$m_1$, $t=0.25$]{\includegraphics[trim={50 10 50 10}, width=0.17\columnwidth]{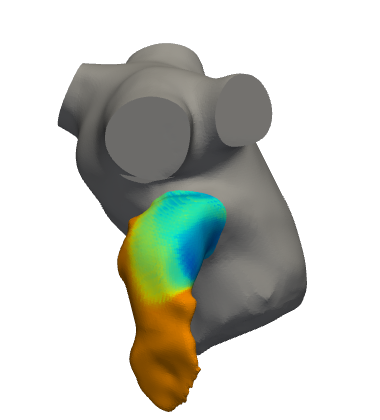}}\,
    \subfloat[$m_1$, $t=0.5$]{\includegraphics[trim={50 10 50 10}, width=0.17\columnwidth]{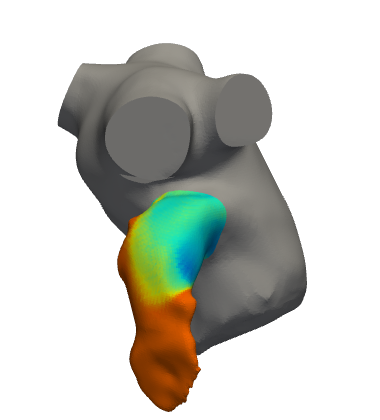}}\,
    \subfloat[$m_1$, $t=0.75$]{\includegraphics[trim={50 10 50 10}, width=0.17\columnwidth]{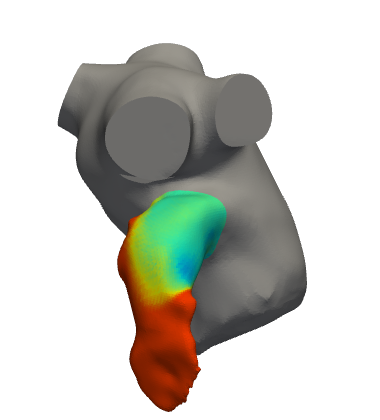}}\,
    \subfloat[$m_1$, $t=1$]{\includegraphics[trim={50 10 50 10}, width=0.17\columnwidth]{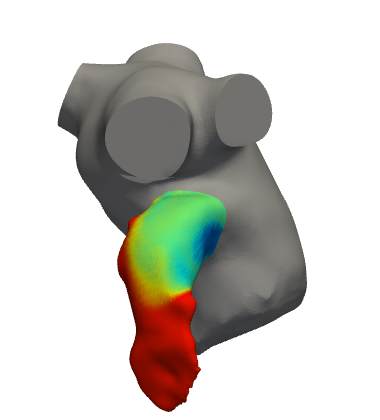}}\,
    \subfloat{\begin{overpic}[abs,unit=1mm,width=0.5cm]{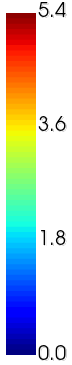}\put(5,12){\tiny\textcolor{black!90}{[s]}}\end{overpic}}\\
    \centering
    \subfloat[$m_1^\rb$, $t=0$]{\includegraphics[trim={50 10 50 10}, width=0.17\columnwidth]{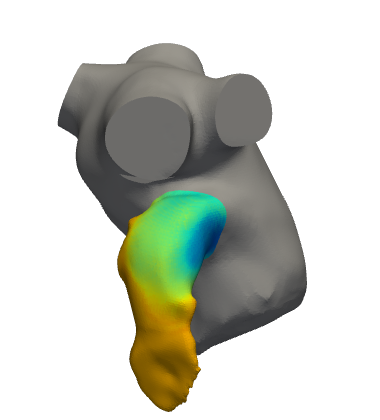}}\,
    \subfloat[$m_1^\rb$, $t=0.25$]{\includegraphics[trim={50 10 50 10}, width=0.17\columnwidth]{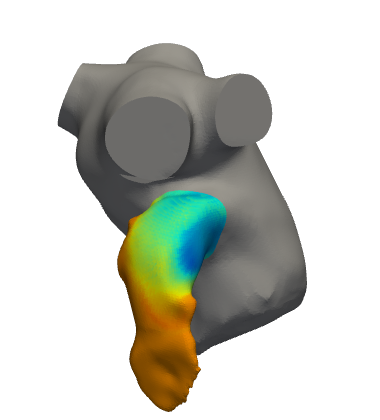}}\,
    \subfloat[$m_1^\rb$, $t=0.5$]{\includegraphics[trim={50 10 50 10}, width=0.17\columnwidth]{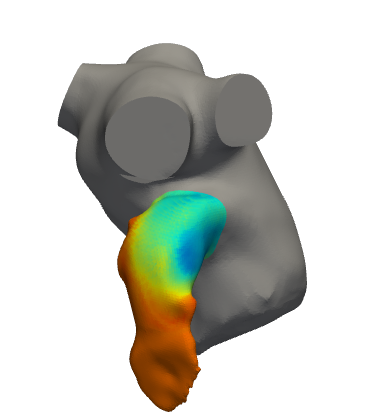}}\,
    \subfloat[$m_1^\rb$, $t=0.75$]{\includegraphics[trim={50 10 50 10}, width=0.17\columnwidth]{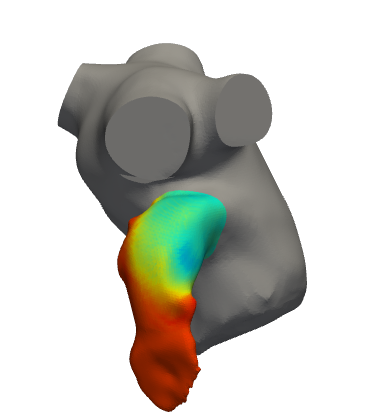}}\,
    \subfloat[$m_1^\rb$, $t=1$]{\includegraphics[trim={50 10 50 10}, width=0.17\columnwidth]{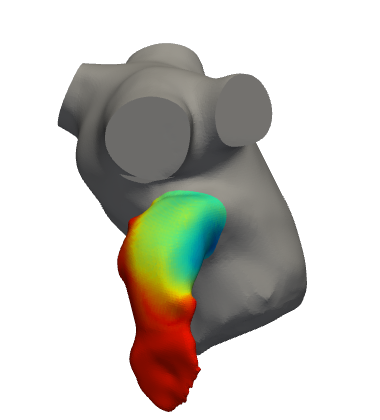}}\,
    \subfloat{\begin{overpic}[abs,unit=1mm,width=0.5cm]{images/m1colorbar.png}\put(5,12){\tiny\textcolor{black!90}{[s]}}\end{overpic}}\\
    \caption{Newtonian case: qualitative comparison between FOM (top) and ROM (bottom) solutions at different times for $m_1$.}
    \label{fig:qualitativeNm1}
\end{figure}

\begin{figure}[h!]
    \centering
    \subfloat[$m_1$, $t=0$]{\includegraphics[trim={50 10 50 10}, width=0.17\columnwidth]{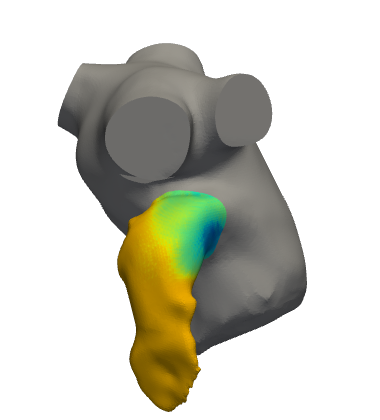}}\,
    \subfloat[$m_1$, $t=0.25$]{\includegraphics[trim={50 10 50 10}, width=0.17\columnwidth]{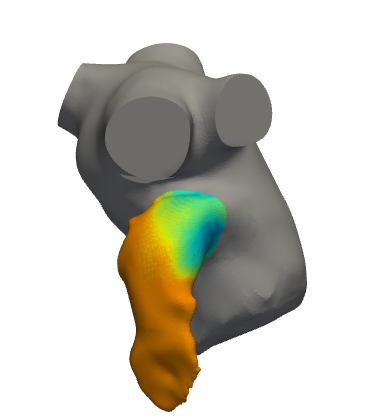}}\,
    \subfloat[$m_1$, $t=0.5$]{\includegraphics[trim={50 10 50 10}, width=0.17\columnwidth]{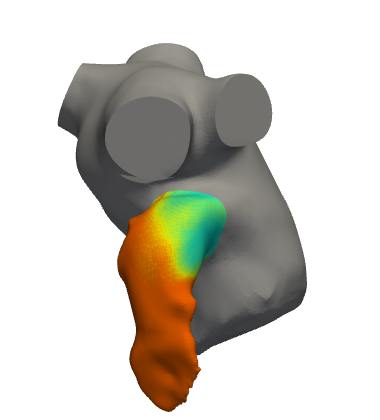}}\,
    \subfloat[$m_1$, $t=0.75$]{\includegraphics[trim={50 10 50 10}, width=0.17\columnwidth]{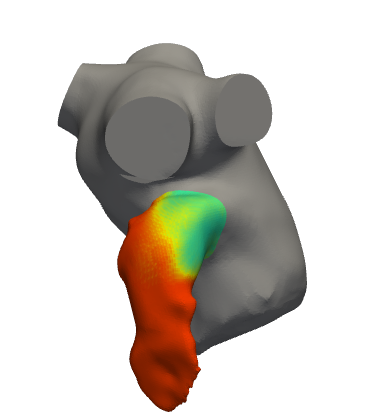}}\,
    \subfloat[$m_1$, $t=1$]{\includegraphics[trim={50 10 50 10}, width=0.17\columnwidth]{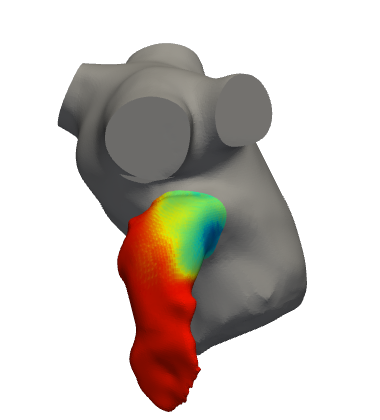}}\,
    \subfloat{\begin{overpic}[abs,unit=1mm,width=0.5cm]{images/m1colorbar.png}\put(5,12){\tiny\textcolor{black!90}{[s]}}\end{overpic}}\\
    \subfloat[$m_1^\rb$, $t=0$]{\includegraphics[trim={50 10 50 10}, width=0.17\columnwidth]{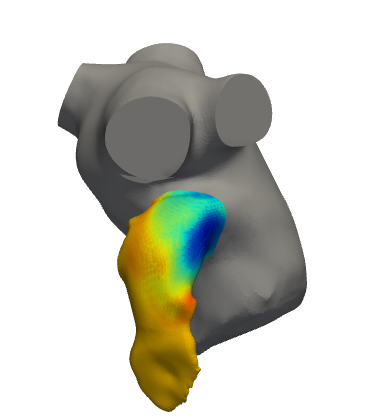}}\,
    \subfloat[$m_1^\rb$, $t=0.25$]{\includegraphics[trim={50 10 50 10}, width=0.17\columnwidth]{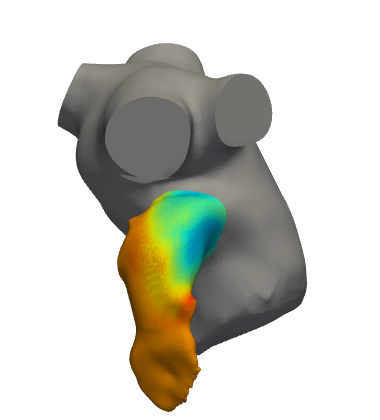}}\,
    \subfloat[$m_1^\rb$, $t=0.5$]{\includegraphics[trim={50 10 50 10}, width=0.17\columnwidth]{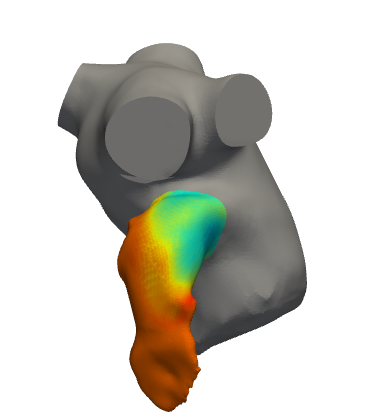}}\,
    \subfloat[$m_1^\rb$, $t=0.75$]{\includegraphics[trim={50 10 50 10}, width=0.17\columnwidth]{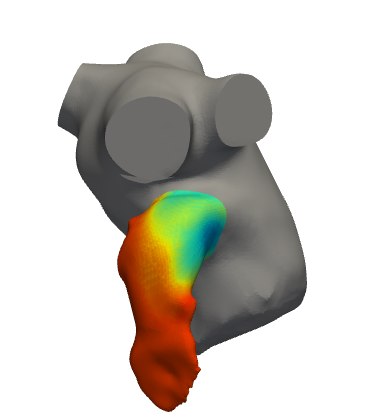}}\,
    \subfloat[$m_1^\rb$, $t=1$]{\includegraphics[trim={50 10 50 10}, width=0.17\columnwidth]{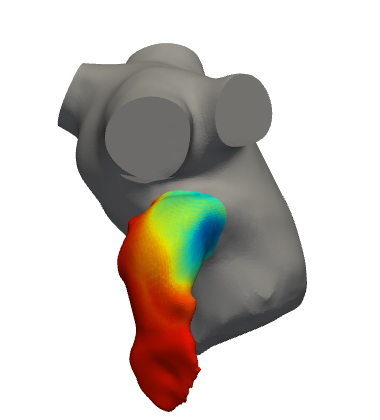}}\,
    \subfloat{\begin{overpic}[abs,unit=1mm,width=0.5cm]{images/m1colorbar.png}\put(5,12){\tiny\textcolor{black!90}{[s]}}\end{overpic}}\\
    \caption{Casson's case: qualitative comparison between FOM (top) and ROM (bottom) solutions, different times for $m_1$.}
    \label{fig:qualitativeCm1}
\end{figure}

\begin{figure}[hpbt!]
    \centering
    \subfloat[$m_2$, $t=0$]{\includegraphics[trim={50 10 50 10}, width=0.17\columnwidth]{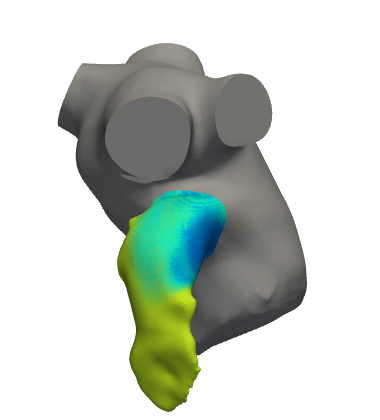}}\,
    \subfloat[$m_2$, $t=0.25$]{\includegraphics[trim={50 10 50 10}, width=0.17\columnwidth]{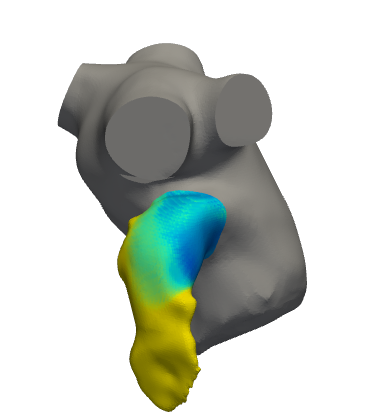}}\,
    \subfloat[$m_2$, $t=0.5$]{\includegraphics[trim={50 10 50 10}, width=0.17\columnwidth]{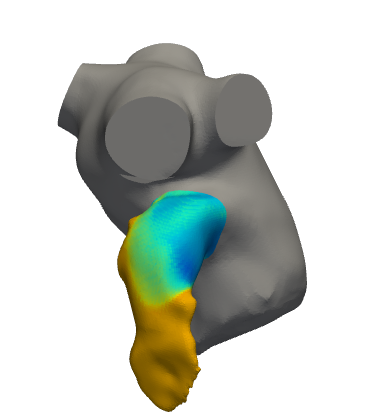}}\,
    \subfloat[$m_2$, $t=0.75$]{\includegraphics[trim={50 10 50 10}, width=0.17\columnwidth]{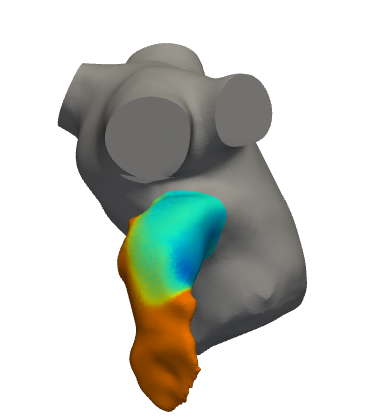}}\,
    \subfloat[$m_2$, $t=1$]{\includegraphics[trim={50 10 50 10}, width=0.17\columnwidth]{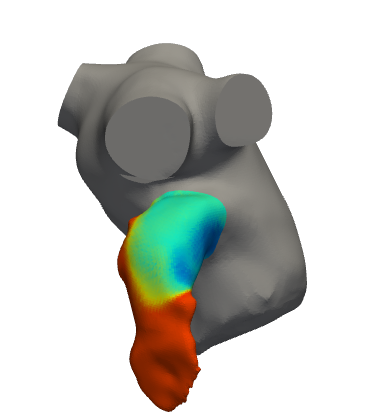}}\,
    \subfloat{\begin{overpic}[abs,unit=1mm,width=0.5cm]{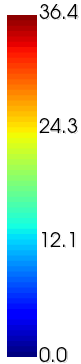}\put(5,10){\tiny\textcolor{black!90}{[s$^2$]}}\end{overpic}}\\
    \subfloat[$m_2^\rb$, $t=0$]{\includegraphics[trim={50 10 50 10}, width=0.17\columnwidth]{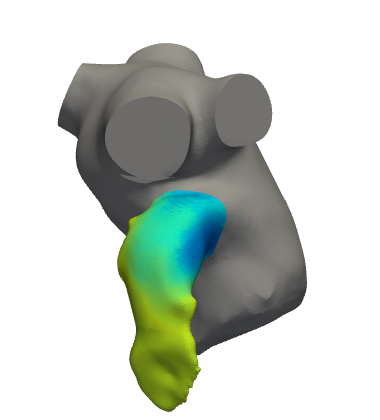}}\,
    \subfloat[$m_2^\rb$, $t=0.25$]{\includegraphics[trim={50 10 50 10}, width=0.17\columnwidth]{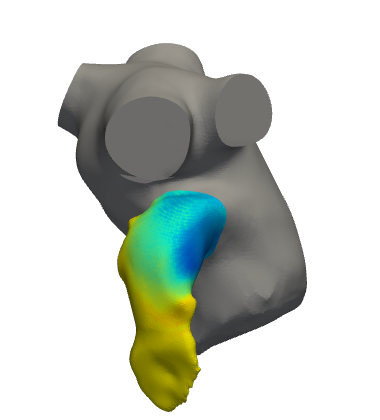}}\,
    \subfloat[$m_2^\rb$, $t=0.5$]{\includegraphics[trim={50 10 50 10}, width=0.17\columnwidth]{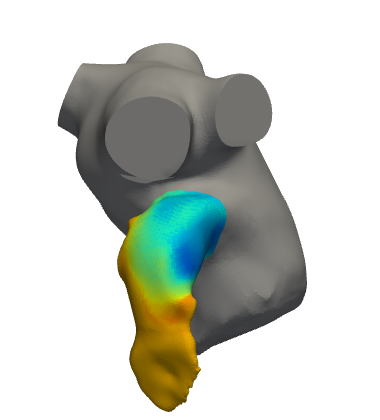}}\,
    \subfloat[$m_2^\rb$, $t=0.75$]{\includegraphics[trim={50 10 50 10}, width=0.17\columnwidth]{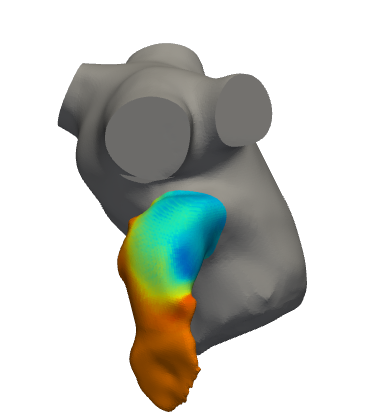}}\,
    \subfloat[$m_2^\rb$, $t=1$]{\includegraphics[trim={50 10 50 10}, width=0.17\columnwidth]{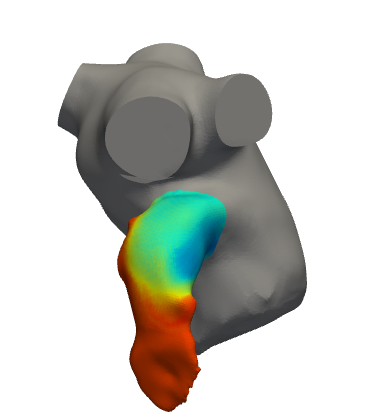}}\,
    \subfloat{\begin{overpic}[abs,unit=1mm,width=0.5cm]{images/m2colorbar.png}\put(5,10){\tiny\textcolor{black!90}{[s$^2$]}}\end{overpic}}\\
    \caption{Newtonian case: qualitative comparison between FOM (top) and ROM (bottom) solutions at different times for $m_2$.}
    \label{fig:qualitativeNm2}
\end{figure}

\begin{figure}[hpbt!]
    \centering
    \subfloat[$m_2$, $t=0$]{\includegraphics[trim={50 10 50 10}, width=0.17\columnwidth]{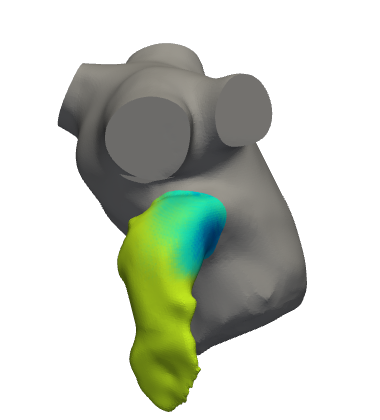}}\,
    \subfloat[$m_2$, $t=0.25$]{\includegraphics[trim={50 10 50 10}, width=0.17\columnwidth]{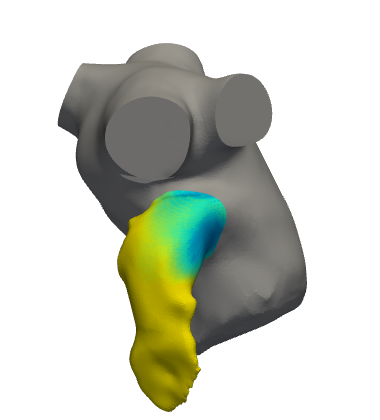}}\,
    \subfloat[$m_2$, $t=0.5$]{\includegraphics[trim={50 10 50 10}, width=0.17\columnwidth]{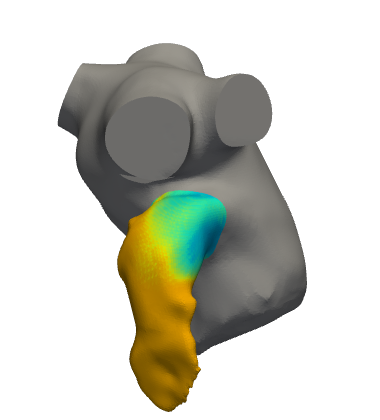}}\,
    \subfloat[$m_2$, $t=0.75$]{\includegraphics[trim={50 10 50 10}, width=0.17\columnwidth]{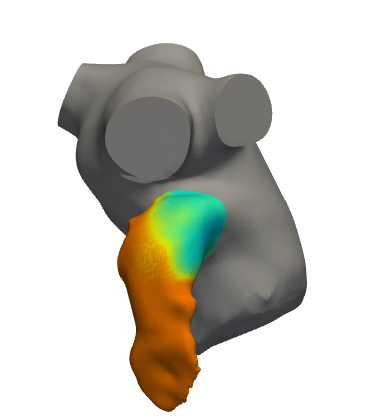}}\,
    \subfloat[$m_2$, $t=1$]{\includegraphics[trim={50 10 50 10}, width=0.17\columnwidth]{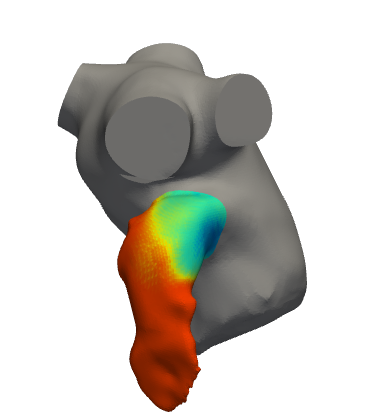}}\,
    \subfloat{\begin{overpic}[abs,unit=1mm,width=0.5cm]{images/m2colorbar.png}\put(5,10){\tiny\textcolor{black!90}{[s$^2$]}}\end{overpic}}\\
    \subfloat[$m_2^\rb$, $t=0$]{\includegraphics[trim={50 10 50 10}, width=0.17\columnwidth]{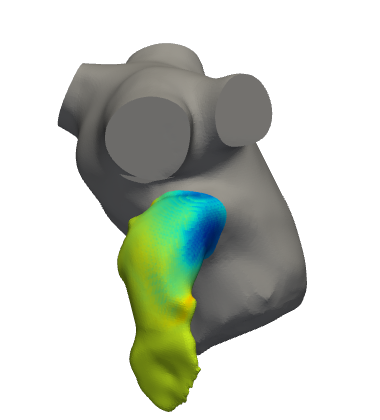}}\,
    \subfloat[$m_2^\rb$, $t=0.25$]{\includegraphics[trim={50 10 50 10}, width=0.17\columnwidth]{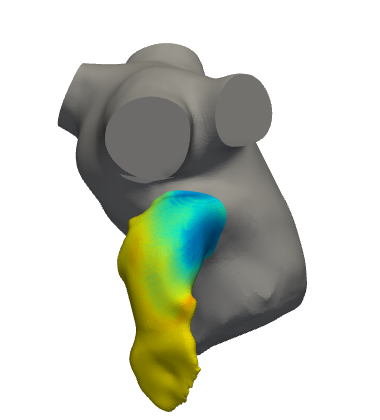}}\,
    \subfloat[$m_2^\rb$, $t=0.5$]{\includegraphics[trim={50 10 50 10}, width=0.17\columnwidth]{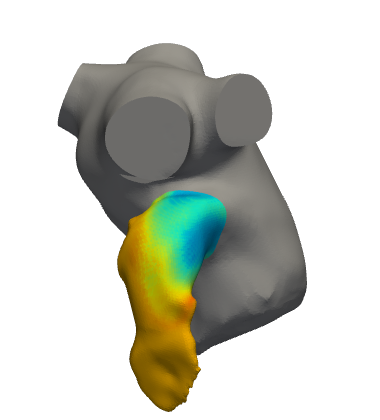}}\,
    \subfloat[$m_2^\rb$, $t=0.75$]{\includegraphics[trim={50 10 50 10}, width=0.17\columnwidth]{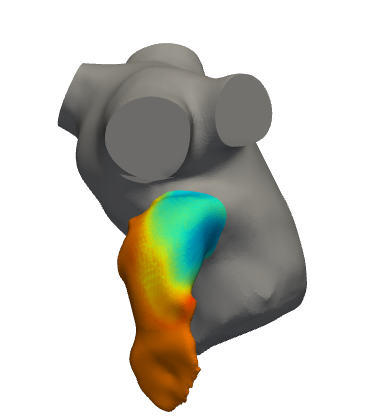}}\,
    \subfloat[$m_2^\rb$, $t=1$]{\includegraphics[trim={50 10 50 10}, width=0.17\columnwidth]{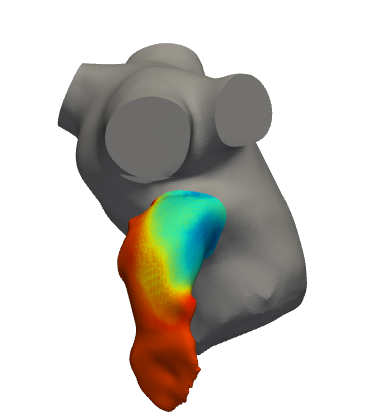}}\,
    \subfloat{\begin{overpic}[abs,unit=1mm,width=0.5cm]{images/m2colorbar.png}\put(5,10){\tiny\textcolor{black!90}{[s$^2$]}}\end{overpic}}\\
    \caption{Casson's case: qualitative comparison between FOM (top) and ROM (bottom) solutions at different times for $m_2$.}
    \label{fig:qualitativeCm2}
\end{figure}

\begin{figure}[hpbt!]
    \centering
    \subfloat[$\varphi$, $t=0$]{\includegraphics[trim={50 10 50 10}, width=0.17\columnwidth]{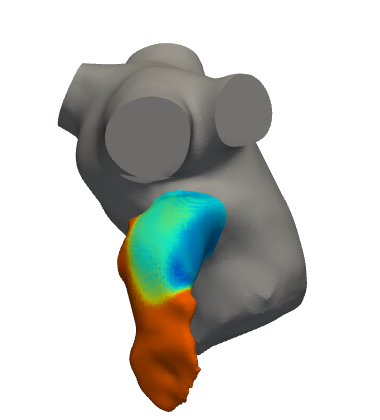}}\,
    \subfloat[$\varphi$, $t=0.25$]{\includegraphics[trim={50 10 50 10}, width=0.17\columnwidth]{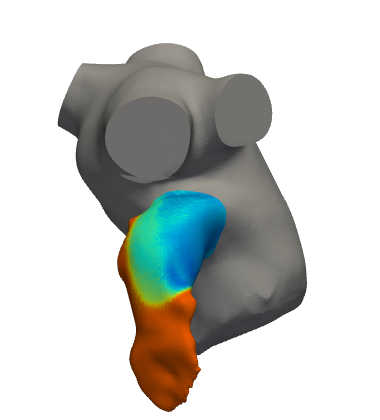}}\,
    \subfloat[$\varphi$, $t=0.5$]{\includegraphics[trim={50 10 50 10}, width=0.17\columnwidth]{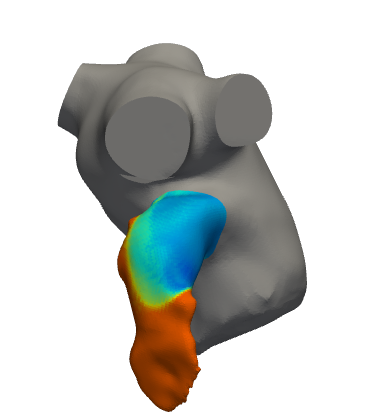}}\,
    \subfloat[$\varphi$, $t=0.75$]{\includegraphics[trim={50 10 50 10}, width=0.17\columnwidth]{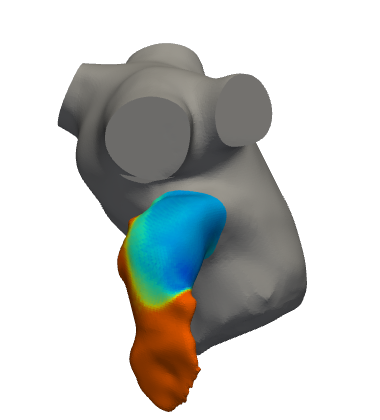}}\,
    \subfloat[$\varphi$, $t=1$]{\includegraphics[trim={50 10 50 10}, width=0.17\columnwidth]{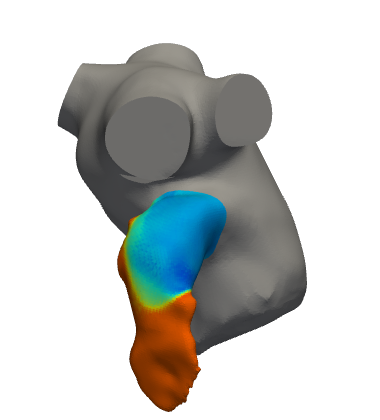}}\,
    \subfloat{\includegraphics[width=0.5cm]{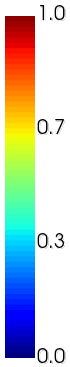}}\\
    \subfloat[$\varphi^\rb$, $t=0$]{\includegraphics[trim={50 10 50 10}, width=0.17\columnwidth]{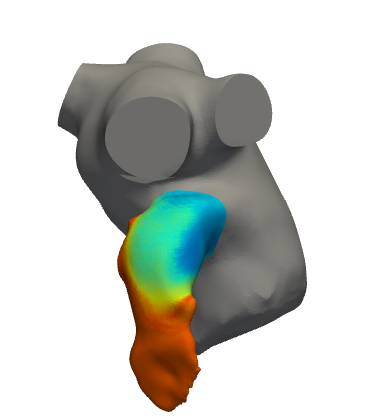}}\,
    \subfloat[$\varphi^\rb$, $t=0.25$]{\includegraphics[trim={50 10 50 10}, width=0.17\columnwidth]{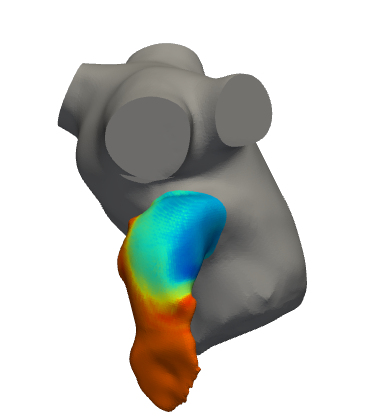}}\,
    \subfloat[$\varphi^\rb$, $t=0.5$]{\includegraphics[trim={50 10 50 10}, width=0.17\columnwidth]{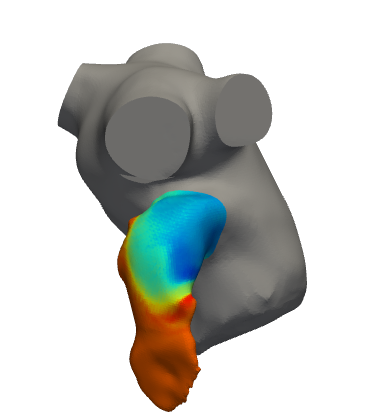}}\,
    \subfloat[$\varphi^\rb$, $t=0.75$]{\includegraphics[trim={50 10 50 10}, width=0.17\columnwidth]{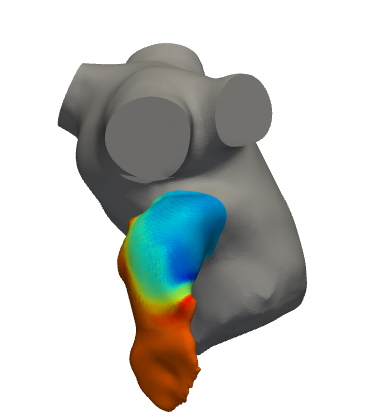}}\,
    \subfloat[$\varphi^\rb$, $t=1$]{\includegraphics[trim={50 10 50 10}, width=0.17\columnwidth]{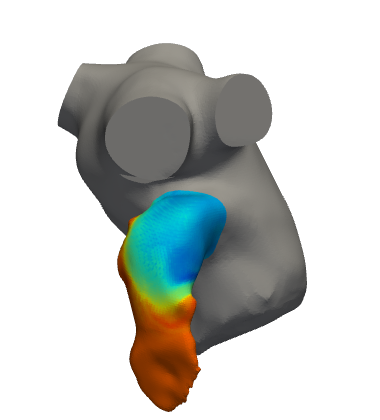}}\,
    \subfloat{\includegraphics[width=0.5cm]{images/wcolorbar.png}}\\
    \caption{Newtonian case: qualitative comparison between FOM (top) and ROM (bottom) solutions at different times for $\varphi$.}
    \label{fig:qualitativeNw}
\end{figure}

\begin{figure}[hpbt!]
    \centering
    \subfloat[$\varphi$, $t=0$]{\includegraphics[trim={50 10 50 10}, width=0.17\columnwidth]{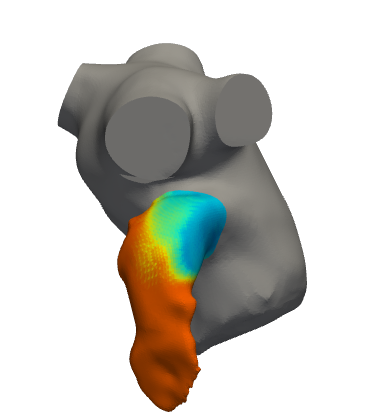}}\,
    \subfloat[$\varphi$, $t=0.25$]{\includegraphics[trim={50 10 50 10}, width=0.17\columnwidth]{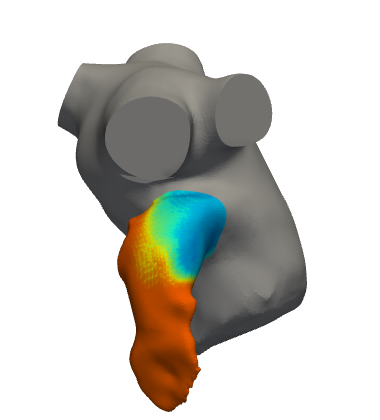}}\,
    \subfloat[$\varphi$, $t=0.5$]{\includegraphics[trim={50 10 50 10}, width=0.17\columnwidth]{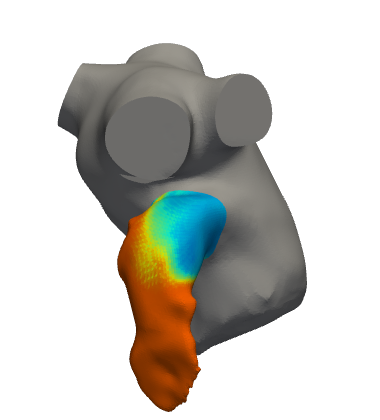}}\,
    \subfloat[$\varphi$, $t=0.75$]{\includegraphics[trim={50 10 50 10}, width=0.17\columnwidth]{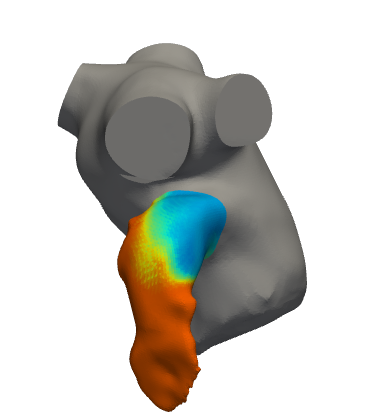}}\,
    \subfloat[$\varphi$, $t=1$]{\includegraphics[trim={50 10 50 10}, width=0.17\columnwidth]{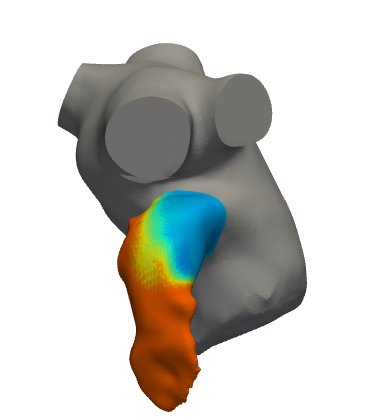}}\,
    \subfloat{\includegraphics[width=0.5cm]{images/wcolorbar.png}}\\
    \subfloat[$\varphi^\rb$, $t=0$]{\includegraphics[trim={50 10 50 10}, width=0.17\columnwidth]{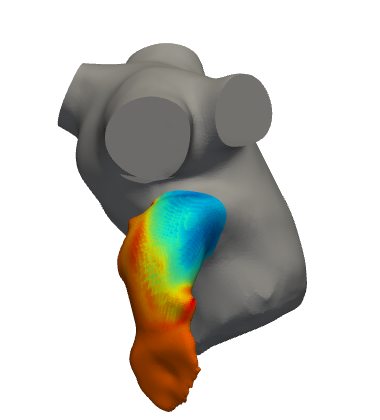}}\,
    \subfloat[$\varphi^\rb$, $t=0.25$]{\includegraphics[trim={50 10 50 10}, width=0.17\columnwidth]{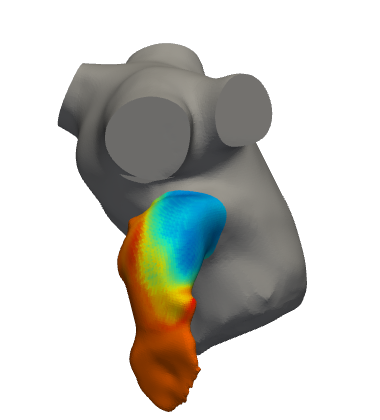}}\,
    \subfloat[$\varphi^\rb$, $t=0.5$]{\includegraphics[trim={50 10 50 10}, width=0.17\columnwidth]{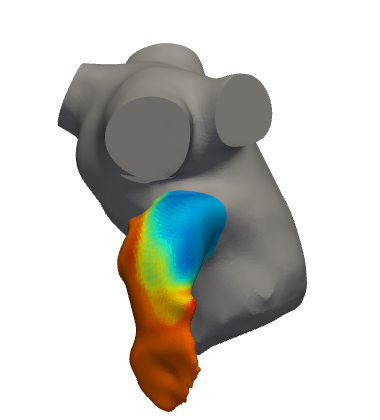}}\,
    \subfloat[$\varphi^\rb$, $t=0.75$]{\includegraphics[trim={50 10 50 10}, width=0.17\columnwidth]{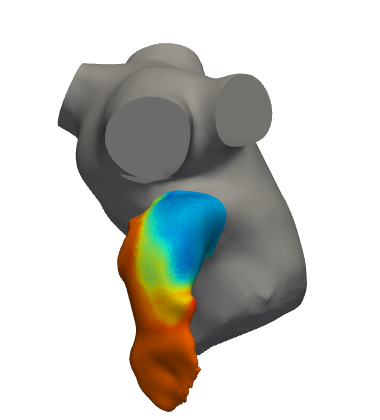}}\,
    \subfloat[$\varphi^\rb$, $t=1$]{\includegraphics[trim={50 10 50 10}, width=0.17\columnwidth]{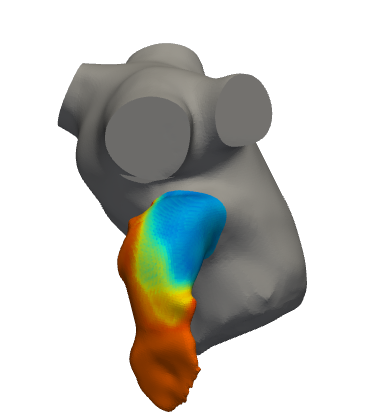}}\,
    \subfloat{\includegraphics[width=0.5cm]{images/wcolorbar.png}}\\
    \caption{Casson's case: qualitative comparison between FOM (top) and ROM (bottom) solutions at different times for $\varphi$.}
    \label{fig:qualitativeCw}
\end{figure}

\begin{figure}[hpbt!]
    \centering
    \subfloat[$\tawss$]{\includegraphics[trim={100 10 100 10}, width=0.17\columnwidth]{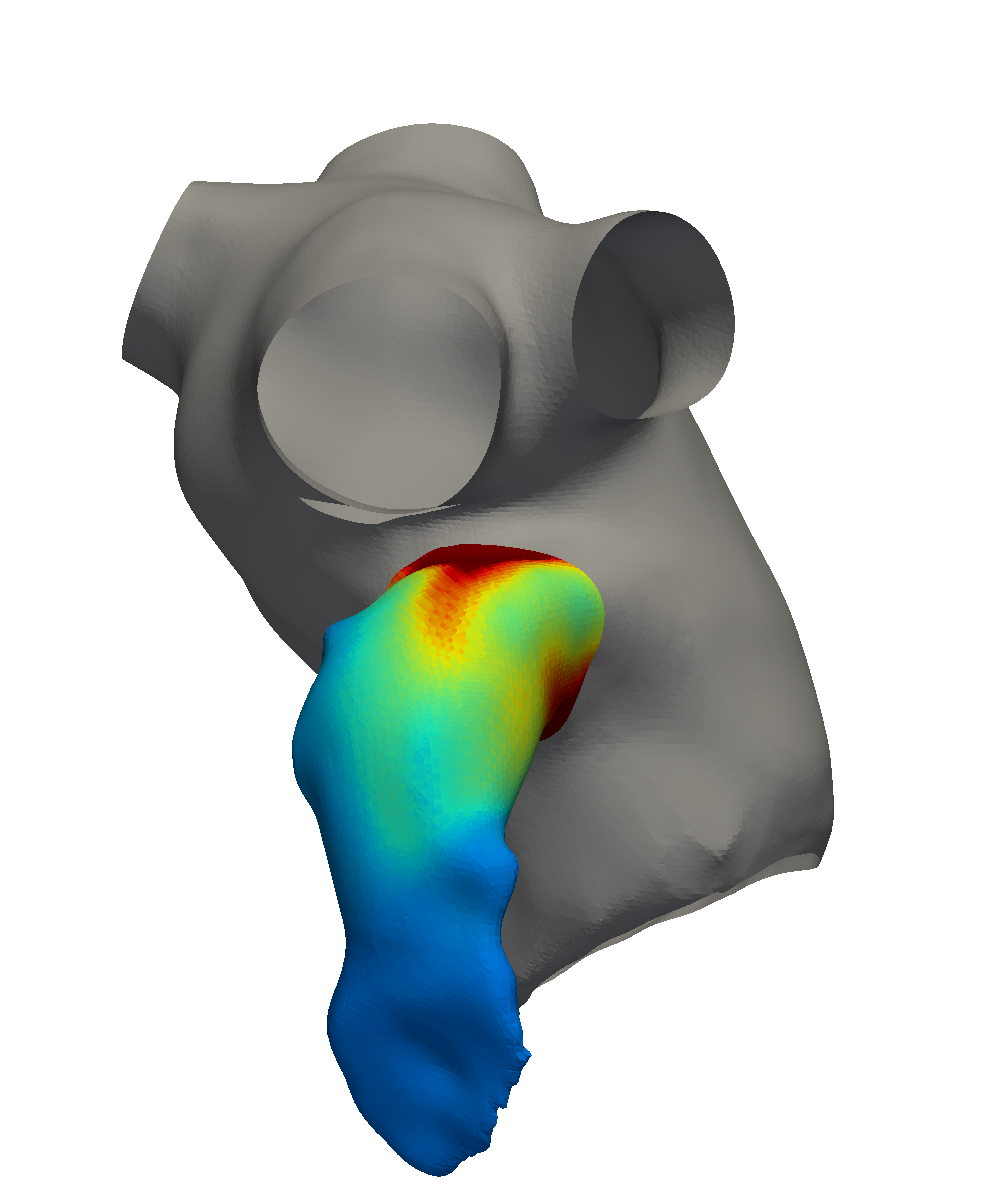}}\,
    \subfloat[$\tawss^\rb$]{\includegraphics[trim={100 10 100 10}, width=0.17\columnwidth]{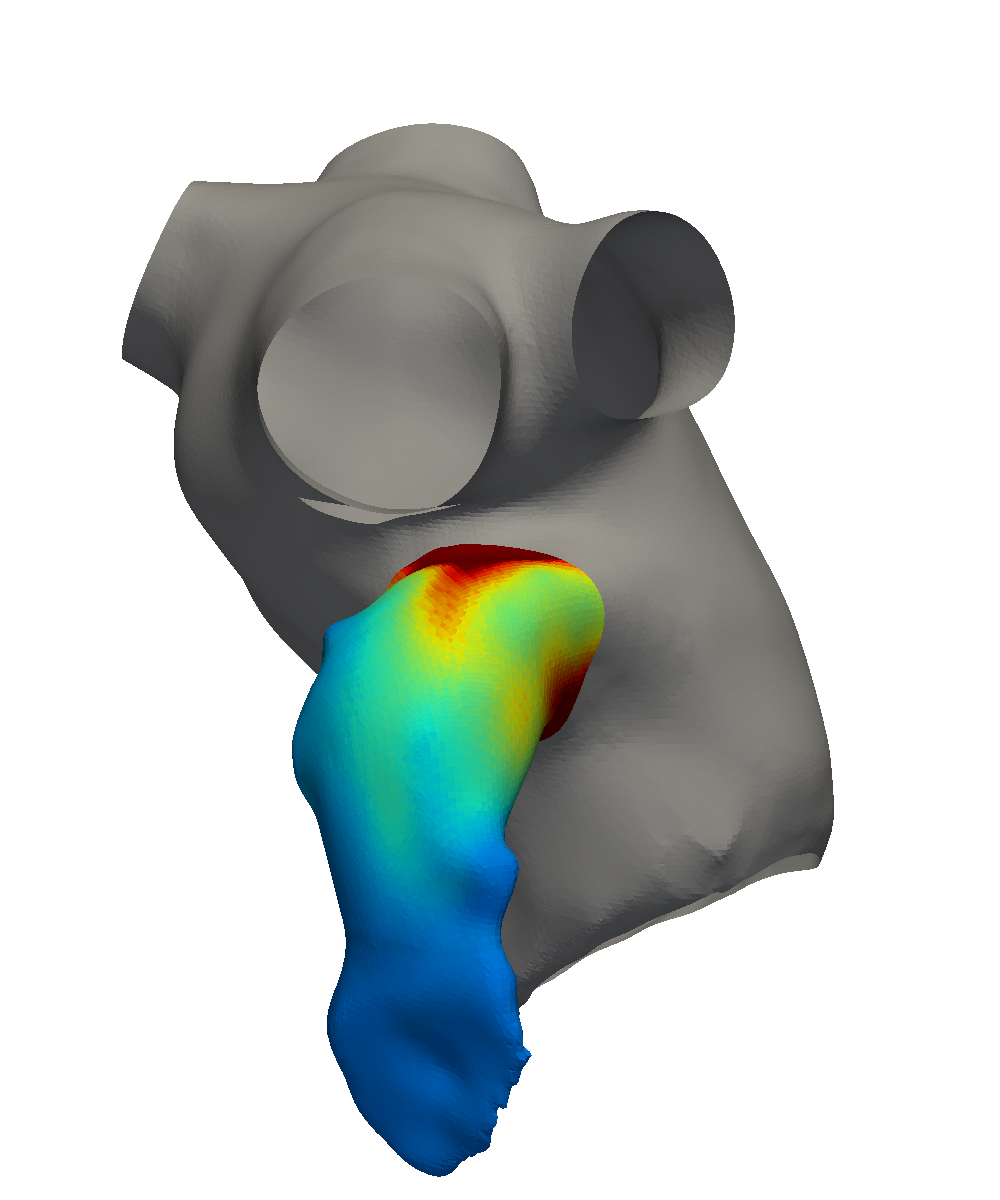}}\,
    \subfloat{\begin{overpic}[abs,unit=1mm,width=0.8cm]{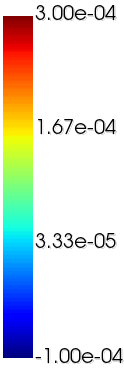}\put(7,11){\tiny\textcolor{black!90}{[Pa]}}\end{overpic}}\,
    \subfloat[$\osi$]{\includegraphics[trim={100 10 100 10}, width=0.17\columnwidth]{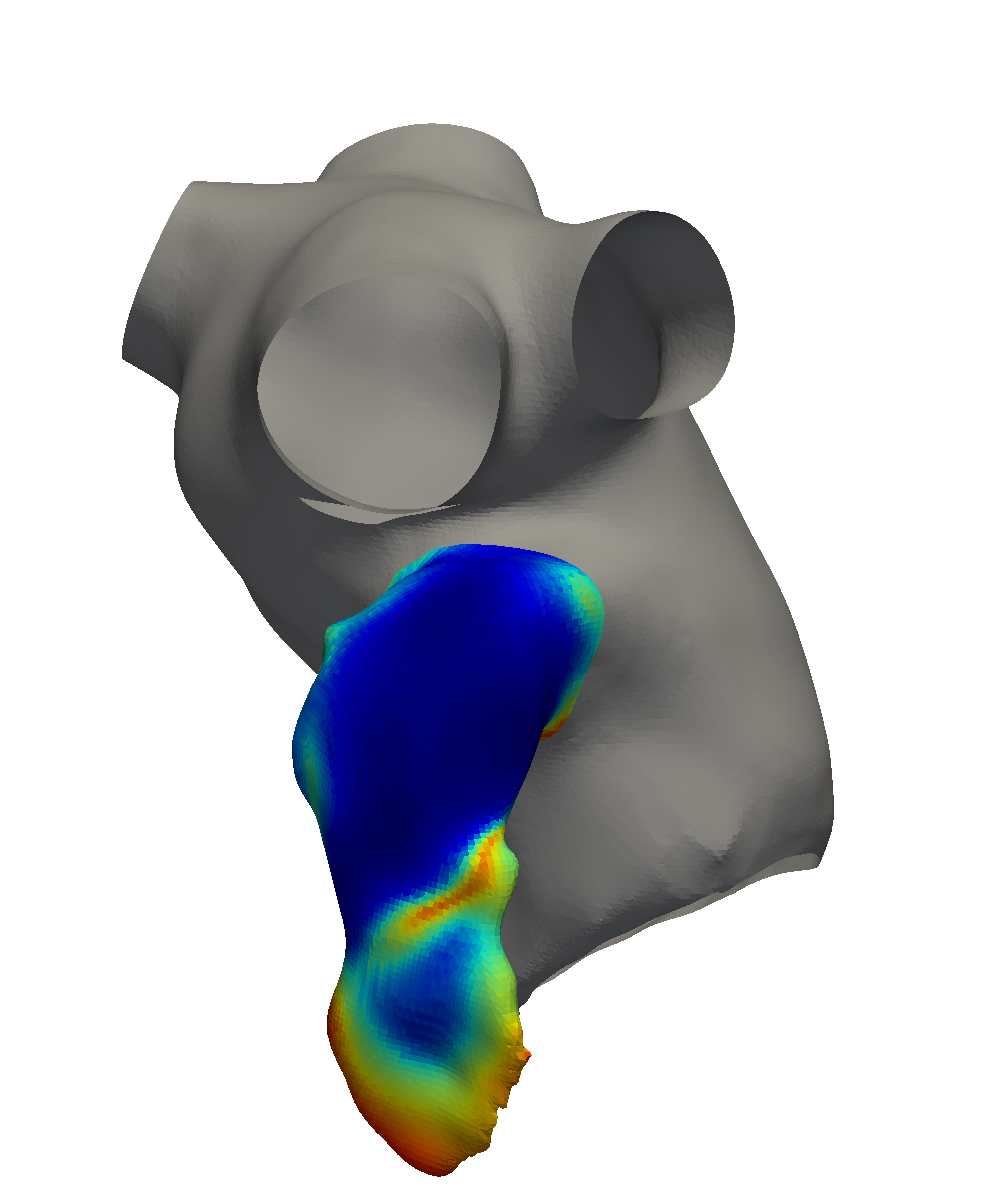}}\,
    \subfloat[$\osi^\rb$]{\includegraphics[trim={100 10 100 10}, width=0.17\columnwidth]{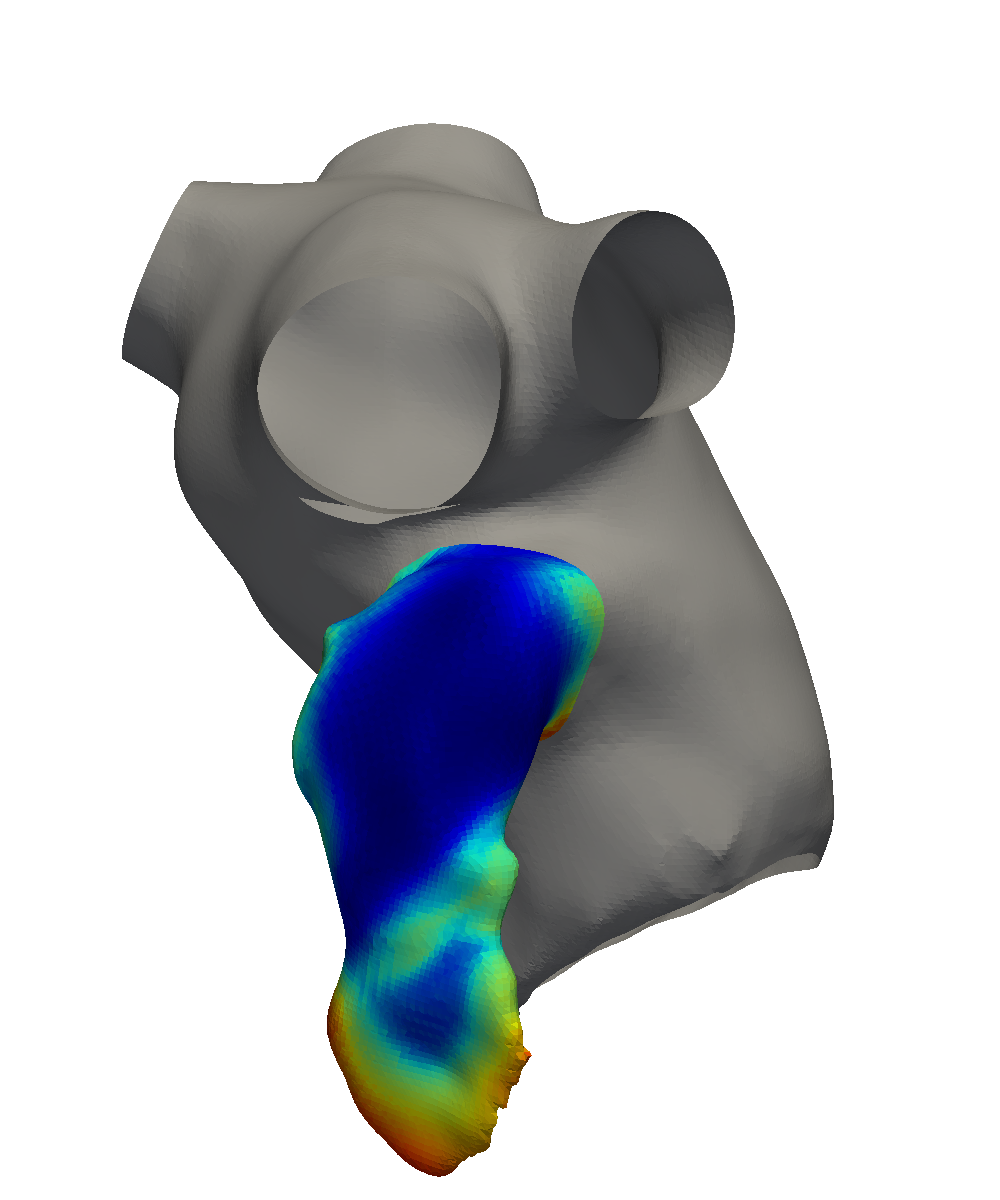}}\,
    \subfloat{\includegraphics[width=0.85cm]{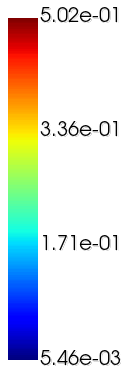}}\,
    \caption{Newtonian case: qualitative comparison between FOM and ROM solutions for $\tawss$ (panels A and B) and $\osi$ (panels D and E).}
    \label{fig:qualitativeTAWSS}
\end{figure}

\begin{figure}[hpbt!]
    \centering
    \subfloat[$\tawss$]{\includegraphics[trim={100 10 100 10}, width=0.17\columnwidth]{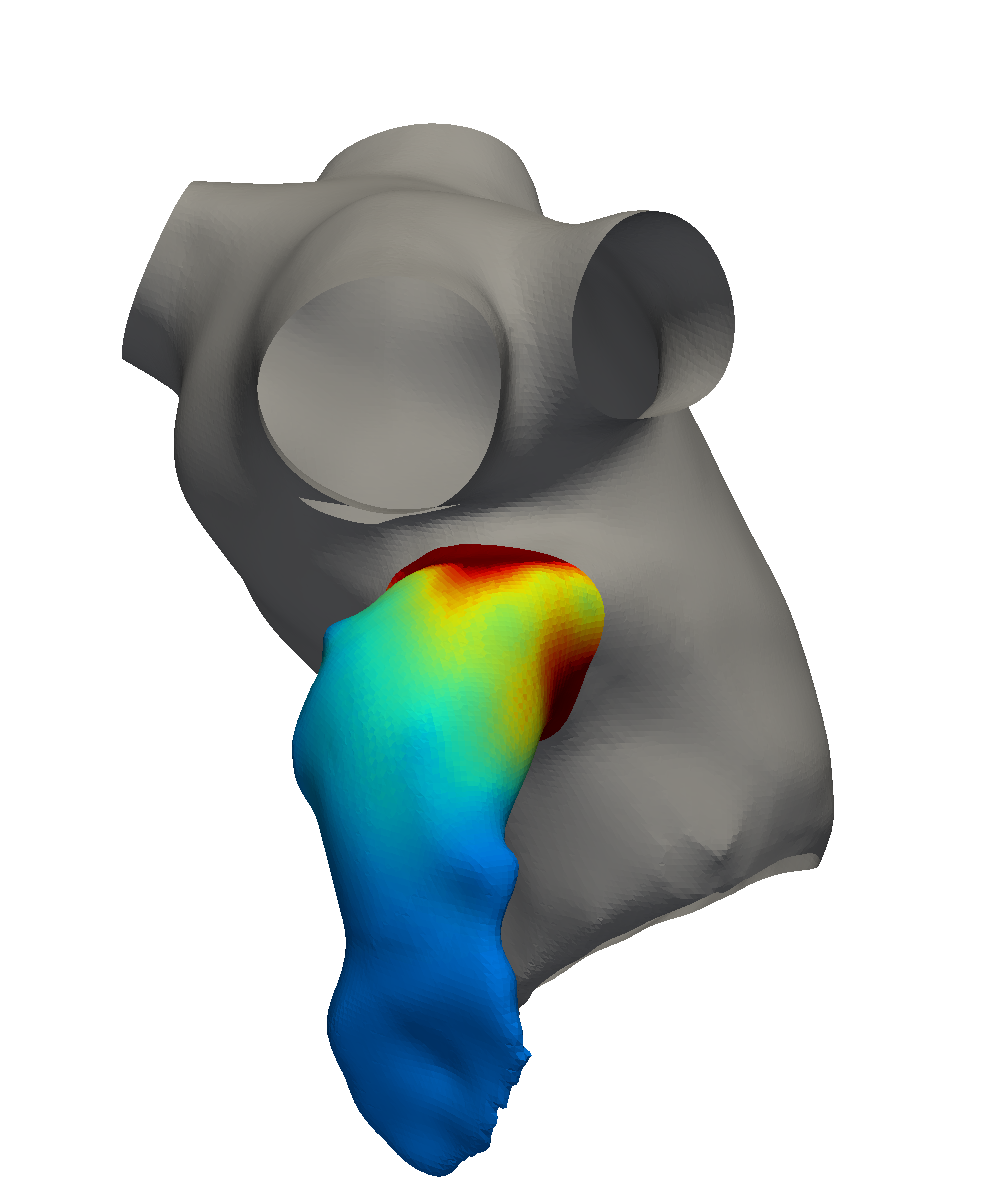}}\,
    \subfloat[$\tawss^\rb$]{\includegraphics[trim={100 10 100 10}, width=0.17\columnwidth]{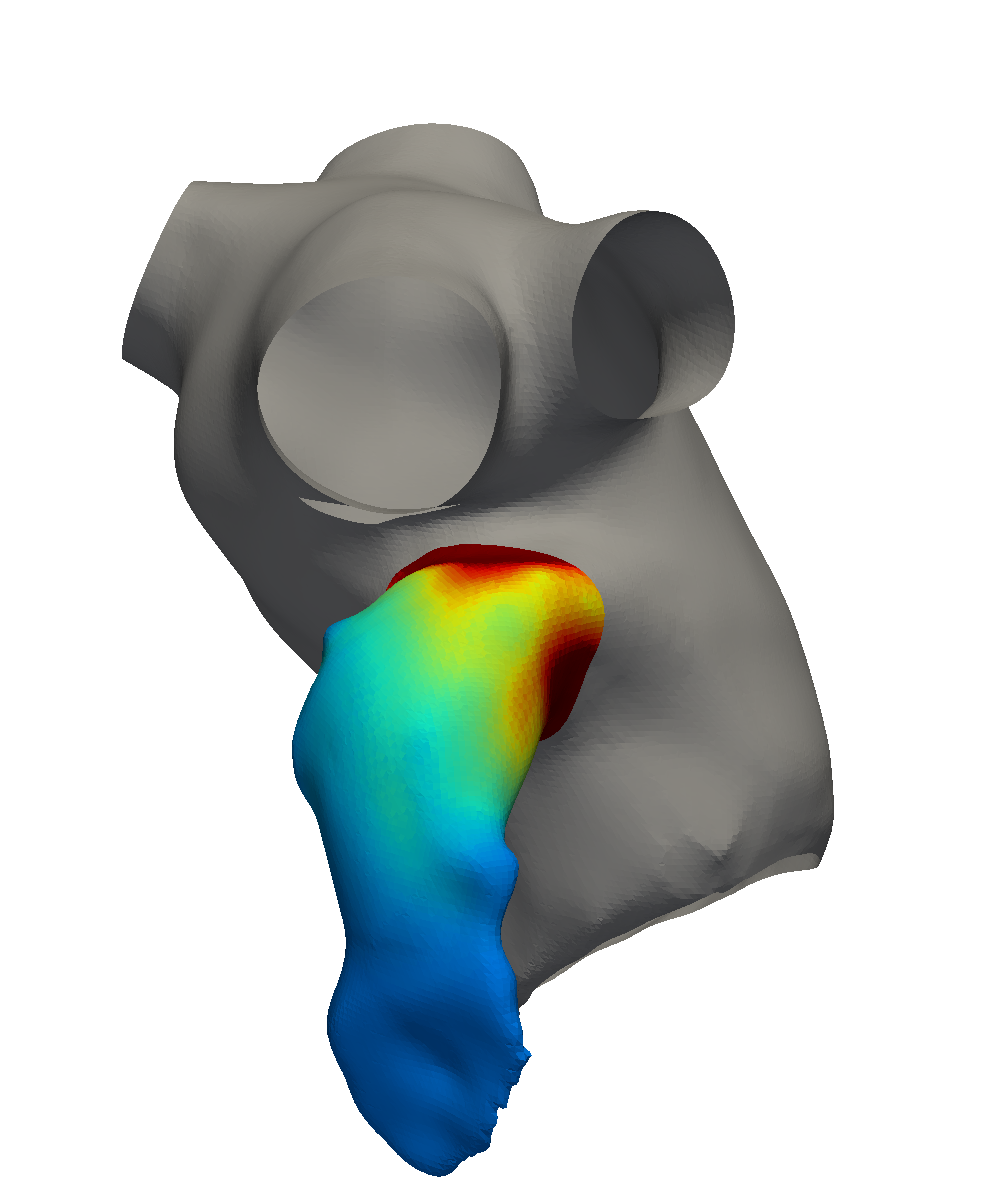}}\,
    \subfloat{\begin{overpic}[abs,unit=1mm,width=0.8cm]{images/TAWSScolorbar.png}\put(7,11){\tiny\textcolor{black!90}{[Pa]}}\end{overpic}}\,
    \subfloat[$\osi$]{\includegraphics[trim={100 10 100 10}, width=0.17\columnwidth]{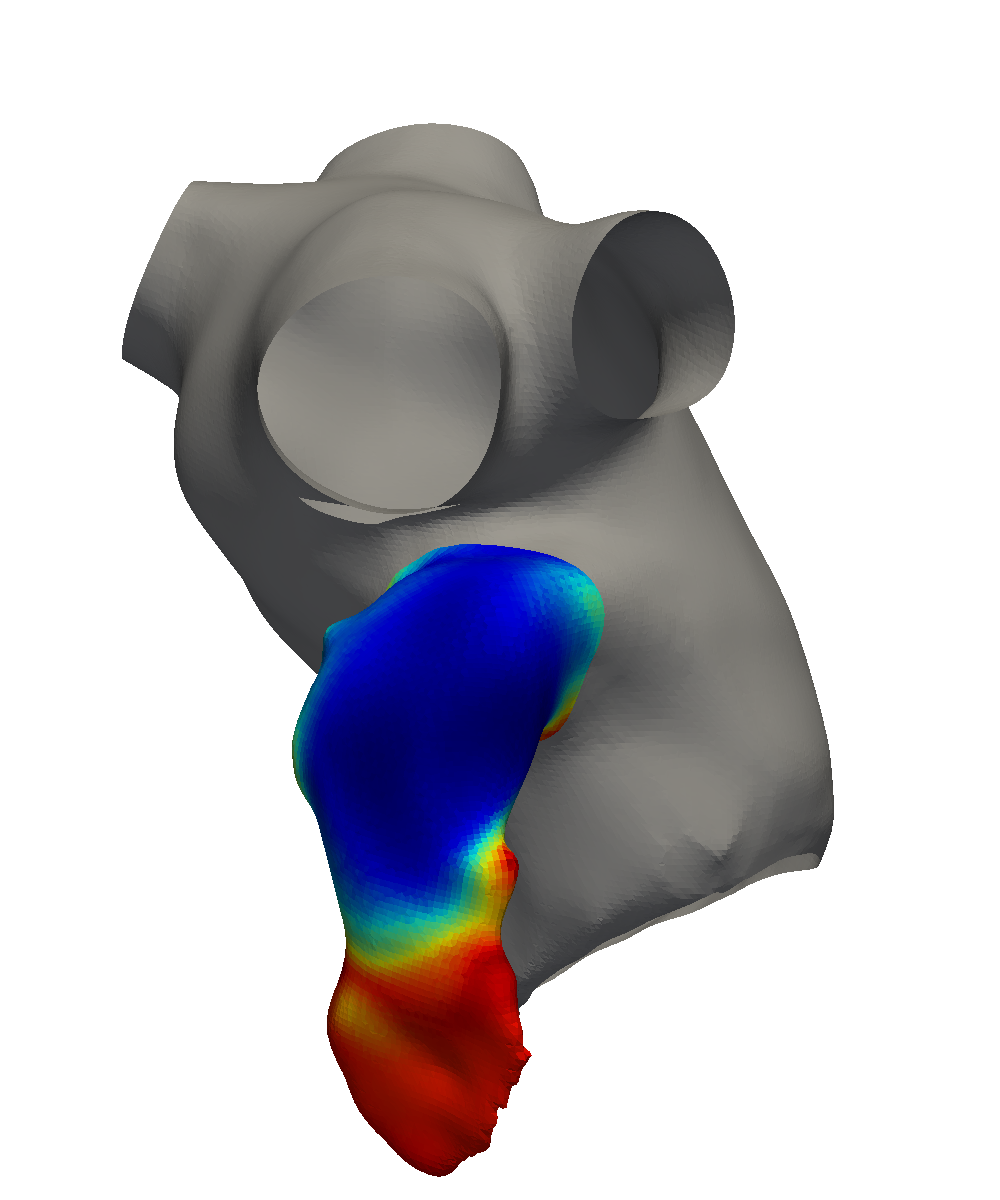}}\,
    \subfloat[$\osi^\rb$]{\includegraphics[trim={100 10 100 10}, width=0.17\columnwidth]{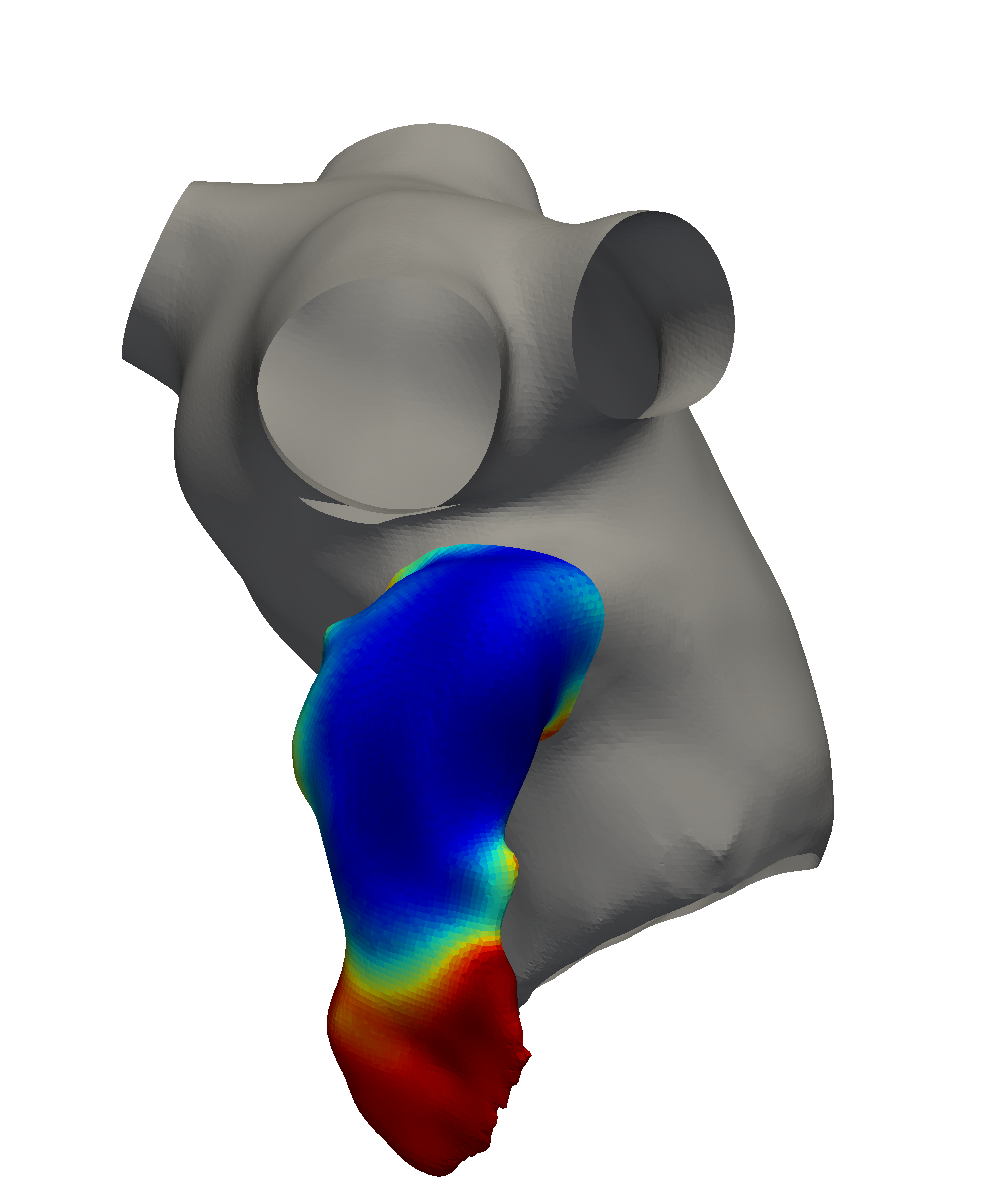}}\,
    \subfloat{\includegraphics[width=0.85cm]{images/OSIcolorbar.png}}\,
    \caption{Casson's case: qualitative comparison between FOM and ROM solutions for $\tawss$ (panels A and B) and $\osi$ (panels D and E).}
    \label{fig:qualitativeOSI}
\end{figure}

In Figures \ref{fig:qualitativeNm1} -- \ref{fig:qualitativeOSI} we show some qualitative comparisons between the FOM and ROM solutions for the Newtonian and Casson's cases. The plots refer to the test parameters values $\mubold^N_3$ and $\mubold^C_{1}$. Figures \ref{fig:qualitativeNm1}, \ref{fig:qualitativeNm2} and \ref{fig:qualitativeNw} (\ref{fig:qualitativeCm1}, \ref{fig:qualitativeCm2} and \ref{fig:qualitativeCw}) show the results obtained for the variables $m_1$, $m_2$ and $\varphi$, respectively, for the Newtonian (Casson) case. %Similarly, Figures \ref{fig:qualitativeCm1}, \ref{fig:qualitativeCm2} and \ref{fig:qualitativeCw} refer to the Casson model. 
For each variable, the FOM simulations (top plots) are very similar to the ROM ones (bottom plots) as expected by the accuracy analysis based on the relative error carried out above. Furthermore, by looking at the patterns obtained, some speculations of clinical interest can be made. Both models reveal an higher momentum $m_1$ in the terminal region of the LAA. Also the washout $\varphi$ and $m_2$ are characterized by a larger magnitude on the tip of the appendage. Therefore we can argue that a higher residence time at the end of the LAA is shown. Note that, although in the Newtonian case a slightly greater variation between the base of the appendage and its tip is shown for all the variables under consideration (compare Figures \ref{fig:qualitativeNm1} and \ref{fig:qualitativeCm1}, Figures \ref{fig:qualitativeNm2} and \ref{fig:qualitativeCm2}, Figures \ref{fig:qualitativeNw} and \ref{fig:qualitativeCw}), it would seem that for this study case the introduction of Casson's model is not justified.
%However, both models reveal an higher momentum $m_1$ in the terminal region of the LAA and, even if some differences can be observed, they are not so much evident to justify the introduction of Casson's model for the study of the blood flow. Also the washout $\varphi$ and $m_2$ reveal an higher residence time on the tip of the appendage. Therefore, with good approximation, we can argue that the blood age distribution has a higher residence time %is higher 
%at the end of the LAA. 
Finally, in agreement with the trend of the relative errors, also the qualitative comparisons of the Casson's case reveal a worse reconstruction of the reduced solution at $t=0$. % and $t=1$ (the boundary of the time interval). 

%m1

%m2

%washout

For what concerns time independent indices, in Figure \ref{fig:qualitativeTAWSS} and \ref{fig:qualitativeOSI} we see that the TAWSS distribution is very well reconstructed by ROM. As expected, it shows greater values where the blood age is lower and the computations provided by Newton and Casson's models are very similar. %if the Newtonian or the Casson's model is adopted. 
However, a more pronounced difference between the two rheologies can be found in the OSI distribution. %This marked discrepancy could disprove the possibility of using the Newtonian model at all. 

%These results are not enough to justify the use of the Casson's model in blood modeling, but they certainly highlight the presence of some differences. 
However, deeper investigations should be conducted, also in direct contact with medical doctors, to establish whether the differences shown by Netwon and Casson's models may be substantial or negligible.

\subsection{Computational cost} In Table \ref{tab:times} we report the computational time taken by the FOM and the ROM. In this case we refer to the entire domain (and not only to the LAA region) in order to provide a fair comparison. All the simulations have been run on the SISSA HPC cluster Ulysses (200 TFLOPS, 2TB RAM, 7000 cores), the FOM ones in parallel using 16 processors while the ROM ones by using one processor only. Each FOM simulation takes roughly 
3.75 hours in terms of wall time, or 60 hours in terms of total CPU time, %to simulate the LA blood flows, 
while the online phase only needs a few seconds. %to approximate the solution, 
Thus the speed-up is of the order of $10^5$ and the ROM is able to practically work in a real-time way. In Table  \ref{tab:times} we also report the estimation of the time required for the computation of the POD basis and the RBF interpolation for sake of completeness. Such results are very promising and could push toward the transfer of ROM techniques
in hospitals and surgery rooms by means of the development of user-friendly digital platforms to be accessed with portable devices, such as a smartphone or a tablet. %to be easy used with portable devices, such as a smartphone or a tablet. 
In this context, we have designed the ATLAS project that allows computations to be run from standard 
web browsers. Further details could be found in \cite{girfoglio2021non}.

% for both Newtonian and Casson's model.

 %\textcolor{red}{Michele: some further comments about the comparison between Newtonian and non Newtonian case. It could be interesting to show not only the qualitative comparison but also the pointwise absolute error}
 
\begin{table}[h!]
    \caption{CPU time taken by the offline and the online phases related to the whole computational domain.}
    \label{tab:times}
    \centering
    \begin{tabular}{c|c|c|c|c|c}\toprule
        \multirow{2}{*}{\bf Model} & \multicolumn{3}{c|}{\bf Offline time} & \bf Online & \bf \multirow{2}{*}{\bf Speed-up} \\ 
        & \bf FOM & \bf POD & \bf RBF & \bf time & \\\midrule
        %Newtonian & 3 h? & 150\,s & 3.6e--03\,s & 1e--01\,s &\\  %APPENDICE
        %Casson & 3 h? & 369\,s & 4.8e--03\,s & 1e--01\,s &       %APPENDICE
        Newtonian & 60\,h & 35\,min & 11\,s & 7\,s & 1e+05\\
        Casson's & 60\,h & 60\,min & 3\,s & 9\,s & 1e+05
        \\ \bottomrule
    \end{tabular}
\end{table}
% \begin{table}[h!]
%     \centering
%     \begin{tabular}{c|c|c|c|c|c|c}\toprule
%         \multirow{2}{*}{\bf Model} & \multirow{2}{*}{\bf Variable} & \multirow{2}{*}{\bf \#modes} & \multicolumn{3}{c|}{\bf Time offline} & \multirow{2}{*}{\bf Time online} \\ 
%         & & & \bf FOM & \bf POD & $\boldsymbol{\knr}$ &\\\midrule
%         \multirow{3}{*}{Newtonian} & $m_1$ & 1 & ? & 50\,s & 1.2e--03\,s & 4.2e--02\,s\\
%         & $m_2$ & 2 &? & 50\,s & 1.2e--03\,s & 4.6e--02\,s\\
%         & $w$ & 5 & ? &  50\,s & 1.2e--03\,s & 4.9e--02\,s\\\midrule
%         \multirow{3}{*}{Casson} & $m_1$ & 1 & ? & 123\,s & 1.6e--03\,s & 4.4e--02\,s\\
%         & $m_2$ & 1 &? & 123\,s & 1.6e--03\,s & 4.4e--02\,s\\
%         & $w$ & 2 & ? & 123\,s & 1.6e--03\,s & 4.6e--02\,s
%         %\begin{rotate}{90}Casson\end{rotate} & $m_1$ & ? & 123\,s & 1e--03\,s & 4e--02\,s
%         \\ \bottomrule
%     \end{tabular}
%     \caption{Technical details of the ROM simulations: number of modes and computational times.}
%     \label{tab:times}
% \end{table}
\section{Conclusions and perspectives}
\label{sec:conclusions}
\addcontentsline{toc}{chapter}{Conclusions}
A data-driven ROM based on POD-RBF technique is adopted in this work for the analysis of the blood flow in a patient-specific domain of LA when AF occurs. Such  approach extracts a reduced basis space from a proper set of high fidelity solutions via POD and adopts RBF to compute the map between parameter space and reduced coefficients. The Newtonian and the Casson's models for the rheology of the blood are employed and compared. We consider a physical parametric framework involving  the scaling factor of the cardiac output (both for Newtonian and Casson's case), the plasma viscosity and the hematocrit (for the Casson's case).

After an expensive offline phase, the POD-RBF approach demonstrated to be able to provide clinically relevant blood flow predictions for the problem at hand at a considerably lower computational cost. This indicates that in perspective such computational tool could be used in hospitals and surgery rooms to support the medical doctors. From a clinical point of view, the results revealed that, unless some differences between the Newtonian model and the Casson's one, the distribution of the mean blood age is higher on the tip of LAA. %quite comparable. 
%Also the TAWSS is not significantly influenced by the blood model. 
%However a remarkable difference between the two models could be found for the OSI concerning the flow direction. 
%This deviation should be better investigated in collaboration with biomedical community to understand deeper its importance and to detect the better model to be employed for realistic simulations. 

%An improvement of considerable relevance for this work could be the exploration of \textcolor{red}{forse piu' hybrid che fully intrusive} intrusive POD-Galerkin ROMs for this type of problems. The aim is to introduce at reduced level some physical information to capture the physiology of the case, by using a lower number of FOM solutions in the offline phase. In addition,
In perspective, an improvement of considerable relevance could be the exploration of the performance of the POD-RBF approach in a geometrical parametric setting by extending to our case study what carried out in \cite{siena2023data}.
We are also going to move towards non-linear ROM methods (see, e.g., \cite{fresca2021pod,milano2002neural,lee2020model}) in order to improve the accuracy as well as the efficiency of our approach.%An idea would be to extend the work 
%carried out in [66–68]. The final goal is to develop a fluid structure interaction problem in order to see if the introduction of this model affects significantly the results. 
%\textcolor{red}{e poi magari possiamo parlare di nonlinear ROM?}

\section{Acknowledgements}
José Sierra-Pallares wants to acknowledge "Movilidad de Investigadores e Investigadoras UVa - Banco Santander 2022" for funding his stay at SISSA Trieste and project number DPI2017-83911-R from Spanish Minitry of Science, Innovation and Universities. Jorge Due\~{n}as-Pamplona wants to acknowledge the "Programa Propio - Universidad Politécnica de Madrid", and the "Ayuda Primeros Proyectos de Investigación ETSII-UPM". We also thank the "Programa de Excelencia para el Profesorado Universitario de la Comunidad de Madrid" for its financial support and the "CeSViMa UPM project" for its computational resources. The acknowledgments are addressed also to the support provided by European Union Funding for Research and Innovation - Horizon 2020 Program - in the framework of European Research Council Executive Agency: H2020 ERC CoG 2015 AROMA-CFD project 681447 "Advanced Reduced Order Methods with Applications in Computational Fluid Dynamics" P.I. Professor Gianluigi Rozza. This work was also supported by the "Gruppo Nazionale per il Calcolo Scientifico" (GNCS - INdAM) and the European Union Funding for Research and Innovation - Horizon Europe Program - in the framework of European Research Council Executive Agency: ERC POC 2022 ARGOS project 101069319 ``Advanced Reduced order modellinG: Online computational web server for complex parametric Systems'' P.I. Professor Gianluigi Rozza. We also thank the PRIN NA FROM-PDEs project.
\appendix
\section{Supporting materials}
\label{appendice}
\subsection{Model validation}
\begin{figure}[h!]
\includegraphics[width=.7\columnwidth]{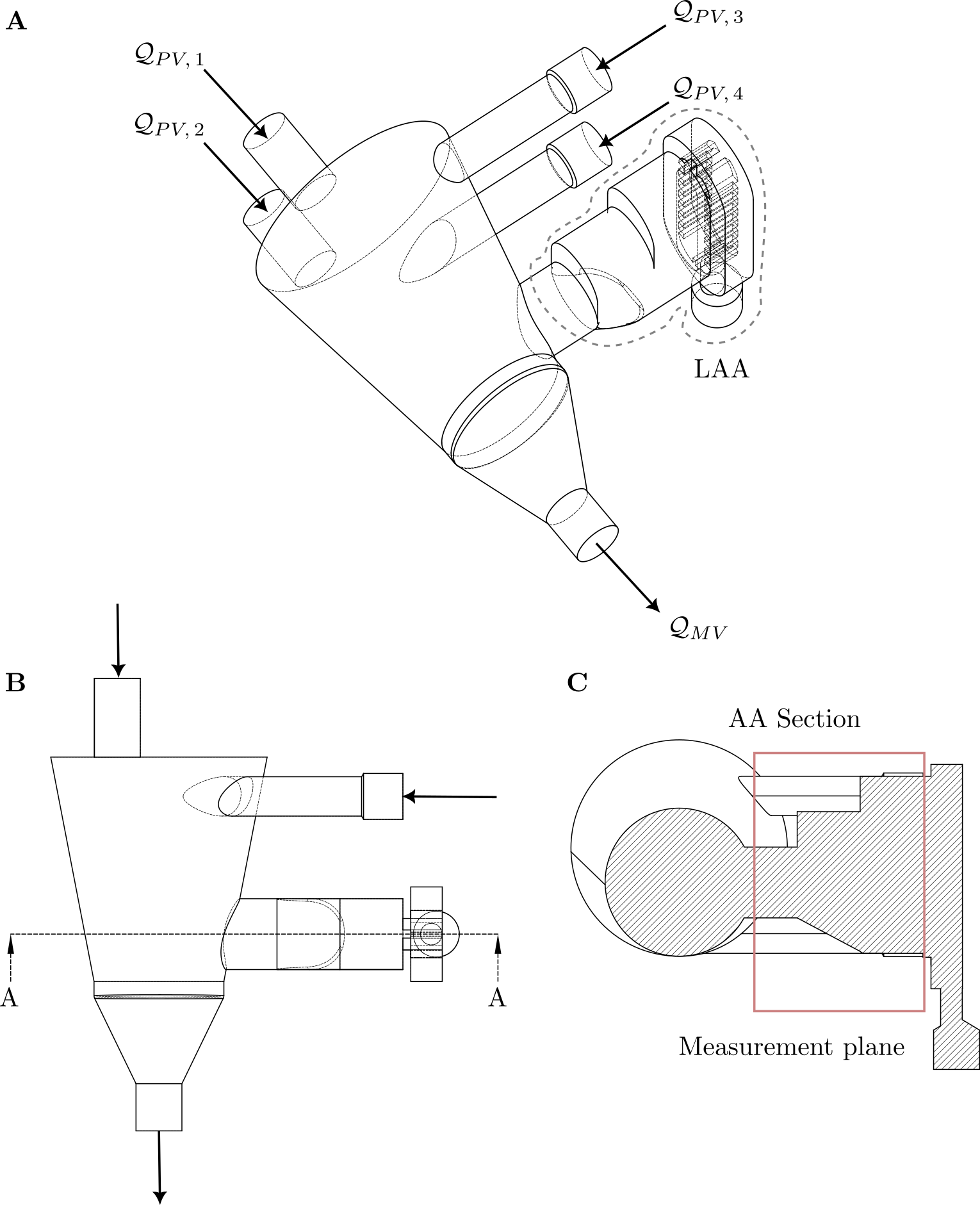}
\caption{Description of validation experiment taken from \cite{DuenasPamplona2021}. \textbf{A}: Isometric view of the in vitro LA with inflows $\mathcal{Q}_{PV,\,i}$, mitral valve outflow $\mathcal{Q}_{MV}$ and LAA volume highlighted; \textbf{B}: Front view of the in vitro LA; \textbf{C}: Detailed view of the measurement section and plane at the in vitro LAA.}
\label{fig:in_vitro_LA}
\end{figure}

\begin{figure}[htbp]
\includegraphics[width=.75\columnwidth]{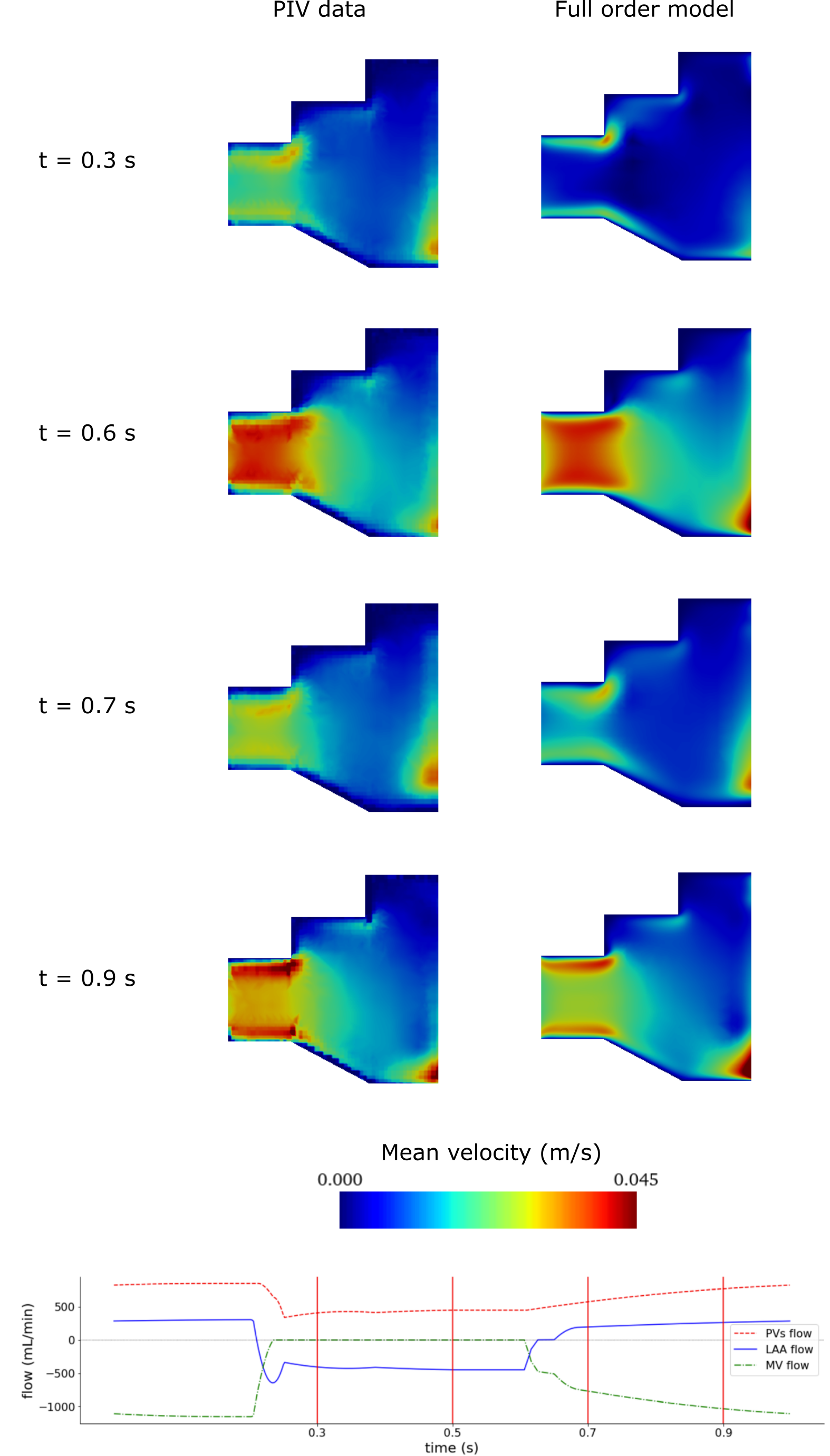}
\caption{Comparison of mean velocity field obtained by experimental measurements (PIV, \cite{DuenasPamplona2021}) Vs. OpenFOAM simulation data (top) and boundary conditions (bottom).}
\label{fig:validation_contours}
\end{figure}

Model validation is a critical step in the development of computational models and simulations, particularly in engineering and scientific research. Model validation involves comparing the model's predictions to experimental or observational data, verifying that the model accurately captures the behavior of the system under study. In this work, the FOM is validated against the experimental benchmark reported by \cite{DuenasPamplona2021}. In that work, an in vitro model mimicking a real left atrium geometry working with a blood mimicking fluid is measured by Particle Image Velocimetry (PIV). Figure \ref{fig:in_vitro_LA} shows the experimental geometry and the measurement plane employed. Figure \ref{fig:validation_contours} shows the experimental measurements of the mean velocity in the in vitro model (top of the figure) along with boundary conditions (bottom of the figure) for mass flow rates and the predictions of our FOM implementation. Mean velocity is well predicted by the FOM, which is in line with the results provided in the benchmark paper \cite{DuenasPamplona2021}. In fact, current OpenFOAM implementation gives almost identical results to those obtained by the authors in that reference. %\gs{Maybe this is too much, and this section can be placed in the supporting materials. We can always say the model is already validated and this is just other implementation of the same model}

\subsection{Mesh convergence analysis}
Mesh convergence analysis is a crucial step in verifying a CFD code, for ensuring that a CFD simulation is properly resolved and that the numerical results are reliable. The Grid Convergence Index (GCI) \cite{roache1998verification} %\gs{maybe citation?} 
is used here since it is a widely accepted method for assessing the mesh convergence of a CFD simulation. The GCI provides an estimate of the order of accuracy and the uncertainty associated with the numerical solution obtained from a given mesh. The GCI is calculated by comparing the solutions obtained from two or more grids with different mesh sizes. 

Table \ref{tab:gci_study} shows the result of a GCI study on four different grids for the patient-specific geometry at hand (see Figure \ref{fig:geometry}), taking a point located at the geometric center of the ostium as sample point. Mean velocity is evaluated in such grid point for four different grid spacing, assuming in all cases a time step of $10^{-3}$s and running the simulation for 5 complete cycles. Results show the solution on Grid 2 is mesh independent, therefore this grid size is used for all the simulations in this study.

\begin{table}[htbp]
\caption{Grid convergence study over 4 grids. $ \bar{\mathbf{v}} $ represents the mean velocity at a point located in the ostium of the LAA and $ \bar{\mathbf{v}}_{\text{extrapolated}} $ its extrapolated value. $ N_{h} $ is the number of grid elements, $ r $ the refinement ration between two successive grids. $\text{GCI}$ is the grid convergence index in percent and its asymptotic value is provided by $ \text{GCI}_{\text{asymptotic}} $, where a value close to unity indicates a grid independent solution. The order achieved in the simulation is given by $ p $ \cite{roache1998verification}.}
\label{tab:gci_study}
\begin{tabular}{@{}l|c|c|c|c|c|c|c@{}}
\toprule
& $ \bar{\mathbf{v}} $ & $ N_{h} $ & $ r $ & $ \text{GCI} $ & $ \text{GCI}_{\text{asymptotic}} $ & $ p $ & $ \bar{\mathbf{v}}_{\text{extrapolated}} $ \\ \midrule
Grid 1 & 4.755e-02 & 1230787 & 1.3 & 7.38\%   & \multirow{3}{*}{ 1.014 } & \multirow{3}{*}{ 1.31 } & \multirow{3}{*}{ 5.04e-02 } \\ 
Grid 2 & 4.63e-02 & 542560  & 1.5 & 10.71\% &                       &                      &                       \\ 
Grid 3 & 4.38e-02 & 174370  & -   & -     &                       &                      &                       \\ \cmidrule(r){1-8} 
Grid 2 & 4.634e-02 & 542560 & 1.5 & 20.83\%   & \multirow{3}{*}{ 1.678 } & \multirow{3}{*}{ 0.75 } & \multirow{3}{*}{ 5.41e-02 } \\ 
Grid 3 & 4.38e-02 & 174370  & 1.3 & 46.49\% &                       &                      &                       \\ 
Grid 4 & 4.68e-02 & 88564  & -   & -     &                       &                      &                       \\ \bottomrule 
\end{tabular}
\end{table}

\bibliographystyle{abbrv}
\bibliography{references}

\begin{thebibliography}{10}

\bibitem{openFoam}
{O}pen{FOAM} {L}ibrary.
\newblock \url{https://openfoam.org/}.

\bibitem{AlSaady1999}
N.~M. Al-Saady, O.~A. Obel, and A.~J. Camm.
\newblock {Left atrial appendage: Structure, function, and role in
  thromboembolism}.
\newblock {\em Heart}, 82(5):547--554, 1999.

\bibitem{atwell2001MCM}
J.~A. Atwell and B.~B. King.
\newblock {Proper orthogonal decomposition for reduced basis feedback
  controllers for parabolic equations}.
\newblock {\em Mathematical and Computer Modelling}, 33(1-3):1--19, 2001.

\bibitem{ballarin2016fast}
F.~Ballarin, E.~Faggiano, S.~Ippolito, A.~Manzoni, A.~Quarteroni, G.~Rozza, and
  R.~Scrofani.
\newblock {Fast simulations of patient-specific haemodynamics of coronary
  artery bypass grafts based on a POD--Galerkin method and a vascular shape
  parametrization}.
\newblock {\em Journal of Computational Physics}, 315:609--628, 2016.

\bibitem{ballarin2017numerical}
F.~Ballarin, E.~Faggiano, A.~Manzoni, A.~Quarteroni, G.~Rozza, S.~Ippolito,
  C.~Antona, and R.~Scrofani.
\newblock {Numerical modeling of hemodynamics scenarios of patient-specific
  coronary artery bypass grafts}.
\newblock {\em Biomechanics and Modeling in Mechanobiology}, 16:1373--1399,
  2017.

\bibitem{balzottidata2022}
C.~Balzotti, P.~Siena, M.~Girfoglio, A.~Quaini, and G.~Rozza.
\newblock {A data-driven reduced order method for parametric optimal blood flow
  control: Application to coronary bypass graft}.
\newblock {\em Communications in Optimization Theory}, 2022(26):1--19, 2022.

\bibitem{Beinart2011}
R.~Beinart, E.~K. Heist, J.~B. Newell, G.~Holmvang, J.~N. Ruskin, and
  M.~Mansour.
\newblock Left atrial appendage dimensions predict the risk of stroke/{TIA} in
  patients with atrial fibrillation.
\newblock {\em Journal of Cardiovascular Electrophysiology}, 22(1):10--15,
  2011.

\bibitem{benim2011simulation}
A.~Benim, A.~Nahavandi, A.~Assmann, D.~Schubert, P.~Feindt, and S.~Suh.
\newblock {Simulation of blood flow in human aorta with emphasis on outlet
  boundary conditions}.
\newblock {\em Applied Mathematical Modelling}, 35(7):3175--3188, 2011.

\bibitem{Benjamin2019}
E.~J. Benjamin, P.~Muntner, A.~Alonso, M.~S. Bittencourt, C.~W. Callaway, A.~P.
  Carson, A.~M. Chamberlain, A.~R. Chang, S.~Cheng, S.~R. Das, et~al.
\newblock Heart disease and stroke statistics—2019 update: a report from the
  american heart association.
\newblock {\em Circulation}, 139(10):e56--e528, 2019.

\bibitem{degruyter1}
P.~Benner, S.~Grivet-Talocia, A.~Quarteroni, G.~Rozza, W.~Schilders, and L.~M.
  Silveira.
\newblock {\em {System- and Data-Driven Methods and Algorithms}}.
\newblock De Gruyter, 2021.

\bibitem{degruyter2}
P.~Benner, W.~Schilders, S.~Grivet-Talocia, A.~Quarteroni, G.~Rozza, and
  L.~Miguel~Silveira.
\newblock {\em {Model Order Reduction: Volume 2: Snapshot-Based Methods and
  Algorithms}}.
\newblock De Gruyter, 2020.

\bibitem{benner2020book}
P.~Benner, W.~Schilders, S.~Grivet-Talocia, A.~Quarteroni, G.~Rozza, and
  L.~Miguel~Silveira.
\newblock {\em Model order reduction: volume 3 applications}.
\newblock De Gruyter, 2020.

\bibitem{Bosi2018}
G.~M. Bosi, A.~Cook, R.~Rai, L.~J. Menezes, S.~Schievano, R.~Torii, and
  G.~Burriesci.
\newblock {Computational Fluid Dynamic Analysis of the Left Atrial Appendage to
  Predict Thrombosis Risk}.
\newblock {\em Frontiers in Cardiovascular Medicine}, 5:1--8, 2018.

\bibitem{buhmann2003book}
M.~D. Buhmann.
\newblock {\em Radial basis functions: theory and implementations}, volume~12
  of {\em Cambridge Monographs on Applied and Computational Mathematics}.
\newblock Cambridge University Press, Cambridge, 2003.

\bibitem{buoso2019reduced}
S.~Buoso, A.~Manzoni, H.~Alkadhi, A.~Plass, A.~Quarteroni, and V.~Kurtcuoglu.
\newblock {Reduced-order modeling of blood flow for noninvasive functional
  evaluation of coronary artery disease}.
\newblock {\em Biomechanics and Modeling in Mechanobiology}, 18:1867--1881,
  2019.

\bibitem{caruso2015computational}
M.~V. Caruso, V.~Gramigna, M.~Rossi, G.~F. Serraino, A.~Renzulli, and
  G.~Fragomeni.
\newblock {A computational fluid dynamics comparison between different outflow
  graft anastomosis locations of Left Ventricular Assist Device (LVAD) in a
  patient-specific aortic model}.
\newblock {\em International Journal for Numerical Methods in Biomedical
  Engineering}, 31(2):e02700, 2015.

\bibitem{chien1967blood}
S.~Chien, S.~Usami, R.~J. Dellenback, M.~I. Gregersen, L.~B. Nanninga, and
  M.~M. Guest.
\newblock {Blood viscosity: influence of erythrocyte aggregation}.
\newblock {\em Science}, 157(3790):829--831, 1967.

\bibitem{Chnafa2014}
C.~Chnafa, S.~Mendez, and F.~Nicoud.
\newblock Image-based large-eddy simulation in a realistic left heart.
\newblock {\em Computers \& Fluids}, 94:173--187, 2014.

\bibitem{demo2018JOSS}
N.~Demo, M.~Tezzele, and G.~Rozza.
\newblock Ezyrb: Easy reduced basis method.
\newblock {\em Journal of Open Source Software}, 3(24):661, 2018.

\bibitem{DiBiase2012}
L.~{Di Biase}, P.~Santangeli, M.~Anselmino, P.~Mohanty, I.~Salvetti, S.~Gili,
  R.~Horton, J.~E. Sanchez, R.~Bai, S.~Mohanty, A.~Pump, M.~{Cereceda Brantes},
  G.~J. Gallinghouse, J.~D. Burkhardt, F.~Cesarani, M.~Scaglione, A.~Natale,
  and F.~Gaita.
\newblock {Does the left atrial appendage morphology correlate with the risk of
  stroke in patients with atrial fibrillation? Results from a multicenter
  study}.
\newblock {\em Journal of the American College of Cardiology}, 60(6):531--538,
  2012.

\bibitem{drapaca2018comparison}
C.~S. Drapaca, Z.~Zhang, and R.~Meng.
\newblock {A Comparison of Constitutive Models of Blood}.
\newblock {\em arXiv preprint arXiv:1808.07977}, 2018.

\bibitem{DuenasPamplona2022}
J.~Due{\~n}as-Pamplona, J.~G. Garc{\'\i}a, F.~Castro, J.~Mu{\~n}oz-Paniagua,
  J.~Goicolea, and J.~Sierra-Pallares.
\newblock Morphing the left atrium geometry: A deeper insight into blood stasis
  within the left atrial appendage.
\newblock {\em Applied Mathematical Modelling}, 108:27--45, 2022.

\bibitem{DuenasPamplona2021a}
J.~Due{\~n}as-Pamplona, J.~G. Garc{\'\i}a, J.~Sierra-Pallares, C.~Ferrera,
  R.~Agujetas, and J.~R. L{\'o}pez-M{\'\i}nguez.
\newblock {A comprehensive comparison of various patient-specific CFD models of
  the left atrium for atrial fibrillation patients}.
\newblock {\em Computers in Biology and Medicine}, 133:104423, 2021.

\bibitem{DuenasPamplona2021}
J.~Due{\~n}as-Pamplona, J.~Sierra-Pallares, J.~Garc{\'\i}a, F.~Castro, and
  J.~Munoz-Paniagua.
\newblock {Boundary-condition analysis of an idealized left atrium model}.
\newblock {\em Annals of Biomedical Engineering}, 49(6):1507--1520, 2021.

\bibitem{eckart1936P}
C.~Eckart and G.~Young.
\newblock The approximation of one matrix by another of lower rank.
\newblock {\em Psychometrika}, 1(3):211--218, 1936.

\bibitem{errill1969rheology}
E.~Errill.
\newblock {Rheology of blood}.
\newblock {\em Physiological Reviews}, 49(4):863--888, 1969.

\bibitem{forti2014IJCFD}
D.~Forti and G.~Rozza.
\newblock {Efficient geometrical parametrisation techniques of interfaces for
  reduced-order modelling: application to fluid-structure interaction coupling
  problems}.
\newblock {\em International Journal of Computational Fluid Dynamics},
  28(3-4):158--169, 2014.

\bibitem{fresca2021pod}
S.~Fresca, A.~Manzoni, L.~Ded{\`e}, and A.~Quarteroni.
\newblock {POD-enhanced deep learning-based reduced order models for the
  real-time simulation of cardiac electrophysiology in the left atrium}.
\newblock {\em Frontiers in Physiology}, 12:1431, 2021.

\bibitem{fung2013biomechanics}
Y.-c. Fung.
\newblock {\em {Biomechanics: mechanical properties of living tissues}}.
\newblock Springer Science \& Business Media, 1993.

\bibitem{GarciaIsla2018}
G.~Garc{\'{i}}a-Isla, A.~L. Olivares, E.~Silva, M.~Nu{\~{n}}ez-Garcia,
  C.~Butakoff, D.~Sanchez-Quintana, H.~{G. Morales}, X.~Freixa, J.~Noailly,
  T.~{De Potter}, and O.~Camara.
\newblock {Sensitivity analysis of geometrical parameters to study
  haemodynamics and thrombus formation in the left atrial appendage}.
\newblock {\em International Journal for Numerical Methods in Biomedical
  Engineering}, 34(8):1--14, 2018.

\bibitem{GarciaVillalba2021}
M.~Garc{\'\i}a-Villalba, L.~Rossini, A.~Gonzalo, D.~Vigneault,
  P.~Martinez-Legazpi, E.~Dur{\'a}n, O.~Flores, J.~Bermejo, E.~McVeigh, A.~M.
  Kahn, et~al.
\newblock {Demonstration of Patient-specific simulations to assess left atrial
  appendage thrombogenesis risk}.
\newblock {\em Frontiers in Physiology}, 12:596596, 2021.

\bibitem{girfoglio2022non}
M.~Girfoglio, F.~Ballarin, G.~Infantino, F.~Nicol{\'o}, A.~Montalto, G.~Rozza,
  R.~Scrofani, M.~Comisso, and F.~Musumeci.
\newblock {Non-intrusive PODI-ROM for patient-specific aortic blood flow in
  presence of a LVAD device}.
\newblock {\em Medical Engineering \& Physics}, 107:103849, 2022.

\bibitem{girfoglio2019finite}
M.~Girfoglio, A.~Quaini, and G.~Rozza.
\newblock {A Finite Volume approximation of the Navier-Stokes equations with
  nonlinear filtering stabilization}.
\newblock {\em Computers \& Fluids}, 187:27--45, 2019.

\bibitem{girfoglio2021non}
M.~Girfoglio, L.~Scandurra, F.~Ballarin, G.~Infantino, F.~Nicolo, A.~Montalto,
  G.~Rozza, R.~Scrofani, M.~Comisso, and F.~Musumeci.
\newblock {Non-intrusive data-driven ROM framework for hemodynamics problems}.
\newblock {\em Acta Mechanica Sinica}, 37:1183--1191, 2021.

\bibitem{Go2001}
A.~S. Go, E.~M. Hylek, K.~A. Phillips, Y.~Chang, L.~E. Henault, J.~V. Selby,
  and D.~E. Singer.
\newblock {Prevalence of Diagnosed Atrial Fibrillation in Adults}.
\newblock {\em Jama}, 285(18):2370, 2001.

\bibitem{Goette2016}
A.~Goette, J.~M. Kalman, L.~Aguinaga, J.~Akar, J.~A. Cabrera, S.~A. Chen, S.~S.
  Chugh, D.~Corradi, A.~D'Avila, D.~Dobrev, et~al.
\newblock {EHRA}/{HRS}/{APHRS}/{SOLAECE} expert consensus on atrial
  cardiomyopathies: definition, characterization, and clinical implication.
\newblock {\em Ep Europace}, 18(10):1455--1490, 2016.

\bibitem{Gonzalo2022}
A.~Gonzalo, M.~Garc{\'\i}a-Villalba, L.~Rossini, E.~Dur{\'a}n, D.~Vigneault,
  P.~Mart{\'\i}nez-Legazpi, O.~Flores, J.~Bermejo, E.~McVeigh, A.~M. Kahn,
  et~al.
\newblock {Non-Newtonian blood rheology impacts left atrial stasis in
  patient-specific simulations}.
\newblock {\em International Journal for Numerical Methods in Biomedical
  Engineering}, 38(6):e3597, 2022.

\bibitem{gunzburger2002SIAM}
M.~D. Gunzburger.
\newblock {\em Perspectives in flow control and optimization}.
\newblock SIAM, 2002.

\bibitem{hesthaven2016book}
J.~S. Hesthaven, G.~Rozza, B.~Stamm, et~al.
\newblock {\em {Certified reduced basis methods for parametrized partial
  differential equations}}, volume 590.
\newblock Springer, 2016.

\bibitem{john2007techniques}
V.~John, I.~Angelov, A.~{\"O}nc{\"u}l, and D.~Th{\'e}venin.
\newblock Techniques for the reconstruction of a distribution from a finite
  number of its moments.
\newblock {\em Chemical Engineering Science}, 62(11):2890--2904, 2007.

\bibitem{johnston2004non}
B.~M. Johnston, P.~R. Johnston, S.~Corney, and D.~Kilpatrick.
\newblock {Non-Newtonian blood flow in human right coronary arteries: steady
  state simulations}.
\newblock {\em Journal of Biomechanics}, 37(5):709--720, 2004.

\bibitem{Khurram2013}
I.~M. Khurram, J.~Dewire, M.~Mager, F.~Maqbool, S.~L. Zimmerman, V.~Zipunnikov,
  R.~Beinart, J.~E. Marine, D.~D. Spragg, R.~D. Berger, et~al.
\newblock Relationship between left atrial appendage morphology and stroke in
  patients with atrial fibrillation.
\newblock {\em Heart rhythm}, 10(12):1843--1849, 2013.

\bibitem{Korhonen2015}
M.~Korhonen, A.~Muuronen, O.~Arponen, P.~Mustonen, M.~Hedman,
  P.~J{\"a}k{\"a}l{\"a}, R.~Vanninen, and M.~Taina.
\newblock Left atrial appendage morphology in patients with suspected
  cardiogenic stroke without known atrial fibrillation.
\newblock {\em PLoS One}, 10(3):e0118822, 2015.

\bibitem{ku1985pulsatile}
D.~N. Ku, D.~P. Giddens, C.~K. Zarins, and S.~Glagov.
\newblock {Pulsatile flow and atherosclerosis in the human carotid bifurcation.
  Positive correlation between plaque location and low oscillating shear
  stress.}
\newblock {\em Arteriosclerosis: An Official Journal of the American Heart
  Association, Inc.}, 5(3):293--302, 1985.

\bibitem{kunisch2002JNA}
K.~Kunisch and S.~Volkwein.
\newblock {Galerkin proper orthogonal decomposition methods for a general
  equation in fluid dynamics}.
\newblock {\em SIAM Journal on Numerical Analysis}, 40(2):492--515, 2002.

\bibitem{Lantz2019}
J.~Lantz, V.~Gupta, L.~Henriksson, M.~Karlsson, A.~Persson, C.~J.
  Carlh{\"{a}}ll, and T.~Ebbers.
\newblock {Impact of Pulmonary Venous Inflow on Cardiac Flow Simulations:
  Comparison with In Vivo 4D Flow MRI}.
\newblock {\em Annals of Biomedical Engineering}, 47(2):413--424, 2019.

\bibitem{Lee2017}
J.~M. Lee, J.-B. Kim, J.-S. Uhm, H.-N. Pak, M.-H. Lee, and B.~Joung.
\newblock Additional value of left atrial appendage geometry and hemodynamics
  when considering anticoagulation strategy in patients with atrial
  fibrillation with low cha2ds2-vasc scores.
\newblock {\em Heart Rhythm}, 14(9):1297--1301, 2017.

\bibitem{Lee2015}
J.~M. Lee, J.~Seo, J.-S. UHM, Y.~J. Kim, H.-J. LEE, J.-Y. KIM, J.-H. SUNG,
  H.-N. PAK, M.-H. LEE, and B.~Joung.
\newblock Why is left atrial appendage morphology related to strokes? an
  analysis of the flow velocity and orifice size of the left atrial appendage.
\newblock {\em Journal of Cardiovascular Electrophysiology}, 26(9):922--927,
  2015.

\bibitem{lee2020model}
K.~Lee and K.~T. Carlberg.
\newblock {Model reduction of dynamical systems on nonlinear manifolds using
  deep convolutional autoencoders}.
\newblock {\em Journal of Computational Physics}, 404:108973, 2020.

\bibitem{Masci2019}
A.~Masci, L.~Barone, L.~Ded{\`{e}}, M.~Fedele, C.~Tomasi, A.~Quarteroni, and
  C.~Corsi.
\newblock {The impact of left atrium appendage morphology on stroke risk
  assessment in atrial fibrillation: A computational fluid dynamics study}.
\newblock {\em Frontiers in Physiology}, 9:1--11, 2019.

\bibitem{mckay2000comparison}
M.~D. McKay, R.~J. Beckman, and W.~J. Conover.
\newblock {A comparison of three methods for selecting values of input
  variables in the analysis of output from a computer code}.
\newblock {\em Technometrics}, 42(1):55--61, 2000.

\bibitem{milano2002neural}
M.~Milano and P.~Koumoutsakos.
\newblock {Neural network modeling for near wall turbulent flow}.
\newblock {\em Journal of Computational Physics}, 182(1):1--26, 2002.

\bibitem{Nedios2014}
S.~Nedios, J.~Kornej, E.~Koutalas, L.~Bertagnolli, J.~Kosiuk, S.~Rolf, A.~Arya,
  P.~Sommer, D.~Husser, G.~Hindricks, et~al.
\newblock Left atrial appendage morphology and thromboembolic risk after
  catheter ablation for atrial fibrillation.
\newblock {\em Heart Rhythm}, 11(12):2239--2246, 2014.

\bibitem{Otani2016}
T.~Otani, A.~Al-Issa, A.~Pourmorteza, E.~R. McVeigh, S.~Wada, and H.~Ashikaga.
\newblock {A Computational Framework for Personalized Blood Flow Analysis in
  the Human Left Atrium}.
\newblock {\em Annals of Biomedical Engineering}, 44(11):3284--3294, 2016.

\bibitem{Polaczek2019}
M.~Polaczek, P.~Szaro, I.~Baranska, B.~Burakowska, and B.~Ciszek.
\newblock Morphology and morphometry of pulmonary veins and the left atrium in
  multi-slice computed tomography.
\newblock {\em Surgical and Radiologic Anatomy}, 41(7):721--730, 2019.

\bibitem{pons2022joint}
M.~I. Pons, J.~Mill, A.~Fernandez-Quilez, A.~L. Olivares, E.~Silva,
  T.~De~Potter, and O.~Camara.
\newblock Joint analysis of morphological parameters and in silico
  haemodynamics of the left atrial appendage for thrombogenic risk assessment.
\newblock {\em Journal of Interventional Cardiology}, 2022, 2022.

\bibitem{roache1998verification}
P.~J. Roache.
\newblock {\em {Verification and validation in computational science and
  engineering}}, volume 895.
\newblock Hermosa Albuquerque, NM, 1998.

\bibitem{aromabook}
G.~Rozza, G.~Stabile, and F.~Ballarin.
\newblock {\em Advanced reduced order methods and applications in computational
  fluid dynamics}, volume~27 of {\em Computational Science \& Engineering}.
\newblock Society for Industrial and Applied Mathematics (SIAM), Philadelphia,
  PA, [2023] \copyright 2023.

\bibitem{saiz2022unsupervised}
M.~Saiz-Viv{\'o}, J.~Mill, J.~Harrison, G.~Jimenez-P{\'e}rez, B.~Legghe,
  X.~Iriart, H.~Cochet, G.~Piella, M.~Sermesant, and O.~Camara.
\newblock Unsupervised machine learning exploration of morphological and
  haemodynamic indices to predict thrombus formation in the left atrial
  appendage.
\newblock In {\em International Workshop on Statistical Atlases and
  Computational Models of the Heart}, pages 200--210. Springer, 2022.

\bibitem{Seo2016}
J.~H. Seo, T.~Abd, R.~T. George, and R.~Mittal.
\newblock A coupled chemo-fluidic computational model for thrombogenesis in
  infarcted left ventricles.
\newblock {\em American Journal of Physiology-Heart and Circulatory
  Physiology}, 310(11):H1567--H1582, 2016.

\bibitem{siena2023data}
P.~Siena, M.~Girfoglio, F.~Ballarin, and G.~Rozza.
\newblock {Data-driven reduced order modelling for patient-specific
  hemodynamics of coronary artery bypass grafts with physical and geometrical
  parameters}.
\newblock {\em Journal of Scientific Computing}, 94(2):1--30, 2023.

\bibitem{siena2023fast}
P.~Siena, M.~Girfoglio, and G.~Rozza.
\newblock {Fast and accurate numerical simulations for the study of coronary
  artery bypass grafts by artificial neural networks}.
\newblock In {\em Reduced Order Models for the Biomechanics of Living Organs},
  pages 167--183. Elsevier, 2023.

\bibitem{Sierra2017}
J.~Sierra-Pallares, C.~M{\'e}ndez, P.~Garc{\'\i}a-Carrascal, and F.~Castro.
\newblock Spatial distribution of mean age and higher moments of unsteady and
  reactive tracers: Reconstruction of residence time distributions.
\newblock {\em Applied Mathematical Modelling}, 46:312--327, 2017.

\bibitem{Vedula2015}
V.~Vedula, R.~George, L.~Younes, and R.~Mittal.
\newblock {Hemodynamics in the left atrium and its effect on ventricular flow
  patterns}.
\newblock {\em Journal of Biomechanical Engineering}, 137(11):1--8, 2015.

\bibitem{vignali2021fully}
E.~Vignali, E.~Gasparotti, S.~Celi, and S.~Avril.
\newblock {Fully-coupled FSI computational analyses in the ascending thoracic
  aorta using patient-specific conditions and anisotropic material properties}.
\newblock {\em Frontiers in Physiology}, 12:732561, 2021.

\bibitem{volkein2011notes}
S.~Volkwein.
\newblock {\em Model Reduction using Proper Orthogonal Decomposition}.
\newblock Lecture notes, University of Konstanz, 2011.

\bibitem{skala2016IO}
V.~\v{S}kala.
\newblock A practical use of radial basis functions interpolation and
  approximation.
\newblock {\em Investigaci\'{o}n Operacional}, 37(2):137--145, 2016.

\bibitem{willcox2002AIAA}
K.~Willcox and J.~Peraire.
\newblock {Balanced model reduction via the proper orthogonal decomposition}.
\newblock {\em AIAA Journal}, 40(11):2323--2330, 2002.

\bibitem{Wolf1991}
P.~A. Wolf, R.~D. Abbott, and W.~B. Kannel.
\newblock {Atrial fibrillation as an independent risk factor for stroke: The
  framingham study}.
\newblock {\em Stroke}, 22(8):983--988, 1991.

\bibitem{Yaghi2020}
S.~Yaghi, A.~D. Chang, R.~Akiki, S.~Collins, T.~Novack, M.~Hemendinger,
  A.~Schomer, B.~Mac~Grory, S.~Cutting, T.~Burton, et~al.
\newblock The left atrial appendage morphology is associated with embolic
  stroke subtypes using a simple classification system: a proof of concept
  study.
\newblock {\em Journal of Cardiovascular Computed Tomography}, 14(1):27--33,
  2020.

\bibitem{Yamamoto2014}
M.~Yamamoto, Y.~Seo, N.~Kawamatsu, K.~Sato, A.~Sugano, T.~Machino-Ohtsuka,
  R.~Kawamura, H.~Nakajima, M.~Igarashi, Y.~Sekiguchi, et~al.
\newblock Complex left atrial appendage morphology and left atrial appendage
  thrombus formation in patients with atrial fibrillation.
\newblock {\em Circulation: Cardiovascular Imaging}, 7(2):337--343, 2014.

\bibitem{zainib2021reduced}
Z.~Zainib, F.~Ballarin, S.~Fremes, P.~Triverio, L.~Jim{\'e}nez-Juan, and
  G.~Rozza.
\newblock {Reduced order methods for parametric optimal flow control in
  coronary bypass grafts, toward patient-specific data assimilation}.
\newblock {\em International Journal for Numerical Methods in Biomedical
  Engineering}, 37(12):e3367, 2021.

\end{thebibliography}

\end{document}